\documentclass{article}
\usepackage[english]{babel}
\usepackage[utf8]{inputenc}
\usepackage{url}
\usepackage{todonotes}
\usepackage{csquotes}
\usepackage[backend=biber, style=nature]{biblatex}
\usepackage[T1]{fontenc}
\usepackage{microtype}
\usepackage{bbm}
\usepackage{textcomp}
\addto\captionsenglish{%
	\renewcommand{\contentsname}%
	{\empty}%
}
\usepackage{pdfpages}
\usepackage{soul}
\usepackage{subfigure}
\usepackage[margin=3cm]{geometry}
\usepackage{epsfig}
\usetikzlibrary{shapes.symbols,patterns} %
\usepackage{pgfplots}
\usepackage[font=footnotesize]{caption}
\usepackage{makecell}
\usepackage{tikz}
\usetikzlibrary{chains}
\usetikzlibrary{fit}
\usepackage[procnames]{listings}
\usepackage{graphicx}
\usepackage{booktabs}
\usepackage{tabularx, colortbl}
\usepackage{xltabular}
\usepackage{multirow}

\usepackage{amsfonts}
\usepackage{amsmath,amsthm,amssymb,dsfont}
\usepackage{enumerate}
\usepackage{nicefrac}
\usepackage[automark,headsepline]{scrlayer-scrpage}
\allowdisplaybreaks    %

\usepackage{algorithm}
\usepackage{algpseudocode}
\definecolor{keywords}{RGB}{255,0,90}
\definecolor{comments}{RGB}{0,0,113}
\definecolor{red}{RGB}{160,0,0}
\definecolor{green}{RGB}{0,150,0}
\definecolor{light-gray}{gray}{0.95}
\usepackage{titlesec}
\titleformat{\section}{\bfseries\large}{\thesection.}{0.5em}{}
\titleformat*{\subsection}{\bfseries}
\titleformat{\subsubsection}{}{\thesubsubsection}{1em}{\itshape}
\titleformat*{\paragraph}{\bfseries}
\clearpairofpagestyles
\usepackage{booktabs}

\usepackage{hyperref}
\hypersetup{colorlinks=true,citecolor=blue,linkcolor=blue,filecolor=blue,urlcolor=blue,breaklinks=true}
\DeclareCiteCommand{\parencite}[\mkbibparens]
  {\usebibmacro{prenote}}
  {\usebibmacro{citeindex}%
    \printtext[bibhyperref]{\usebibmacro{cite}}}
  {\multicitedelim}
  {\usebibmacro{postnote}}
\usepackage{authblk} %
\floatname{algorithm}{Protocol}

\theoremstyle{plain}

\newcommand*{\E}{\mathbb{E}}

\newcommand*{\ee}{\mathrm{e}}

\newcommand*{\cA}{\mathcal{A}}

\newcommand*{\cC}{\mathcal{C}}
\newcommand*{\cD}{\mathcal{D}}
\newcommand*{\cE}{\mathcal{E}}

\newcommand*{\cL}{\mathcal{L}}
\newcommand*{\cI}{\mathcal{I}}
\newcommand*{\cJ}{\mathcal{J}}

\newcommand*{\cN}{\mathcal{N}}
\newcommand*{\cM}{\mathcal{M}}
\newcommand*{\cO}{\mathcal{O}}

\newcommand*{\cY}{\mathcal{Y}}

\newcommand*{\N}{\mathbb{N}}

\newcommand*{\R}{\mathbb{R}}

\newcommand*{\I}{\mathbb{I}}

\title{Will national renewable costs continue declining?\thanks{\textbf{Acknowledgments:} This work was supported by NSF APTO Grant Award 2403999. For helpful comments and feedback, we thank François Lafond, Benjamin Wagenvoort, Brendon Tankwa, Cameron Hepburn, and seminar and conference participants at the seventeenth annual meeting of the Integrated Assessment Modelling Consortium (IAMC), 2024.}}
\author[1,2]{C. Lennart Baumgärtner}
\author[1,2,3,4]{J. Doyne Farmer}
\affil[1]{Institute for New Economic Thinking at the Oxford Martin School, University of Oxford, Oxford, OX1 3UQ, UK}
\affil[2]{Smith School of Enterprise and the Environment, University of Oxford, Oxford, OX1 3QY, UK}
\affil[3]{Macrocosm Inc, 235 E. 4th St., Brooklyn NY 11218}
\affil[4]{Santa Fe Institute, 1399 Hyde Park Road, Santa Fe NM 87501}
\date{\today}

\begin{document}
\maketitle
\begin{refsection}[library_2.bib]

\section*{Summary}
Since the 1980s, the global average solar photovoltaic (PV) cost has decreased by about two orders of magnitude while the global average onshore wind cost has decreased by about one order of magnitude. But global averages are only part of the story: For both solar and wind, costs vary between countries by about an order of magnitude. We curate a comprehensive database of national costs and build predictive models for future costs by decomposing the levelized cost of electricity (LCOE) into its components at the national level and modeling them separately. For solar we find that since 1990 the module cost and the balance of system (BOS) cost have both declined roughly exponentially at rates of 12\% per year. In contrast, wind turbine cost has declined roughly exponentially by 4\% per year, while BOS cost has not declined at all. This suggests that total global wind cost will approach a floor cost of about 35 USD/MWh, reaching about 43 USD/MWh in 2050, whereas global solar cost will continue to decline exponentially, reaching about 8 USD/MWh in 2050. For solar, about half of the cross-sectional variations in national LCOE costs is due to variations in capacity factors, though capital costs also play an important role. For wind, the cross-sectional variations are split approximately equally between capacity factor, capital costs, and investment costs. National BOS costs revert to global costs with a timescale of about 5-7 years, while variations in other factors are more persistent. We develop a simple model for predicting the deviations from global costs and show that it makes reasonable predictions with predictable errors. Finally, we compare our predictions to recent projections of integrated assessment models, and show that China has only accounted for about $0.05\%$ of the overall decline in solar energy costs.

\section*{Introduction}

The rapid decline of the costs of solar and wind energy over the last sixty years or so has changed views about the feasibility of the energy transformation, injecting a rare note of optimism into the climate debate. In hindsight, it is clear that this could have been predicted much earlier by extrapolating past statistical trends; a few did this, but their voices received limited attention.\supercite{farmer_us_2010, farmer16, way22} Paying more attention to such forecasts would have made it possible to anticipate which technologies were likely to be cheap and plan and invest accordingly.\supercite{creutzig17, bolinger22, way22}

Today, with significantly more granular data available for renewable costs, we can improve on these simple statistical models and investigate if and when future renewable costs may deviate from past trends. A number of studies have investigated granular trends in renewable costs. Examples include more granular assessment of cost declines in solar modules,\supercite{nemet06, kavlak18} balance of system,\supercite{elshurafa18, kothari_drivers_2023} wind turbines,\supercite{elia_wind_2020} operation and maintenance,\supercite{wiser_assessing_2019, riva_iea_2018, steffen20a, raupach-sumiya_regional_2015} or financing.\supercite{steffen18}
Furthermore, increasing attention is being paid to understanding empirical and theoretical drivers of cost declines in renewables, and how these may translate to other technologies.\supercite{wilson_granular_2020, malhotra_accelerating_2020}
Nonetheless, there is an important need for better out-of-sample statistical testing, as there has been a tendency toward overfitting and in hindsight many previous forecasts (that didn't do this) are wrong.\supercite{baumgartner_need_2024}

Here we develop reliable forecasts for solar and wind costs by decomposing national costs into their components and studying their statistical properties in detail. Leveraging much of the previous academic work in this space, we have aggregated a novel dataset of renewable costs in 65 countries for solar PV and 68 countries for onshore wind. The dataset spans time-series data for different national cost components over the past 70 years (though albeit some of these are sparsely populated). Our dataset contains approximately 20,000 data points and is the largest academic dataset currently available. Due to data availability, the dataset is skewed towards developed economies. There are, however, a number of emerging markets represented that do not exhibit statistically different properties to developed counterparts.

National costs of renewables vary widely, so forecasts need to take this into account.\supercite{lutz_economic_2015} As we show here, the deviations across countries can be understood by decomposing the costs of renewables into their components. This turns out to also have implications for global costs; as we will see, extrapolating the revealed trends suggests that wind power will drop in cost more slowly and reach a floor, whereas solar will continue to drop exponentially for some time. This suggests that twenty years from now, solar energy will be substantially cheaper than wind energy.

The standard way to compare the costs of capital-intensive renewable technologies with those of more operationally intensive technologies such as fossil fuels is based on the \textit{levelized cost of electricity} (LCOE), which is defined as the average annualized cost of generating one unit of electricity.\supercite{irena_renewable_2024, bolinger22}
Here we decompose the formula for LCOE into its different components and develop a forecasting model for each component separately. This allows us to forecast national renewable electricity costs and better understand the drivers that determine them.

It is worth mentioning that the LCOE is not without limitations. The LCOE has been criticized for dismissing the intermittency of renewable electricity sources.\supercite{wang_cost_2022, smil_examining_2016, ueckerdt_system_LCOE_2023} Energy systems with large shares of renewables also require sufficient short- and long-term energy storage to provide reliable power in the absence of sunshine or wind.\supercite{way_empirically_2022, creutzig17, luderer_impact_2021} Forecasting LCOE components is therefore helpful in applications where the LCOE is inappropriate.

\section*{Decomposition of the LCOE}
We consider the LCOE $\cL_{j}$ of country $j$. 
Over its lifetime an asset produces electricity, which we measure in kWh, and incurs costs measured here in 2020 US dollars.
The LCOE translates the fixed and ongoing costs for renewable electricity into an effective ongoing cost per unit of electricity produced. 
We follow the standard approach of Bolinger et al.\supercite{bolinger22} and decompose the LCOE of an asset $i$ in country $j$ and construction year $t$ as
\begin{equation}\label{eqn: LCOE definition}
  \cL_{ijt} \equiv \frac{\cI_{ijt} \cdot \cC_{ijt} + \cO_{ijt}}{\cY_{ijt}}\ \text{ (USD / kWh).}
\end{equation}
Note that this equation is algebraically equivalent to the LCOE formula used by IRENA\supercite{irena_renewable_2024} and others that discount future electricity generation and  operational expenditures\supercite{hernandez_analytical_2013}.
As explained in more detail below, the components of $\cL_{ijt}$ are the total investment cost $\cI_{ijt}$ (USD/kW), the capital recovery factor $\cC_{ijt}$ (\%/year), the annual operating and maintenance cost $\cO_{ijt}$ (USD/kW/year), and the capacity factor $\cY_{ijt}$ (kWh/kW/year), also called annual electricity yield. Table \ref{tbl: Notation} in the Supplementary Information summarizes our notation.
While ideally this analysis should be done at the level of individual electricity-generating assets, since data is not available at this level, we are forced to operate at the country level. For convenience, we omit the indices $ijt$ for the rest of this paper. We use the national average components ($\cI, \cC, \cO, \cY$) to forecast the national average LCOE $\cL$. While \eqref{eqn: LCOE definition} does not hold exactly for the national averages, we show in Supplementary Information \ref{app:model_LCOE} that it is a good approximation.

We perform detailed statistical analysis on each component of the LCOE. In the main text, we present key results and omit the statistical details. We refer interested readers to the Supplementary Information, particularly Section \ref{app:model}.

The \textit{investment cost} $\cI$ includes both equipment costs and soft costs, such as installation and permitting costs, that occur before electricity production begins. It is typically measured in cost (USD) per electric capacity (kW). 
The investment cost can be further divided into the equipment cost and the Balance of System cost (BOS). 
The largest single piece of equipment for solar energy is the photovoltaic module, and for wind, it is the turbine; in either case we define the cost as $\cM$. The Balance of System (BOS) cost $\cJ$ aggregates all other equipment and soft costs included in the total investment cost $I$, so that we can decompose the investment cost as 
\begin{equation}\label{eqn:I}
  \cI = \cM + \cJ
\end{equation}
for both solar and wind 

The \textit{capital recovery factor} $\cC$ depends on the assets' lifetime and the weighted average cost of capital (WACC). After an initial investment $\cI$, the asset operator makes annual payments to its investors (who may or may not be the same entity as the asset operator). As a result, investors recover their initial investment plus a premium. $\cI \cdot \cC$ is the annual recovery of the original investment, expressed in USD/kW/year.

The \textit{operating and maintenance cost} $O$ is the annual expense of operating the renewable asset, averaged over the asset's lifetime. For renewables, it is typically measured in USD/kW/year (as opposed to USD/kWh for fossil fuels) since there is no variable cost that depends on fuel consumption.

The \textit{capacity factor} $\cY_{j}$ is the annual electricity generation (kWh) per electric capacity (kW). It accounts for variable environmental characteristics, such as sunshine hours and average wind speed, and also includes asset downtime due to maintenance and other operational constraints. 

For renewables the (overnight) total investment cost $\cI$ is much bigger than the (annual) operating cost $\cO$, typically by about two orders of magnitude. As we show, while the capital recovery factor and the capacity factor drive substantial variations in costs, they are essentially modifiers of the investment and operating costs, so that the LCOE goes to zero when $\cI$ and $\cO$ go to zero, regardless of $\cC$ and $\cY$. 

\begin{figure}
  \centering
  \includegraphics[clip, width=\textwidth]{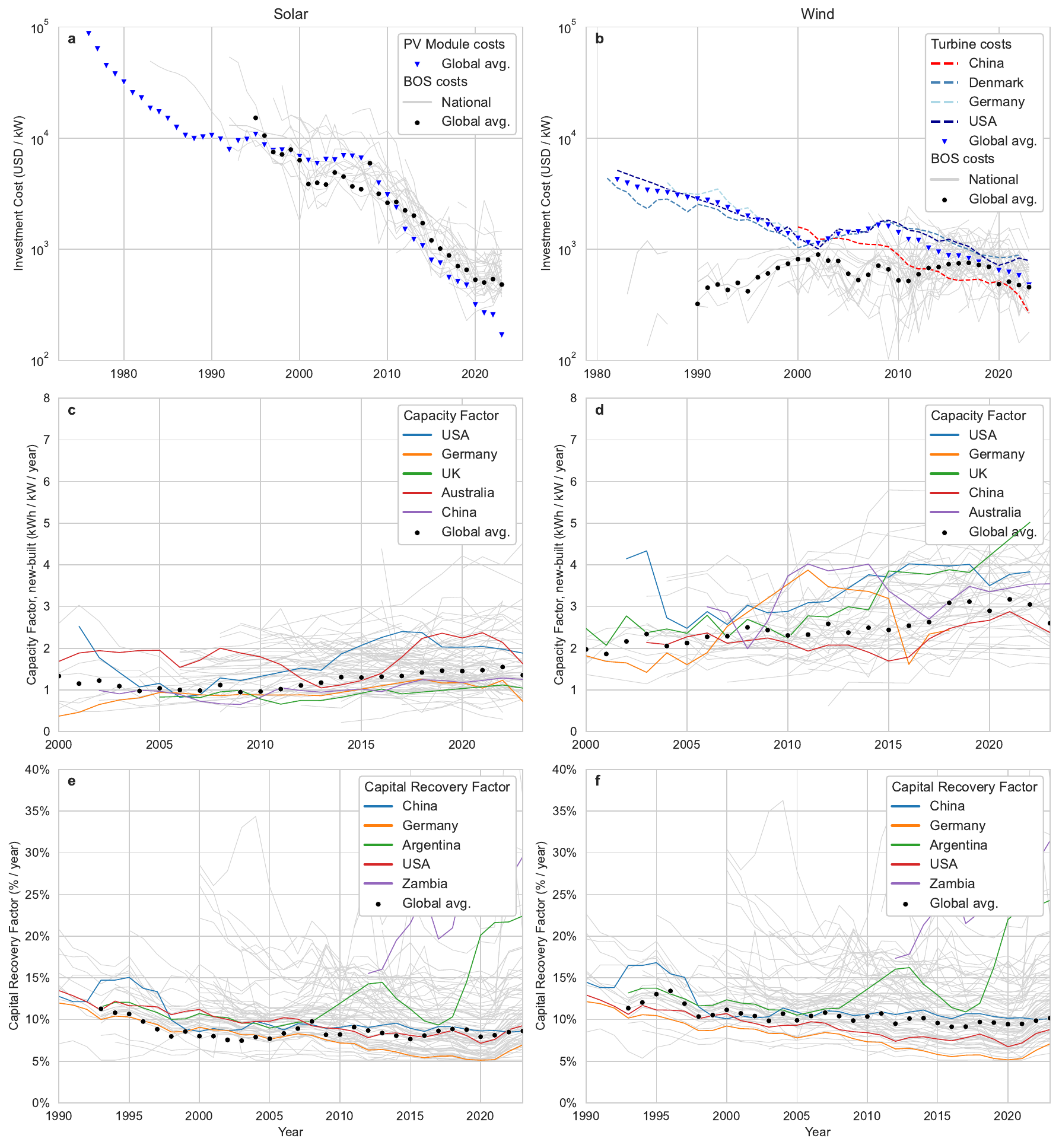}	
  \caption{\textbf{Historical evolution of solar (left) and wind (right) LCOE components.} 
  {\bf (a, b)} Investment cost components over time. Gray lines show national BOS costs. Black dots indicate the global average BOS cost. Blue triangles show the global average PV module costs for solar and turbine costs for wind. For wind, we see that turbines manufactured in China have been cheaper than western-manufactured alternatives after around 2005. All cost data in this paper is inflation- and currency-adjusted to 2020 USD.
  {\bf (c, d)} National capacity factors over time. We observe large variations and temporal fluctuations for national capacity factors driven mostly by external weather factors. Wind has higher capacity factors than solar in most countries. The global average, indicated by black dots, appears to be increasing in both technologies. This is driven primarily by additional countries entering the dataset, not by a long-term increase in national capacity factors.
  {\bf e, f} National capital recovery factors (CRFs) over time. CRFs are very similar for solar and wind since they both depend largely on the national financial market. Changing market conditions over time and between countries cause large fluctuations and cross-sectional variation.  (We neglect $\cO$ because it is two orders of magnitude smaller than the total cost, which is negligible here, and because we were unable to find good data.)
  }
  \label{fig:Components}
\end{figure}

The important components of the LCOE for wind and solar are shown in Figure \ref{fig:Components}, which makes key properties of the components of the LCOE apparent from visual inspection alone. (See also the summary in Table \ref{tbl:Component Models}).

\begin{itemize}
    \item In panel 1(a) we see that solar investment costs have dropped by almost three orders of magnitude since 1975, at an average rate of 12\% per year. Since 2000, when data on BOS costs first became available, the module cost and BOS cost of solar have been comparable, and have dropped at similar rates. National costs vary substantially, with a standard deviation that is currently about 306 USD/kW, which has also dropped roughly exponentially over time.
    \item In panel 1(b) wind investment costs have also dropped, though at a much slower average rate of about 3\% per year. Together with declines in the cost of capital, this is the main driver of LCOE declines. In 1980 wind was more than an order of magnitude cheaper than solar; it is now, on average, slightly more expensive. The standard deviation of wind investment costs is about 437 USD/kW and has not declined. While turbine costs have declined by an average of 4\% per year since 1980, they vary based on origin, with Chinese-built turbines currently about a factor of 3 cheaper than western-built turbines. Most importantly, while BOS costs are highly variable, with a standard deviation of about 400 USD/kW over a span of almost three decades, \textit{average BOS costs have not declined}.\supercite{key_scaling_2022}
    \item  In panels 1(c) and 1(d) the cross-sectional standard deviation of the capacity factors varies by about 0.5 MWh/kW for wind and solar. The average has remained roughly constant for solar and increased slightly for wind. However, it is important to note that capacity factors are not realized values but rather are estimates. Thus trends may reflect changes in weather, unplanned downtime or forecasting methodology rather than real increases in achieved lifetime capacity factors. (If early projects occupy the most favorable sites, one would expect capacity to decrease, but this is the opposite of what is observed). We do not find a statistically significant long-term trend in national capacity factors. See Supplementary Information \ref{app:cf} for details.
    \item In panels 1(e) and 1(f) we see that capital recovery factors vary substantially, with a cross-sectional standard deviation of about 3.4\% for solar and 5.7\% for wind. This variation is driven by bond yields, which are generally higher and more volatile in less developed countries. As a result, capital recovery factors are generally lower for developed countries and higher for less developed countries where investors require higher returns on their investments. Capital recovery factors declined  by about 1\% per year in most countries between the 1990s and 2020, which was a period of declining interest rates, but this has changed since 2022 due to the sharp increase in interest rates, making it clear that this is not a reliable trend. (The high volatility in Argentina's interest rates illustrates this point).
\end{itemize}

\begin{table}
  \begin{centering}
  \footnotesize
  \begin{tabularx}{\textwidth}{>{\raggedright\arraybackslash}p{1.7cm} >{\raggedright\arraybackslash}p{2.2cm} >{\raggedright\arraybackslash}p{3cm} X}
    \textbf{Technology} & \textbf{Component} & \textbf{Cost range} & \textbf{Model} \\
    \midrule[1.5pt]
    Solar & Module cost $\cM$ & 260 USD / kW & Global forecast for costs decline based on Wright's law, conditional on global Solar deployment\\ \midrule[0.2pt]
    Solar & BOS cost $\cJ$ & 300 - 900 USD / kW & Global average BOS cost declines are forecast based on (global) Wright's law and national deviations are forecast using a mean-reverting process\\ \midrule[0.2pt]
    Wind & Turbine cost $\cM$ & 400 - 900 USD / kW & Regional forecast for cost decline in either Western or Chinese turbines, based on Wright's law, conditional on manufacturing location and global wind deployment\\ \midrule[0.2pt]
    Wind & BOS cost $\cJ$ & 200 - 1500 USD / kW & No long-term trend. Both global average BOS costs and national deviations are based on mean-reverting processes\\ \midrule[0.2pt]
    Solar \phantom{PV} Wind & Capital recovery factor $\cC$ & 5 - 20 \%/year & We assume the asset lifetime and the weighted average cost of capital (WACC), which depends on future risk-free rates and technology-specific markups, are given.\\ \midrule[0.2pt]
    Solar \phantom{PV} Wind & Operating \& Maintenance $O$ & $I \cdot 1 \%$/year (Solar),\phantom{P} $I \cdot 3 \%$/year (Wind) & Constant forecast relative to the total investment cost $I$, no long-term trend relative to $I_j$ \\ \midrule[0.2pt]
    Solar \phantom{PV} Wind & Capacity Factor $\cY$ & 10 - 25 \% (Solar),\phantom{PV} 25 - 45 \% (Wind) \hspace{0.5cm} of 8240 hours / year & No long-term trend. We use an average for each country and technology, which depends on geography and the local electricity system. \\
    \midrule[1.5pt]
  \end{tabularx}
  \caption{\textbf{Overview of Solar and Wind component models.} Typical component values are given for the year 2023. We have chosen different models for the components of solar and wind because their historical behavior is very different. For example, future solar BOS costs are forecast using Wright's law and a model for the national deviations, while future wind BOS costs are forecast using a mean-reverting process.  The reasons for these choices are explained in the text.}\label{tbl:Component Models}
  \end{centering}
\end{table}
Table \ref{tbl:Component Models} presents typical values for each component of LCOE and each technology in 2023. We see that BOS costs are the dominant component of the investment cost for solar energy, as much as 3.5 times module costs, whereas for wind, BOS costs and turbine costs are comparable. BOS costs vary cross-sectionally by as much as a factor of 7.5 for wind and a factor of 3 for solar. The capital recovery factor varies by as much as a factor of 4 due to differences in the financial environment. The capacity factor varies by more than a factor of 2 for both solar and wind. In summary, all the factors of LCOE except for solar module cost and operating and maintenance vary widely. 

\section*{Forecasting the components of LCOE}
The fact that the components of renewable technology costs have followed different trends suggests that it is useful to forecast each of them separately and then combine the forecasts. This is useful for global forecasting and helps to understand why national costs vary so much. 
For this purpose it is more convenient to work with logarithms. 

We assume that annual operating and maintenance costs are a constant proportion of the (overnight) total investment costs, $\cO = \gamma \cdot \cI$, where $\gamma \approx 1\%$ for solar and 3\% for wind.\supercite{riva_iea_2018, bolinger_utility-scale_2022, wiser_assessing_2019} This assumption is justified for two reasons. 
First, the cross-sectional variation in $\cO$ is small. As we will see later, we can explain 99\% of the cross-sectional variation in the LCOE without accounting for cross-sectional variation in $\cO$. Second, assuming $\cO = \gamma \cdot \cI$ is consistent with forecasts of $\cO$ made by Steffen et al.\supercite{steffen20a}, who forecast operating and maintenance costs in Germany using a learning curve. As described in the Supplementary Information \ref{app:model_onm}, since operating and maintenance costs have declined at a similar  rate total investment costs (which we forecast with a learning curve), this produces essentially the same results.

Substituting and taking the log of Equation \eqref{eqn: LCOE definition} yields the additive expression
\begin{equation}\label{eqn: LCOE definition_2}
  \log \cL \approx \log \cI + \log(\cC + \gamma) - \log\cY.
\end{equation}
Figure \ref{fig:Components} suggests that these terms behave differently through time and that the behavior of wind differs from that of solar, so that there is value in considering them separately, and in decomposing the investment cost $\cI$ into its components $\cM$ and $\cJ$.

Moore's and Wright's laws provide the two best statistical methods discovered so far to forecast average global costs based on historical data.\supercite{nagy13, farmer16, lafond_forecasting_2025, meng21, lafond_how_2018} Moore's Law predicts that costs will decline approximately exponentially over time, while Wright's Law predicts an approximate power law relationship between costs and cumulative deployment of a technology, where the word ``approximate'' is used to indicate that these relationships are noisy. Moore's and Wright's laws have been tested on about 60 technologies and shown to provide better forecasts than several alternatives \supercite{meng21, farmer16, lafond_how_2018}. Nagy et al.\ \supercite{nagy13} showed that Moore's law and Wright's law forecasts are similar because in all the data sets studied so far, production increases roughly exponentially while costs decline roughly exponentially, which makes Wright's and Moore's laws equivalent. This will likely change once wind and solar become mature technologies and the rapid exponential growth of their deployment slows down. In this case, their forecasts will differ, but with our current state of knowledge, the best method of forecasting remains an open question.\supercite{lafond20}

\begin{figure}
  \centering
  \includegraphics[clip, width=\textwidth]{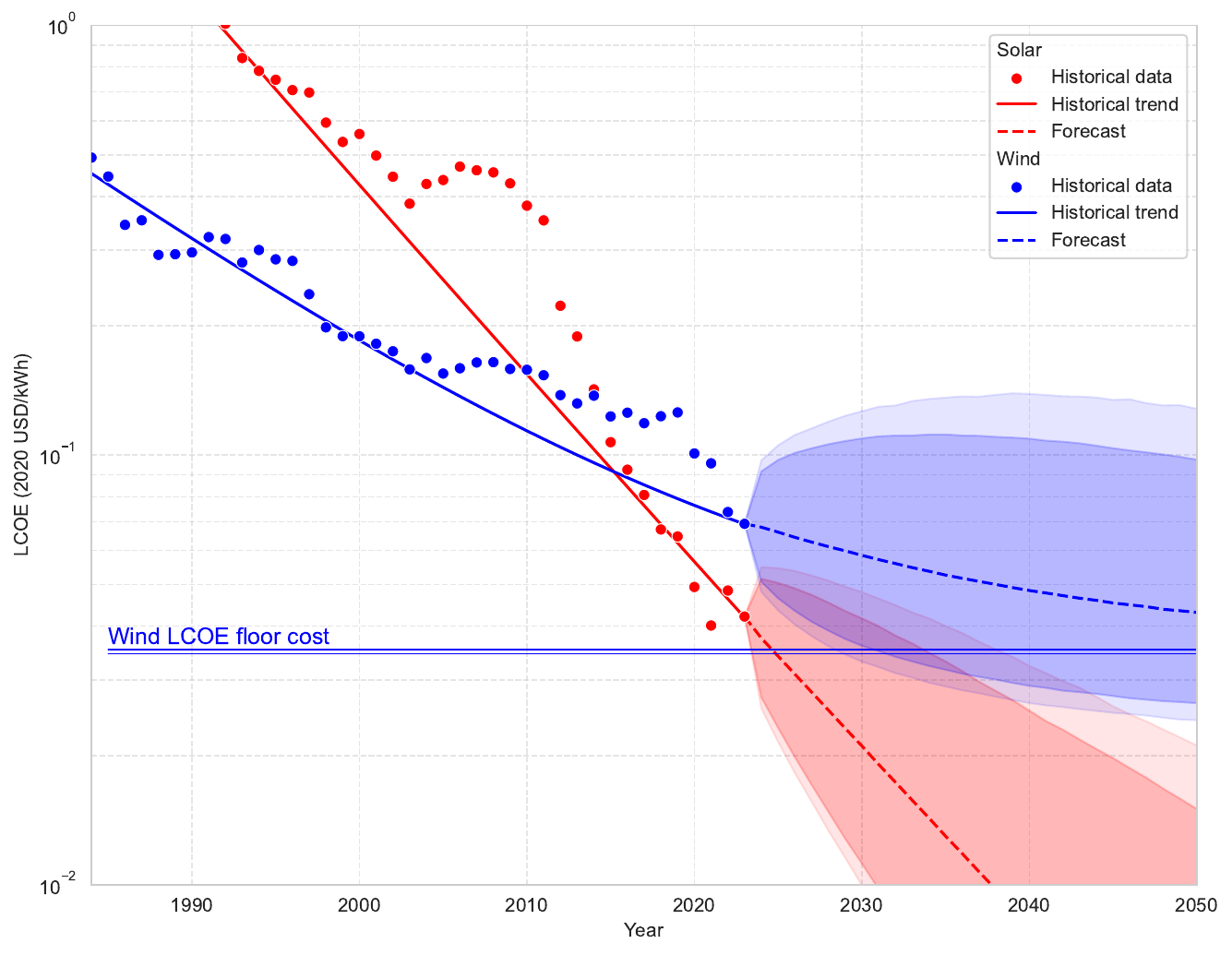}	
  \caption{\textbf{Historical evolution and forecast of global average solar (red) and wind (blue) LCOE.} 
  Dots show the historical evolution of global average solar and wind LCOE. The solid line indicates the historical long-term trend based on decreasing PV modules, wind turbines, and solar BOS costs. The dotted line indicates our long-term forecast for each technology, with shaded areas indicating the 50\% and 90\% confidence intervals. For wind, we forecast a floor cost of around 35 USD/MWh due to the constant long-term trend in BOS costs.
  }
  \label{fig:Forecast}
\end{figure}

If we assume that the trends observed in Figure \ref{fig:Components} persist, then in the future both module costs and BOS costs for solar will drop roughly exponentially according to Moore's law, as they have in the past. In contrast, wind turbine costs will drop roughly exponentially, but its BOS costs will remain roughly constant. This implies that wind will reach a floor cost of about 35 USD/MWh. The resulting forecasts for solar and wind are compared in Figure \ref{fig:Forecast}.
We forecast that solar costs will continue to drop roughly exponentially, reaching a median cost of about 8 USD/MWh in 2050, whereas the rate of decline in overall wind costs slows down, reaching a cost of 43 USD/MWh in 2050, and eventually plateauing at its floor cost of about 35 USD/MWh. Thus, in 2050, solar energy will become about a factor of five cheaper than wind energy. We could also have used Wright's Law, but then we would have needed to postulate a scenario for future production. If we had assumed that deployment continues at current roughly exponential rates, then the forecasts would have been the same as they are under Moore's law. This is unlikely because under current rates of exponential deployment solar and wind would become the sole providers of useful energy around 2036 due to their rapid deployment rates and the large efficient gains in a fully electrified energy system.\supercite{baumgartner_need_2024, way22} 
Thus, before this point is reached, deployment must slow down. In this case the forecasts of the two methods differ, and under Wright's law we expect their costs in 2050 to be somewhat higher than predicted above. As shown in the SI, under plausible decarbonization scenarios from Way et al.,\supercite{way22} the difference has little effect on wind, and increases median solar costs to 12-15 USD/MWh in 2050, so the difference is small compared to the uncertainty in the forecasts.

We want to stress that, although it has become standard in the literature to use Wright's law to forecast technological change, there are strong dissenters in favor of Moore's law;\supercite{funk_rapid_2015} a study of technological change during World War II indicates that stimulating demand causes prices to drop under Wright's law, but it also shows that Wright's law and an overall exponential trend have roughly the same predictive power.\supercite{lafond20} Hence here for simplicity we use Moore's law, and present more detailed evidence using Wright's law in the appendix.

The trends we observe here are plausible for more fundamental reasons. The use of larger turbines and higher hub heights has been a major factor in making wind energy cheaper, next to reduced material need, manufacturing scale, and labor productivity.\supercite{elia_wind_2020}
The lack of decline in BOS costs for wind observed here suggests that as turbines become larger, they also become more difficult to install, offsetting cost reductions from learning.\supercite{key_scaling_2022} Solar photovoltaic panels, in contrast, have maintained a similar size and there are no such confounding factors, so as solar farms become larger, a combination of economies of scale and learning make the installation costs and possibly the other components of BOS costs drop. It is nonetheless remarkable that module costs and BOS costs for solar have decreased at such similar rates. 

Section \ref{app:model} of the  Supplementary Information provides a granular assessment of the statistical robustness of our findings. In particular, we have performed a number of in-sample and out-of-sample statistical tests to validate our forecasts. For solar, a simple learning curve model, such as the one of Way et al.,\supercite{way_empirically_2022} performs similarly to the forecasts presented here; it is also consistent with our findings. For wind, our results are materially different from simpler learning curve models due to the presence of a floor cost for BOS costs. We test this result extensively. In particular, we test our conclusion by removing the latest data points (in a rolling window approach) and comparing our forecasts on the reduced dataset to the latest data. Our results are statistically equivalent to the actual data, giving confidence in the conclusions.
 
To assign errors to these forecasts, we use the method developed by Farmer and Lafond,\supercite{farmer16} which was originally applied to module costs alone but has also been tested in forecasting the costs for many technologies. This method takes into account errors caused by parameter estimation and the noisy fluctuations causing deviations from pure exponential behavior. It has been tested on more than 50 time series and has been shown to do a good job of predicting forecasting errors. It has so far produced good predictions of future renewable costs, with forecasts typically within 95\% confidence intervals.\supercite{way22} Here, we assume that the capacity factor and cost of capital remain constant on average, which is plausible since the capacity factor is largely determined by geography and the cost of capital is largely determined by the financial climate in each country.

Our forecasts can be used to determine the impact of different exogenous financial conditions on local renewable costs. This is particularly important for developing countries, where high financing costs can either hinder renewable adoption or foreign financing mechanisms may be used to accelerate a clean electricity grid.\supercite{lonergan_improving_2023, polzin19, mulugetta_africa_2022, komendantova_-risking_2019}

As a reality check, we regress the components of the logarithm of the LCOE on the right side of Equation \eqref{eqn: LCOE definition_2} against the total LCOE, which is obtained from other sources (though we removed a few obvious outliers). 
The regression explains about 99\% of the variance in the national LCOE data, and the regression coefficients are all within a 95\% confidence interval of one, indicating that the data are fairly consistent (see Section \ref{app:model_LCOE} in the Supplementary Information). 
This further confirms that operating and maintenance costs do not significantly contribute to cross-sectional variation.
The fact that we do not explain 100\% of the variance is because there are some non-zero residuals, which have a heavy-tailed distribution, indicating data errors. Thus, while the data are not perfectly consistent, the errors caused by the inconsistencies are relatively small in relation to the expected forecasting errors.

To provide some insight into the variation in national prices at a given time we study the cross-sectional deviations of the national costs from the global average. It is more convenient to work in terms of logarithms, so we define the variables $\lambda_{jt} = \log \cL_{jt}$, $\iota_{jt} = \log \cI_{jt}$, $\kappa_{jt} = \log(\cC_{jt} + \gamma)$ and $\upsilon_{jt} = \log\cY_{jt}$, which implies that $\lambda_{jt} = \iota_{jt} + \kappa_{jt} + \upsilon_{jt}$. Taking the variance of both sides gives
\begin{equation}\label{eqn:LCOE_var}
  Var(\lambda_{jt}) = Var(\iota_{jt}) + Var(\kappa_{jt}) + Var(\upsilon_{jt}) + \text{cross covariance terms}, 
\end{equation}
or in other words, the variance of the logarithm of the LCOE is the sum of the variance of the log investment costs plus the variance of the log capacity factors plus the variance of the log capital costs, plus cross-covariance terms.
Because the individual terms of the LCOE decomposition depend on very different factors, we expect the cross-covariance terms to be close to zero. The cost of capital is determined by economics and politics, the capacity factor is largely determined by weather, and the investment cost is determined by technology and installation costs. This is confirmed: When we estimate the cross-covariance terms we find that they are all within two standard deviations of zero. We can compute the fractional contributions of each term by dividing each term on the right by $Var( \lambda_{jt})$, see Table \ref{tbl:Variances}. This implies that for solar about 50\% of the variance in cost is due to the capacity factor, 25\% due to investment cost and 25\% due to the cost of capital. In contrast, for wind only 30\% is due to the capacity factor, 30\% is due to investment cost and about 40\% is due to the cost of capital. 

\begin{table}
    \begin{minipage}{\linewidth}
        \centering
        {\it Solar (2022)}\\[.2cm]
        \begin{tabularx}{\textwidth}{>{\raggedright\arraybackslash}p{3.5cm} X >{\raggedright\arraybackslash}p{2.5cm} >{\raggedright\arraybackslash}p{3cm}}
            \toprule
            {\bf Component} & {\bf Cross-sectional mean} & {\bf $\log \cL$ share} & {\bf Variance share}\\ \midrule
            $\cL$ & 97\hspace{.1cm} USD/MWh & (100\%) & (100\%)\\ \midrule
            $\cI$ & 985\hspace{.1cm} USD/kW & 42\% & 25\%\\ 
            $\cY$ & 1.2 \hspace{.1cm} MWh/kW & 43\% & 50\%\\
            $\cC + \gamma$ & 8\% + 1\% & 15\% & 25\%\\ \midrule
        \end{tabularx}
    \end{minipage}
    \\[.5cm]
    \begin{minipage}{\linewidth}
        \centering
        {\it Wind (2022)}\\[.2cm]
        \begin{tabularx}{\textwidth}{>{\raggedright\arraybackslash}p{3.5cm} X >{\raggedright\arraybackslash}p{2.5cm} >{\raggedright\arraybackslash}p{3cm}}
            \toprule
            {\bf Component} & {\bf Cross-sectional mean} & {\bf $\log \cL$ share} & {\bf Variance share}\\ \midrule
            $\cL$ & 112 \hspace{.1cm} USD/MWh & (100\%) & (100\%) \\ \midrule
            $\cI$ & 1440\hspace{.1cm} USD/kW & 43\% & 32\%\\ 
            $\cY$ & 2.1 \hspace{.1cm} MWh/kW & 45\% & 27\%\\
            $\cC + \gamma$ & 9\% + 3\% & 12\% & 41\% \\ \midrule
        \end{tabularx}
    \end{minipage}
    \caption{\textbf{Cross-sectional standard deviation and variance of LCOE and its components.} For the year 2022, we compute the cross-sectional mean and standard deviation of the LCOE and its components (unweighted). We can further use Equation \eqref{eqn:LCOE_var} to estimate the share of the variance that each component contributes to the variance of the logarithm of the LCOE. We leave out the cross-variance terms, contributing less than 10\% for solar and less than 5\% for wind.}
    \label{tbl:Variances}
\end{table}

\section*{Forecasting national costs}

To construct a model for predicting national costs based on historical data, we build on previous work in forecasting global costs \supercite{farmer16, lafond_how_2018}. We use already well-tested methods to make global forecasts and construct new models for forecasting national deviations from global forecasts. For each component of the LCOE, we pool the data for each country and build a single model for all countries. Our approach is in contrast to previous studies that built separate models for each country.\supercite{elshurafa18, helveston_quantifying_2022, huenteler16, kothari_drivers_2023} Building separate models for each country means that ratio of parameters to number of data points is high, which causes overfitting.
To make optimal forecasts one needs to strike the right balance in the bias-variance tradeoff. Bias refers to whether or not a model is capable of fitting the data, and variance refers to errors in parameter estimation, which result in overfitting. In situations like this where data is limited, variance is usually by far the dominant source of error, and it is essential to keep models simple.\supercite{baumgartner_need_2024}

We forecast each component of the LCOE as follows:

\textit{Module costs.} We take advantage of the fact that module costs are essentially global\supercite{  iea_photovoltaic_power_systems_program_pvps_trends_2022} and make use of methods developed in previous studies,\supercite{farmer16, lafond_how_2018} as discussed above. (There are a few cases where costs are not global, such as Japan, which produces and sells a majority of its solar panels domestically,\supercite{raupach-sumiya_regional_2015} but these are exceptions). Since the methods mentioned above have been validated by making out-of-sample forecasts for many different technologies, we do not need to test them here.

\textit{Turbine costs.} Wind turbines are manufactured primarily in the West (US \& Europe) and China, with Chinese turbines now costing around one-third of Western turbines.\supercite{irena_renewable_2024} We can use either the previous global forecasts\supercite{farmer16, lafond_how_2018} for the average cost of Western vs.\ Chinese turbines or differentiate between the two locations and forecast them separately. However, we cannot identify the preferred method due to low statistical significance of any comparison. For simplicity, we apply the global forecasts here.

\textit{Balance of system costs.} Forecasting BOS costs is complicated by the fact that BOS costs behave very differently for solar and wind. As noted, solar BOS costs closely track module costs, whereas wind BOS costs fluctuate, but the global average is roughly flat. This means that we need to take different approaches for solar and wind.

We begin with solar. As seen in Figure \ref{fig:Components}, national solar BOS costs closely track global BOS costs. While the cross-sectional variation between countries is substantial, it is much smaller than the variation in time, which spans almost two orders of magnitude. This makes the national BOS costs highly correlated. To cope with this problem we first take logarithms and then decompose the national costs into the global average and the national deviations from the global average. Since the global average behaves very similarly to the module costs, we use the same methods to forecast it as noted above. 

Because they are essentially stationary, to forecast the national deviations of the BOS costs for solar or wind, or the global BOS costs for wind, we use an AR(1) model of the form \begin{equation}
  x_{j,t+1} = \rho x_{jt} + \sigma n_{jt},
\end{equation}
where $\rho$ is a reversion parameter, with $\rho\approx 0.85$ for the national deviations for solar, $\rho\approx 0.7$ for the national deviations of wind and $\rho\approx 0.5$ for global wind costs. $n_{jt} \sim N(0,1)$ is Gaussian noise with mean zero and standard deviation one and $\sigma$ controls the noise level. For both solar and wind $x_t$ is the logarithmic deviation of the BOS cost at time $t$, i.e. $x_{jt} = \log \cJ_{jt} - \frac{1}{N_t} \sum_{j=1}^{N_t} \log \cJ_{jt}$.
Note that there are no country-specific parameters. This means that all countries revert to the same global mean, which is roughly what we observe.

\textit{Capacity factors} are given mostly by geography, so they can be taken as country-specific constants, corresponding to their historical average for each country. We can either estimate the parameter based on historical electricity generation (subject to noise from annual weather patterns) or ex-ante estimates by project developers (subject to noise and biases from different estimation methods).  
For wind, we estimate the capacity factors based on historical electricity generation since ex-ante estimation baises can be large. For solar, there is little difference between both methods. Here, we use ex-ante estimates by project developers for slightly higher out-of-sample accuracy.
We ignore the rising trend of capacity factors as we believe it is spurious, as demonstrated in the Supplementary Information \ref{app:cf}.

\textit{Capital recovery costs} are determined by the national government bond yields. The best way to treat this depends on the purpose of the forecast. For someone who wants to know hypothetically how costs might change if bond yields change, it is appropriate to make conditional forecasts treating this as an exogenous variable. However, for our purposes here we want to make unconditional forecasts, so we assume bond yields will remain constant at their current national levels (if we had better forecasts we could get rich by trading on them).
Note that previous work on capital recovery costs, particularly Egli et al.,\supercite{egli18} identify a small learning effect in financing costs, contributing 1\% of solar and 4\% of onshore wind cost reductions in Germany between 2000-2005 and 2017. This is not a statistically significant effect for our LCOE forecasts. If we extrapolate the 4\% learning effect over the next 25 years, the net effect would be at most 8\% of our wind forecast (and much less than that for solar). This is well within our statistical uncertainty bands. (See also Supplementary information \ref{subsec:data_crf}.)

\begin{figure}
  \centering
  \includegraphics[clip, width=\textwidth]{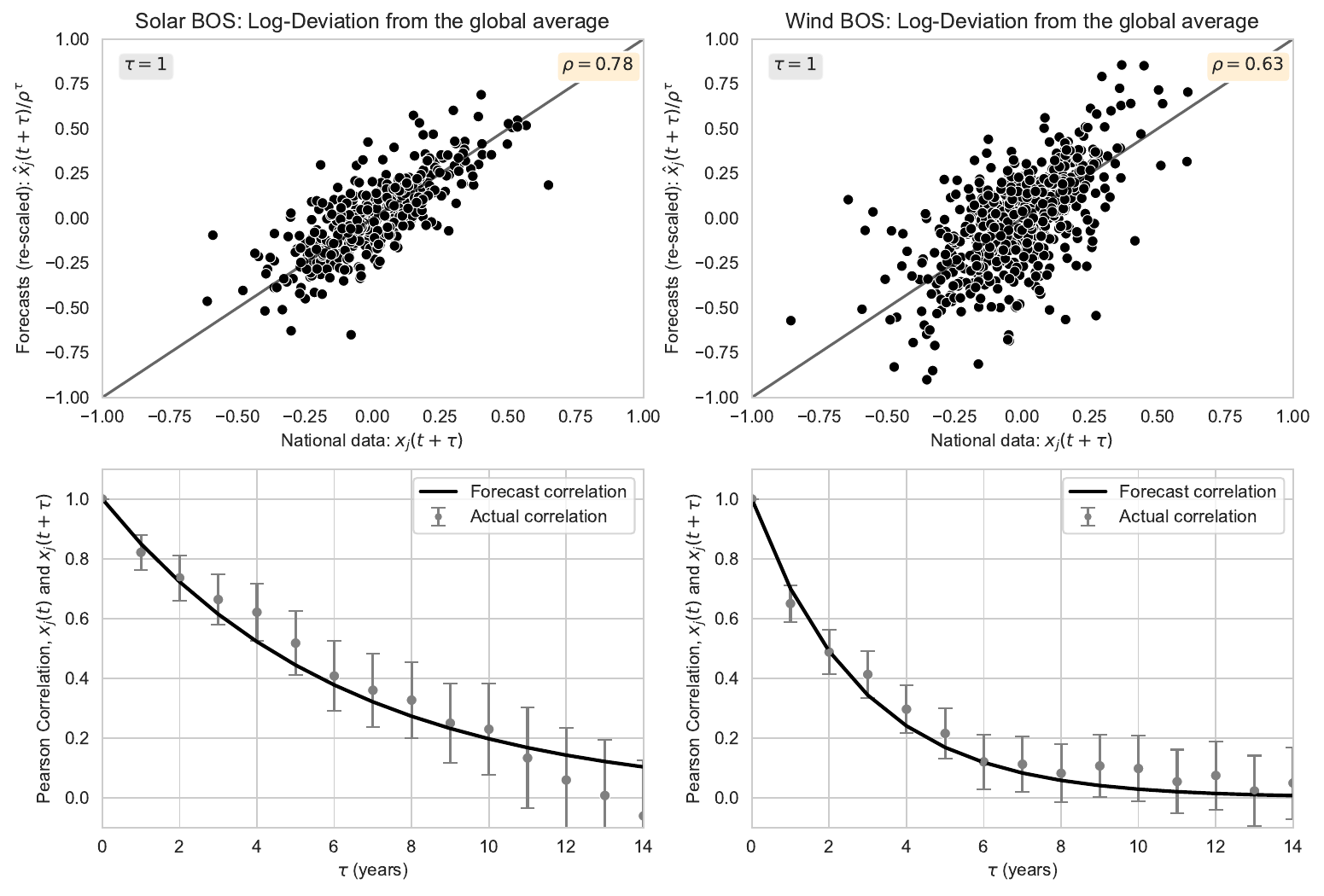}	
  \caption{\textbf{Forecast and actual deviation of national BOS costs from the global average for solar (left) and wind (right).} 
  {\bf Top panels.} X-axis shows the national (log-10) deviations from the global average $x_j$ for times $t + \tau$, with $\tau = 2$. Y-axis shows the forecast for $x_j(t+\tau)$ made at times $\tau$, $\hat{x}_j(t + \tau)$, re-scaled to match the cross-sectional variance of the data. Data and forecast are highly correlated with Pearson correlation coefficients of $\sim 0.78$ for solar and $\sim 0.63$ for wind.
  {\bf Bottom panels.} Correlation between $x_j(t)$ and $x_j(t + \tau)$ with 95\% confidence intervals (dots) and $\hat{x}_j(t+\tau)$ (black line) for different values of $\tau$.
  }
  \label{fig:BOS_Scatter}
\end{figure}

To test our models we divide the data into training sets (also called the in-sample data) and test sets (also called the out-of-sample data). We fit the parameters of our model on the training set and validate the model on the test set. We take 5-year moving windows as training sets (e.g., 2000-2004) and the remaining data in the future relative to this point as the test set (e.g., 2005-2025). We do not attempt to validate the global components of the model, as we have only two very short time series, so any tests would be meaningless, and the global models have already been tested on many technologies in previous work\supercite{lafond_how_2018, way22}.
In Figure \ref{fig:BOS_Scatter} we compare predicted vs. actual values for the deviations of the national BOS costs from the global average. As expected given our use of an AR model, the correlations decay exponentially with a time constant of $1/(1-\rho) = 6.6$ years for solar and $5$ years for wind. This is because national deviations are not persistent -- they revert to the global mean. A plausible interpretation is that this is due to learning spillover effects across countries.

To quantify the reversion to the global mean, we compute the number of degrees of freedom, which is a measure of the effective number of independent variables in the time series (See Supplementary Information \ref{app:model_bos_pv}). If we use the raw BOS costs for solar we find only 5 degrees of freedom. In contrast, the logarithmic deviations from the national average have about 15 degrees of freedom, meaning that tests of forecast performance in the validation set are more statistically significant. This number is still significantly below the 42 countries available for our sample for solar BOS costs due to remaining correlations in the data (for example, we expect neighboring countries to be correlated). The number of degrees of freedom is also subject to substantial measurement error due to gaps in the national data.
The situation for wind is simpler, due to the fact that for wind the national BOS cost time series have about 17 degrees of freedom, and 22 if we subtract the global average -- this is expected given that the BOS costs of wind are flat, so the deviations and the total costs are similar.

\section*{Comparison of our forecasts with Integrated Assessment Model projections}

Integrated Assessment Modeling (IAM) is one of the most prominent methods used to make conditional renewable cost forecasts. IAMs play a dominant role in understanding national and global climate mitigation pathways and are widely used in IPCC reports.\supercite{shukla_climate_2022} We compare our forecasts for total investment costs $I$ to around 570 scenario projections from 13 IAMs, based on the AR6 IPCC report database.\supercite{byers_edward_ar6_2022} We make conditional forecasts for $I$, depending on the global renewable capacity in each scenario in 2040. This allows us to compare our forecasts with the cost projection in each scenario.

\begin{figure}
  \centering
  \includegraphics[clip, width=\textwidth]{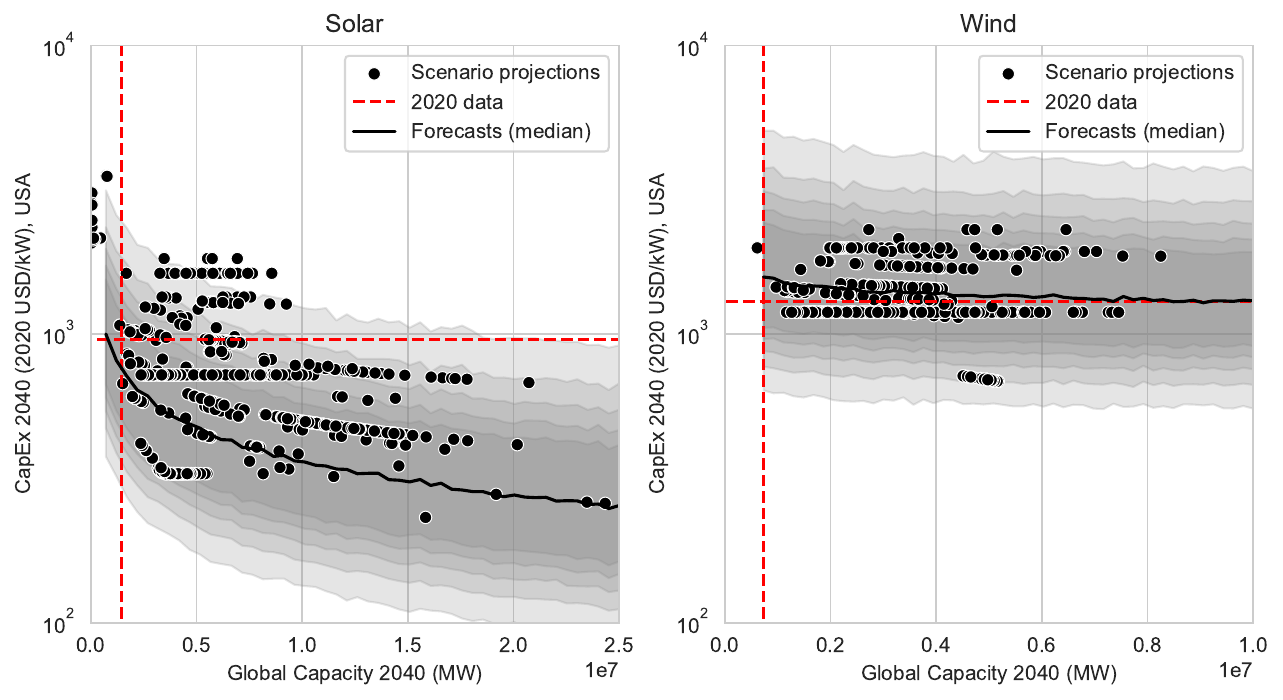}	
  \caption{\textbf{Solar and wind total investment cost projections reported in IAM scenarios against probabilistic projections from our model.} 
  The dots show the total investment cost projections for the US in different IAM scenarios. The x-axis shows the associated global solar / wind capacity. We compare them with the investment cost projections obtained from our model. The gray areas show 50-90\% confidence intervals in 10\% steps. The solid black line shows the median cost from our model. The red dotted lines show the global capacity and investment cost in the US for 2020. For wind, the IAM projections generally align well with the median forecast. For solar, over 85\% of projected IAM costs are above the median costs forecast by our model (and those that are below are only slightly below). In both cases, the probabilistic forecasts show significantly more uncertainty around the median forecasts than the IAM projections represent.
  }
  \label{fig:IAM_proj_USA}
\end{figure}

Figure \ref{fig:IAM_proj_USA} shows the scenario projections as dots, and our probabilistic forecasts in 10\% confidence interval steps.
For solar we see that several scenarios expect less capacity in 2040 than what is operational today, particularly scenarios from IMACLIM 1.1. Secondly, 85\% of all scenarios display costs above our median forecasts, indicating a positive bias in the scenario ensemble.
Some scenarios, such as those of REMIND 2.1 and REMIND-MagPie 2.1 align closely with our 50\% confidence intervals.
There are, however, only very few scenarios (WHICH 5.0, IMACLIM 3.2) have costs within the lower percentiles of our forecasts. About $11\%$ of the ensemble are below the 25\% percentile of our forecast and no model has future costs below the 15\% percentile.

For wind we see that the scenario costs align well with our median forecast. The distribution of scenario costs is approximately symmetric around the median, with 38\% of scenarios projecting costs higher than our forecasts. However, most scenarios are within the 50\% confidence interval of our forecasts, with only $16\%$ of all scenarios having costs below our 25\% percentile. One exception is WHICH 5.0, which shows significantly lower costs than other scenarios.

There are several reasons for this discrepancy between our forecasts and the IAM scenarios. As noted by others,\supercite{farmer15, peters_ar6_2023} scenario ensembles may be biased and frequently under-estimate the uncertainty associated with long-term economic trends. In the Supplementary Information \ref{app:IAM}, we also find that many scenarios overstate the current cost of renewables, particularly solar.  This has significant implications for future cost projections. If we correct for these initial values, we find that our forecast median aligns well with the majority of IAM scenarios. This indicates that any discrepancy between the median forecasts likely stems from outdated input data rather than different model dynamics.

\section*{Discussion}

Our work here shows the value of decomposing costs into their components.  Our most important result is that the BOS costs for solar have dropped in tandem with module costs, while the BOS costs of wind have remained roughly constant. This indicates that if trends continue, wind costs will reach a plateau while solar costs will continue to drop exponentially. A plausible explanation for this is that the physical nature of solar panels has hardly changed, and solar farms are easy to build, while the cost declines in wind turbines are substantially due to using bigger turbines, which makes them harder to install.  

Our second main result is that the difference between national BOS costs tend to revert to the global mean.  This suggests that there is substantial learning transmitted across countries that occurs on a timescale of a few years.

These results also provide new insights for the theoretical literature aiming to understand driving forces between different learning rates in technologies.\supercite{malhotra_accelerating_2020, wilson_granular_2020} In particular, it shows how one technology, such as onshore wind, can be decomposed into multiple sub-technologies, like turbines and BOS costs. These sub-technologies may have different learning rates. As a result the overall technology may exhibit a change in learning rate over time as the relative importance of each factor changes. Wind turbines are standardized technologies with high learning rates from economy of scale effects. In contrast, the constancy of BOS costs suggests that wind installations have been difficult to standardize, making learning more difficult. For solar both modules and their installation has become simpler and highly standardized, yielding sustained high learning rates.

Unlike investment costs or capacity factors, the cost of capital depends on economic factors that are potentially under the control of policymakers in governments or international organizations like the IMF. The cost of capital accounts for about 25\% of the variance for solar and 40\% for wind, so the differences are substantial. This also means that global changes in interest rates, such as the higher interest rates now, can have a substantial effect on cost competitiveness.

As a side note, many people think that China is the main reason that solar energy became so cheap. China has certainly played a role in reducing costs, but the IEA only mentions Chinese solar module manufacturing entering global markets from 2007 onward. While the cost of solar electricity now is roughly 10,000 times cheaper than it was in 1957, it declined by a factor of about 1,000 between 1957 and 2005. This should be compared to its drop by a factor of 10 since 2006 \supercite{iea_photovoltaic_power_systems_program_pvps_trends_2008}. Chinese solar modules are currently about a factor of 6 cheaper than in protected markets such as Japan, i.e. roughly speaking, China accounts for about half of the recent factor of 10 reduction in global prices\supercite{iea_photovoltaic_power_systems_program_pvps_trends_2022}. This indicates that China only accounts for about (10/2)/10,000 = 0.05\% of the total cost decline in solar energy prices.

In the past the projections of integrated assessment models for renewable costs were very poor, substantially overestimating future renewable costs. This is in contrast to time series-based forecasts, which have done quite well.\supercite{way22} The good news is that the projections of integrated assessment models are now much more closely aligned with time series-based forecasts. Nevertheless, some challenges remain, such as keeping scenarios up to date and ensuring that the ensemble of scenarios reflects the uncertainty around future technology costs.\supercite{guivarch_using_2022, peters_ar6_2023}
We hope that the improved forecasts we give here will be useful in honing IAM forecasts in the future.

\section*{Methods}

\subsection*{Data collection and cleaning}

Validating forecasts is essential,\supercite{baumgartner_need_2024}, but achieving statistical significance requires sufficient data. To maximize the available data and develop a widely applicable model, we apply and validate the same methods across a novel sample of 102 countries for solar and 111 countries for onshore wind. Here, we provide a brief overview of the dataset. Additional details are given in Supplementary Information \ref{app:data}.

\begin{figure}
  \centering
  \includegraphics[clip, width=\textwidth]{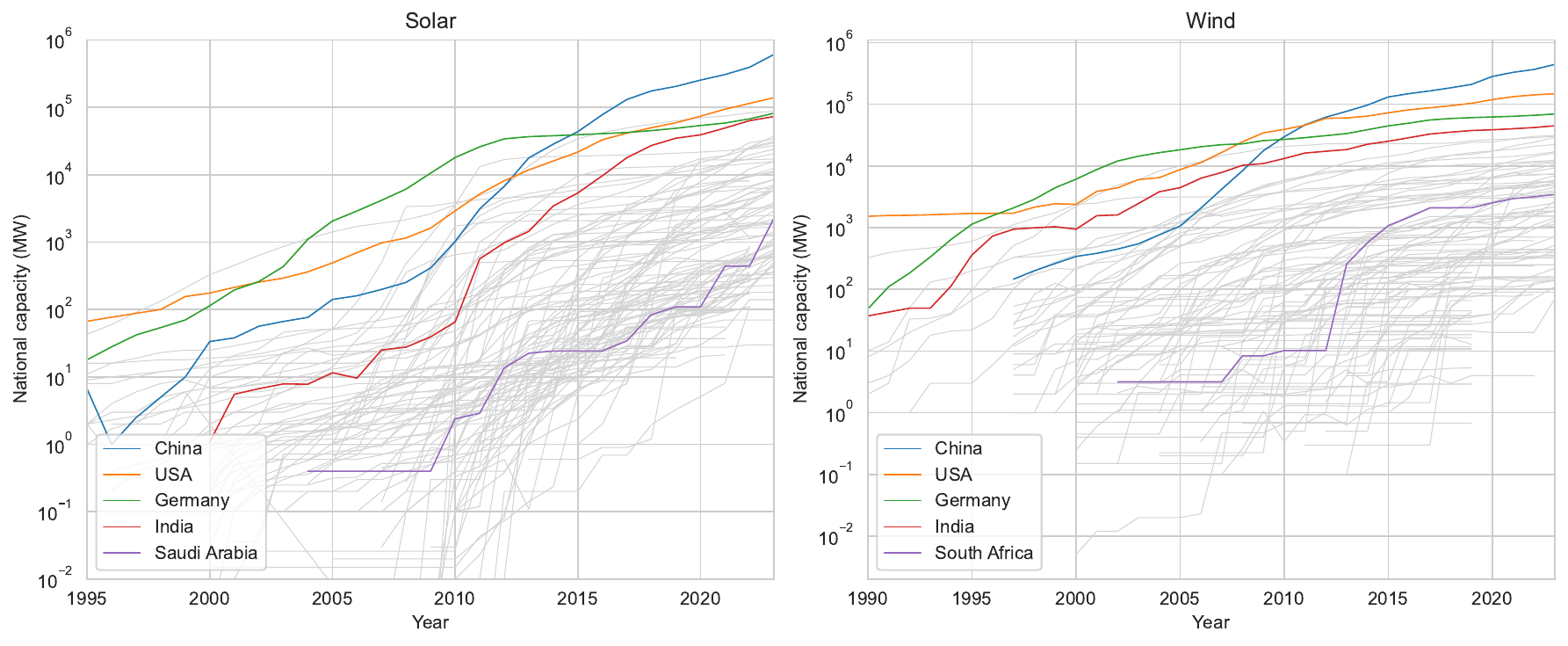}	
  \caption{\textbf{Solar (left) and onshore wind (right) capacities for different countries and years} 
  Over the last 40 years, solar and wind capacities have grown approximately exponentially across the world. While there is significant fluctuation between countries and years, there hat not significant departure from exponential growth so far.\supercite{baumgartner_need_2024}
  }
  \label{fig:capacity}
\end{figure}

We collected and extensively cross-validated national average data for LCOE $\cL_j$, total investment cost $I_j$, cost of capital, electric capacity, and electricity generation by manually scraping the extended scientific and gray literature. Next to the investment cost shown in Figure \ref{fig:Components} and LCOE shown in Figure \ref{fig:Forecast}, Figure \ref{fig:capacity} shows the capacity of installed solar and wind projects for different countries. We see that national capacities have grown by up to 3 orders of magnitude for wind (e.g., South Africa) and 4 orders of magnitude for solar (e.g., Saudi Arabia) in the last two decades alone. So far, there is no statistically significant departure from the exponential trend.\supercite{baumgartner_need_2024}

Due to the diversity of sources, the data is noisy and subject to inconsistent data-cleaning and aggregation procedures. This particularly applies to the LCOE $\cL_j$. For example, data providers sometimes fill data gaps  using ad-hoc assumptions. 
To reduce inconsistencies we perform several corrections on the LCOE data. These corrections are largely modelled on IRENA's data \supercite{irena_renewable_2024} since it is the largest single source of LCOE data in our dataset and arguably the most prominent source of global and country-level data.

{\it Cost of capital.} We harmonize the LCOE data by using a consistent cost of capital across time and countries. For data prior to 2021, IRENA applies simplified cost of capital assumptions. After 2021, they use more granular data.\supercite{irena_cost_2023} We therefore correct the data to obtain a consistently granular cost of capital throughout time and across countries. The impact of this correction on our model is minimal since we take the cost of capital as given.

{\it Capacity factor.} For solar, IRENA's capacity factor closely aligns with other sources. For wind, we find significant discrepancies due to different estimation methods. IRENA largely relies on ex-ante estimates by project developers. These can be materially different from ex-post estimates based on the actual electricity generated (see also Appendix \ref{app:cf}). To harmonize our data, we use the capacity factor based on electricity generation for wind. This correction can be a limitation to our LCOE forecasts, which is why we validate the capacity factor forecasts independently.

{\it Operating and maintenance cost.} While IRENA does not provide detailed information on their operating and maintenance cost prior to 2010, we identify some inconsistencies across the dataset. We harmonize these inconsistencies throughout the data by using our assumptions for the operating and maintenance costs that are based on an independent dataset not used in any other LCOE forecasts. This limits data correction impacts on the LCOE forecasts but does not avoid them completely. However, since the operating and maintenance costs constitute only a small part of the LCOE, the resulting noise is small compared to the variance in other LCOE components.

\subsection*{LCOE decomposition}

Decomposing the LCOE $\cL_{jt}$ into its components is a non-trivial task due to the nonlinearity of equation \eqref{eqn: LCOE definition}. Equations \eqref{eqn: LCOE definition} and \eqref{eqn: LCOE definition_2} hold precisely for individual assets but not for the national average LCOE $\cL_{jt}$. As a result, \eqref{eqn: LCOE definition_2} requires an additional error term to reflect the impact of the national averaging:
\begin{equation}\label{eqn:lcoe_lin}
\log\cL_{jt} = \log\cI_{jt} + \log(\cC_{jt} + \gamma) - \log\cY_{jt} + \log\cE_{jt},
\end{equation}
where $\log\cE_{jt}$ is an additional random variable characterized by an unknown distribution.

We can use our dataset on the LCOE and its components to determine the distribution of $\log\cE_{jt}$. We find that $\log\cE_{jt}$ does not exhibit significant autocorrelation. Furthermore, the sample mean and sample variance of $\log\cE_{jt}$ do not significantly change over time and across countries. Although some countries and years display notably larger values of $\log\cE_{jt}$, their difference is not statistically significant due to the small sample size. Consequently we model $\log\cE_{jt}$ as an independent and identically distributed (i.i.d.) random variable, independent of indices $j$ and $t$.

\subsection*{Solar module and wind turbine component model}
For both solar modules and wind turbines, we employ the experience curve model developed by Lafond et al.\supercite{lafond_how_2018} to forecast global average costs, conditioned on the global cumulative deployment of each respective technology.  For solar modules this approach is straightforward, as the majority of modules are manufactured in China and traded internationally.  For wind turbines Western-manufactured turbines are still widely used, impacting investment costs and the LCOE. Nonetheless, due to data limitations, we cannot determine with statistical significance whether a locally-specific experience curve model would yield improved forecasts compared to the global alternative. Therefore, we adopt the simpler global model.  We do not independently validate these models, as Lafond et al.\supercite{lafond_how_2018} have already extensively validated the approach across a dataset encompassing over 50 technologies.

\subsection*{Balance of system component model}

We use regression analysis to confirm that solar Balance of System (BOS) costs exhibit a statistically significant experience rate, similar in magnitude to that observed for solar modules. In contrast, we find no significant cost reduction or experience exponent for wind BOS components over the past four decades.

We next examine the cross-correlation structure present within our dataset. Both solar and wind national time series demonstrate significant cross-correlations, which are particularly strong for solar, resulting in only nine effective degrees of freedom.  There is also substantial and statistically significant autocorrelation in the data at both global average and national scales. This autocorrelation indicates that an autocorrelated forecasting model is needed for predicting future national deviations from global averages.

For solar, we compare various forecasting models aimed at predicting future national costs. However, due to substantial cross-sectional correlations, our out-of-sample tests do not yield statistically significant differences between models. Nonetheless, drawing upon theoretical insights from the time-series forecasting literature, we identify a global average model with autocorrelated national deviations as the most appropriate method for projecting future solar BOS costs.  For both solar and wind technologies, we formulate the model in terms of national deviations from global averages, which greatly increases the statistical significance of our results, and compare forecast percentiles against a theoretical prior derived from surrogate data.

\subsection*{Capacity factor component model}

There are two methods to estimate the capacity factor $\cY_{jt}$. 
Using our dataset, there are two options to estimate the trend-stationary capacity factor. 1) We can estimate the ex-post capacity factor from historical electric capacity and generation. 2) We can use our data for $\cL_{jt}$, $I_{jt}$, $\cC_{jt}$ and $\gamma$ in Equation \eqref{eqn:lcoe_lin} to recover a priori capacity factors used in LCOE estimates from the literature.
The LCOE-based method yields an ex-ante estimate provided by data suppliers, whereas the electricity-based approach provides an ex-post estimate derived from recent historical generation, serving as a reference for anticipating ex-ante capacity factors.  For solar the differences between the two estimation methods are minor. We apply the ex-ante capacity factor derived from the LCOE-based approach due to its marginally superior out-of-sample predictive performance.  For wind the two methods differ substantially. The LCOE-based approach indicates a significant upward trend in capacity factors, whereas the electricity-based method reveals no such trend\textemdash a discrepancy previously documented in the literature.\supercite{boretti_trends_2020, bolinger_utility-scale_2018, boccard_capacity_2009} Given our objective of accurately predicting actual electricity generation relative to installed capacity, we adopt the electricity-based method, treating future capacity factors as constants specific to each country.

\subsection*{LCOE forecasts}

Utilizing the decomposition of the LCOE into its components and applying the forecast models developed for each component, we generate forecasts for national-level LCOE values. Additionally, we forecast the global average LCOE, as illustrated in Figure \ref{fig:Forecast}, based on projections of global average components.

For solar modules and wind turbines, we directly employ global forecasts derived from experience curves. The future costs are contingent upon projected global cumulative deployment levels for each technology. In the forecasts presented in Figure \ref{fig:Forecast}, we assume an annual deployment growth rate of 30\% for solar and 20\% for wind, which are consistent with historical trends. Alternatively, scenario-based deployment projections, such as those outlined by Way et al.\supercite{way22} can be applied.

Similarly, solar BOS costs are forecast using the experience curve model. For wind BOS costs, we employ an AR(1) model to project global average costs, resulting in a stochastic floor-cost estimate in the investment cost $I_j$ of roughly 450 USD/MW. This corresponds to a floor cost in LCOE $\cL_j$ of roughly 35 USD/MWh.

Regarding the capacity factor and capital recovery factor, we estimate current global averages and assume these values remain constant over time. This assumption is appropriate under the condition that the distribution of future renewable energy deployment mirrors historical patterns. However, the increasing affordability of solar energy may facilitate greater adoption in lower-income regions, potentially altering the distribution of global capacity factors and capital recovery factors, thereby influencing future global average LCOE forecasts.\supercite{polzin21, lonergan_improving_2023, egli_bias_2019}

\subsection*{IAM scenario comparison}
We apply our forecasts for the total investment costs $I_j$ to different IAM scenarios from the AR6 IPCC report database.\supercite{byers_edward_ar6_2022} The purpose of this exercise is to compare the forecasts for future national investment costs to the costs projected by respective IAM scenarios and identify potential issues with the latter.

Our forecasts for $I_j$ are conditional on global solar and wind deployment. Since we require annual deployment scenario data, we consider the global solar and wind deployment in 2040 and interpolate linearly with the global solar and wind deployment in 2020, our base year. The reason for this approach over using the 5-year scenario data directly is that a majority of scenarios show a capacity in 2020 that is below what was actually observed and total investment costs that are larger than today's costs. If we were to use the scenario data as it is, it would not be a fair comparison since our forecasts are sensitive to these input values.
Next to linear interpolation, we can also interpolate the log-deployment. We omit from showing the results of this analysis here since there is no material difference from the linear interpolation.

For each possible value of future renewable deployment, we forecast the associated total investment costs and compare them to the IAM scenario projections. Figure \ref{fig:IAM_proj_USA} shows our forecasts and the scenario projections, conditional on the global solar and wind deployment.
For wind, we see that the IAM projections align well with our forecasts. For solar, there is less alignment, with 85\% of scenario costs above our median forecast. The reason for this is that future solar costs are sensitive to their current costs. IAM scenarios are quickly outdated, leading to these discrepancies. In some cases, the scenarios even project renewable capacity in 2040 to be lower than today.
In both cases, our forecasts show substantially more uncertainty than the set of IAM scenarios. For solar, over 90\% of scenario costs are above our 25th forecast percentile, for wind it is over 80\% of scenarios.

\newpage
\printbibliography
\newpage
\end{refsection}

\newcommand{\beginsupplement}{%
        \setcounter{figure}{0}
        \setcounter{table}{0}
        \setcounter{equation}{0}
        \renewcommand{\thefigure}{S\arabic{figure}}%
        \renewcommand{\thesection}{S\arabic{section}}%
        \renewcommand{\thetable}{S\arabic{table}}
        \renewcommand{\theequation}{S\arabic{equation}}
     }
\clearpage
\beginsupplement
\begin{refsection}[library.bib]
 
\begin{centering}
    \LARGE \bfseries 
    Supplementary Information
    
\end{centering}

\vspace*{-0.7cm}

\tableofcontents
\newpage

\section{Notation overview}

\begin{xltabular}{\textwidth}{>{\raggedright\arraybackslash}p{2cm} X X >{\raggedright\arraybackslash}p{2cm}}
\textbf{Variable} & \textbf{Name} &\textbf{Description} & \textbf{Unit}\\
\toprule
$\cL$ & Levelized Cost of Electricity (LCOE) & Average annual cost to produce 1 kWh of electricity & $USD / kWh$ \\
$I$ & Initial investment cost & Initial overnight investment cost of a new wind/solar PV asset & $USD / kW$\\
$O$ & Operating \& Maintenance cost (O\&M) & Expected annual O\&M cost of a new wind/solar PV asset, averaged over the asset lifetime & $USD / kW / year$\\
$\cC$ & Capital Recovery Factor & Annualized return required to meet the cost of debt and equity for the initial investment $I_0$, averaged over the asset lifetime (present value) & \%$/year$ \\
$\cY$ & Capacity Factor & Annual yield of a wind/solar PV asset & $kWh / kW / year$\\
$\cD$ & Weighted average Cost of Capital (WACC) & Annual return on the initial investment $I$, averaged over investors, weighted by the share of investment & \%$/year$ \\
$T$ & Asset lifetime & Financial \& operating lifetime of the wind/solar PV asset & years\\
$Y$ & Electricity generation & Electricity generated within a given year & $kWh / year$ \\
$\Bar{Y}$ & Average annual electricity generation & Annual electricity generated, averaged over the asset lifetime & $kWh / year$ \\
$X$ & Peak-capacity & Peak power capacity of a newly built wind/solar PV asset & $kW$\\
$\gamma$ & O\&M cost over initial investment cost & Simplified notation for $\gamma = O / I$ & \%$/year$ \\
$\cM$ & Module (solar PV) or turbine (wind) investment cost & Initial investment costs for the solar modules or wind turbines of a new asset, per peak capacity & $USD / kW_{peak}$\\
$\iota$ & Inverter loading ratio & Loading ratio or capacity of solar PV inverter & $kW_{peak} / kW_{nameplate}$ \\
$\cA$ & Set of LCOE components & This set includes LCOE components $A \in \cA = \{ I, \cC, O, \cY + \gamma \}$ & n.a.\\
$\cE$ & Average LCOE error & Multiplicative error term between the average national LCOE and the LCOE of the national averages & n.a.\\
\bottomrule
\end{xltabular}
\captionof{table}{\textbf{Overveiew of notation used throughout this article.}}\label{tbl: Notation}
\setlength\parindent{15pt}

\newpage
\section{Description of national data collection and cleaning}\label{app:data}

\subsection{Capacity and generation data}
\emph{Capacity} denotes the AC capacity of an electricity generation asset connected to the public or private grid (in kW). \emph{Generation} denotes the electricity generated by this asset (in kWh). We collected annual data on the operational capacity, as well as the total electricity generated by this capacity.
We did so by consolidating different datasets, particularly those of BP\supercite{BP22}, IEA\supercite{ieaWorldEnergyBalances}, and IRENA\supercite{irena22}. While these sources are mostly consistent with each other, their geographical coverage differs. Wherever these sources disagree, we have selected the data points most consistent with other data points and other reports from the literature.

\subsection{LCOE data}\label{subsec:data_lcoe}
The \emph{levelized cost of electricity (LCOE)} denotes the cost of a single asset to produce one unit of electricity, averaged over the asset's lifetime. We consolidated a dataset of (capacity-) weighted averages of national LCOE. This data was obtained by manually scraping the extended scientific and grey literature.

While this diverse set of sources increases the available data, it also increases measurement noise and inconsistencies. In particular, different sources may use different assumptions to clean their dataset. We correct these inconsistencies where possible, as described below. 

\subsubsection{LCOE data corrections}
Since we are interested in the LCOE at the year of installation, the LCOE is an ex-ante estimation over the asset lifetime. This means that LCOE is frequently calculated using synthetic and uniform assumptions about operating and maintenance costs, electricity production, and capital recovery. The LCOE data can be inconsistent across sources if these assumptions differ.
We perform several corrections on the LCOE data. These corrections are largely modelled on IRENA's data \supercite{irenaRenewablePowerGeneration2021}.

\paragraph{Cost of Capital}
The \emph{cost of capital} or \emph{weighted average cost of capital (WACC)} denotes the annual returns a project developer needs to pay debt and equity holders for financing the initial total investment costs.
Before 2021, IRENA applied simplified cost of capital assumptions to its project-level data.\supercite{irenaCostFinancingRenewable2023} While the methodology has been significantly improved since then, previous years have not been updated. We, therefore, need to correct the cost of capital before 2021 to include country- and technology differentiation.
After 2021, IRENA applied country- and technology-specific WACCs. These may still differ from the WACC we collected. To avoid artificially increasing the noise in our model based on these explicit assumptions, we harmonize the LCOE data to be consistent with our WACC data.

This means that we are taking the WACC as an exogenous variable rather than making explicit forecasts for it. Given that the WACC is mostly driven by the wider macroeconomic environment, forecasting renewable-specific WACCs is out of scope for this work.

\paragraph{Capacity Factor}
The \emph{capacity factor} denotes the yield factor between electric capacity and the average annual electricity generation of an asset. In other words, it determines how much electricity (kWh) is generated per capacity (kW) in each year, averaged over the asset lifetime.
When calculating the LCOE of a newly commissioned renewable electricity asset, IRENA relies largely on the capacity factor reported by the project developers. As discussed in Section \ref{app:cf}, this expected capacity factor can differ from the realized capacity factors used in our analysis. The reported capacity factor frequently exceeds what we observe in the market. This is particularly the case for wind electricity, where higher capacity factors from increased rotors and hub heights can have less impact on electricity generation than expected. 
We, therefore, correct the capacity factor for wind using the reported national average capacity factor. For solar, the reported capacity factors do not differ significantly from the ones calculated by our method and do not need to be corrected.

\paragraph{Operating and Maintenance cost}
The \emph{operating and maintenance cost} (O\&M) denotes the annual expense of producing electricity with a renewable asset, per operational capacity, averaged over the asset lifetime.
IRENA applies strong assumptions about the O\&M costs in their dataset. These are published for data post 2010. For data prior to 2010 we can calculate the implied O\&M costs based on other data they provide, as well as the LCOE formula they use.
IRENA's O\&M cost assumptions before 2010 differ from other data sources and our findings in Section \ref{app:model_onm}. We replace their assumptions with our findings to avoid artificially increasing the noise in our forecasts.

We thus do not validate our forecasts for the O\&M costs prior to 2010. After 2010, our forecasts are based on a simple relationship with the total investment costs described later. This simplified approach reflects the relatively small contribution of O\&M costs to solar LCOE, poor data quality, high degree of cross-sectional correlation, and high degree of correlation with the total investment cost. If more data becomes available on O\&M costs in the future, this would be a promising avenue of research to further improve our methods.

\subsubsection{Different estimation methods for LCOE}
Different data providers may estimate the LCOE differently. For example, the LCOE can be estimated from Equation (1) in the main text or based on interviews with industry experts. Even within Equation (1), we observe differences, such as some sources applying a different discount rate to the electricity generation to represent differences in future social preferences.

In our dataset, we do not distinguish between these methods and do not correct for any resulting differences. While this significantly increases the available data points, it also increases the associated noise. 
We have also chosen to use proxies for LCOE in some cases, namely the values of commercial Power-Purchase Agreements (PPAs) and governmental (all-in) Feed-In Tariffs (FITs). Both of these represent the price of renewable electricity rather than the cost of electricity generation. However, we have found that the deviations between PPA/FIT and LCOE are usually smaller than the noise within different LCOE estimations (within a given country and year). That does not mean that these values are generally substitutable. For example, early FITs are frequently lower than the actual LCOE due to the political reality of these policy measures. We have, therefore, triangulated different estimation methods to exclude extreme values and inconsistent data points. This leaves us with 669 data points for solar PV and 1002 data points for wind. While this is still a relatively small number of data for our purpose of statistical forecasting and testing, it is significantly larger than what was previously used in the literature. For example, Elshurafa et al.\supercite{elshurafa18} analyze 347 data points for national solar BOS costs (although they differentiate commercial and residential solar, almost doubling their dataset), Kothari et al.\supercite{kothariDriversBalanceofSystemCosts2023} use 112 data points for national BOS costs, Riva et al.\supercite{rivaIEAWindTCP2018} analyze 52 data points for wind O\&M costs.

\subsubsection{Different LCOE formulae}
There exist two commonly used formula for the LCOE $\cL_{ijt}$. In particular, some research applies a discount rate to future electricity generated by a renewable asset:
\begin{equation}
    \cL_{ijt} \equiv \frac{I_{ijt} + \sum_{\tau = 1}^T \frac{O_{ijt}}{(1 + \cD_{ijt})^\tau}}{\sum_{\tau = 1}^T \frac{\cY_{ijt}}{(1 + \cD_{ijt})^\tau}},
\end{equation}
where $T$ is the asset lifetime, $O_{ijt}$ is the operating and maintenance cost in year $\tau$, $\cY_{ijt}$ is the electricity generated in year $\tau$, and $\cD_{ijt}$ is the discount rate set equal to the weighted average cost of capital (WACC) discussed later. All terms are set relative to the capacity of the asset. 
We fix $O_{ijt}$ and $\cY_{ijt}$ for all years $\tau$, so that we can write
\begin{equation}
    \cL_{ijt} \equiv \frac{I_{ijt} * \sum_{\tau = 1}^T \frac{1}{(1 + \cD_{ijt})^\tau} + O_{ijt}}{\cY_{ijt}},
\end{equation}
where $\sum_{\tau = 1}^T \frac{1}{(1 + \cD_{ijt})^\tau}$ is exactly the capital recovery factor $\cC_{ijt}$, proving the equivalence of both formulae.

\subsection{Total investment costs}
\emph{Total investment costs} denote the total overnight costs of purchasing and installing a new renewable electricity asset. It includes the costs for equipment, labour, permitting, land, grid connection, and other soft costs. It typically does not include the cost of integrating that asset into the public grid by providing additional system flexibility to balance out variable renewable generation. As with LCOE, we consider the (capacity-) weighted average total investment cost per capacity.
The process of collecting data on total investment costs is similar to that of LCOE. The only difference is that we are not using proxies for this data collection and do not make any corrections to the raw data. Unlike LCOE data, investment costs are more commonly reported by project developers and rely on fewer assumptions.

\subsection{Module and Turbine costs}
\emph{Module costs} denote the total costs of purchasing a new solar PV module, \emph{Turbine costs} the costs for a new wind turbine. It includes the costs of production, transport, import tariffs and supplier margin. We consider the capacity-weighted average cost in both cases. PV module costs are generally reported per DC capacity. We can use the inverter loading ratio to convert this to AC capacity to make them comparable to wind turbine costs.

The national cost data for PV modules is based on the annual \emph{Trends in Photovoltaic Applications} Report by the IEA's Photovoltaic Power Systems Programme (PVPSP).\supercite{ieaphotovoltaicpowersystemsprogrampvpsTrendsPhotovoltaicApplications2022} Since national module costs are not reported consistently, we combine data from the different annual reports going back to 2002. To obtain the longest possible global time series, we use globally aggregated data from Our World in Data (OWID) that combines different sources since 1975.\supercite{SolarPhotovoltaicPanel}

Turbine data is based on IRENA,\supercite{irenaRenewablePowerGeneration2021} who combine different turbine selling prices depending on the market. We calculate the global weighted average cost by considering the share of Western (Europe and the US, largely dominated by Vestas turbines) and Chinese manufacturers over time. The manufacturing share is based on granular studies of the turbine market.\supercite{gosensProspectsGlobalMarket2014, woodmackenzieChinaLeadsGlobal2024}

\subsection{Balance of system cost}
Unlike capacity, generation, LCOE and total investment costs, balance of system (BOS) costs are rarely observable directly, apart from some exceptions such as Elshurafa et al.\supercite{elshurafa18} This means that we need to estimate the BOS costs from total investment costs and module/turbine costs. Using the equation from the main text, we can write
\begin{equation}
    \cI_{jt} = \cM_{jt-\tau} + \cJ_{jt}.
\end{equation}
The factor $\tau \ge 0$ is a lag between the procurement of modules/turbines and the procurement of BOS costs. 
For solar modules, we do not need to pay special attention to this, with $\tau \approx 0$ since project planning and development typically only requires a few months.\supercite{beattieHowLongDoes2021, cruceResidentialSolarAdoption2022}
For wind turbines, we do need to consider project development timelines. Turbine procurement is typically conducted 1-2 years before an asset's operation date. \supercite{u.s.departmentofenergyLandBasedWindEnergy2022} We, therefore, need to take a procurement time difference of $\tau = 1.5$ into account. We smooth the reported turbine costs over time and apply this lag to calculate $\cJ$. As we show below, our results are not sensitive to this assumption in $\tau$.

\subsection{Capacity factor}

The capacity factor $\cY_{ij}$ gives the relationship between a renewable asset $i$'s electricity output and the installed capacity. We can define it as
\begin{equation}
    \cY_{ijt} = \frac{Y_{ijt}}{X_{ijt}},
\end{equation}
where $Y_{ijt}$(kWh/year) is the average annual electricity production of the asset with nameplate capacity $X_{ijt}$(kW) that started operation in year $t$.

Since the capacity factor is an ex-post average over an asset's lifetime, it needs to be estimated ex-ante at the beginning of a renewable asset's life. This means that there are different ways of estimating an asset's capacity factor. For example, a wind farm's capacity factor can be modeled based on the expected wind speeds in the farm's location or based on historical capacity factors for similar wind farms.

In our case, we can estimate the capacity factor either from the LCOE data we collected or from the capacity and generation data in our dataset. In Section \ref{app:cf} we compare the out-of-sample accuracy of both methods.
For wind, we find that the method based on the LCOE is less accurate. We thus use the electric capacity and generation to estimate the capacity factor. For solar, we find that both methods are comparable, with slightly lower errors from estimating the capacity factor based on the LCOE.

\subsection{Operating and maintenance costs}
Operating \& Maintenance costs $O_{ijt}$ are the costs accrued over the lifetime of a renewable project after the initial investment. They include maintenance and repair costs, operation management, commercial management, and site management. \supercite{steffen20a}
Compared to fossil-fuel operations, operating costs are a small contribution to the LCOE for renewables. As a result, Operating \& Maintenance costs have received significantly less attention than other cost components, and the available empirical data is limited. \supercite{steffen20a, wiserAssessingWindPower2019, wiserBenchmarkingUtilityScalePV2020, irenaRenewablePowerGeneration2021, rivaIEAWindTCP2018} 

Estimations for Operating \& Maintenance costs between these sources differ substantially. For example, IEA reports estimate O\&M costs at around 3.0 - 3.5\% of total investment costs in German onshore wind in 2012, \supercite{rivaIEAWindTCP2018, comissionderegulationdelenergieCoutsRentabiliteEnergies2014} while Raupach-Sumiya et al.\ calculate almost 11\%. \supercite{raupach-sumiyaRegionalEconomicEffects2015} 
In section \ref{app:model_onm} we combine both sources and make our own estimates of O\&M costs relative to the total investment costs. Our estimates for solar are around 1\%, for wind around 3\%.

\subsection{Capital recovery factor}\label{subsec:data_crf}

The Capital Recovery Factor (CRF) is determined by the after-tax weighted average cost of capital $\cD_{ijt}$ (or WACC) and the asset's lifetime $T_i$. The WACC represents the share of the initial investment $I_{ijt}$ that covers the expected annual returns of equity- and debt-investors.
We calculate the CRF from the WACC as \supercite{egli18}
\begin{equation}\label{eqn:crf}
    \cC_{ijt} = \frac{1}{\sum_{t=1}^{T_i} \frac{1}{(1+\cD_{ijt})^t}} = \frac{(1+\cD_{ijt})^{T_i} \cD_{ijt}}{(1+\cD_{ijt})^{T_i} - 1}.
\end{equation}
Furthermore, we can write the country-average WACC $\cD_{jt}$ for a technology $k$ and country $j$ as a sum of the country-specific risk-free rate $r_{jt}$ and technology-specific risk premium $p_k$: \supercite{polzin21}
\begin{equation}\label{eqn:wacc}
    \cD_{jt}^k = r_{jt} + p_k.
\end{equation}

We can infer the risk-free rate from macroeconomic data. Specifically, we use long-term (10 and 7-year) government bond yields from Bloomberg \supercite{bloombergBloombergTerminal2022} and Reuters Eikon. \supercite{reutersRefinitivEikon2022} For countries without data, we use proxies from IMF. \supercite{internationalmonetaryfundimfIMFData2022} In rare cases (for example, years with government defaults), we infer bond yields based on previous years and neighbouring countries.

Technology risk premia $p_k$ may be subject to learning and could decline over time. The literature, however, diverges about the size of that effect. Egli et al.\ \supercite{egli18} suggest that financing experience contributed around 1\% of solar PV and 4\% of onshore wind cost reductions in Germany between the 2000-2005 average and 2017. Kempa et al.\ \supercite{kempaCostDebtRenewable2021} attribute all cost-of-debt reductions to more stringent environmental policies and the maturity of the national banking sector instead of learning. While learning effects are likely, they appear to be small and not significant with respect to our stochastic forecasts. We therefore consider $p_k$ as fixed over time.

Steffen\supercite{steffen20} provides one of the most comprehensive literature reviews and country- and technology-specific estimates of the after-tax WACC for renewables of 47 different countries. 
Unfortunately, despite our simplifying assumption, the data is not available for all countries in our sample.
We, therefore, need to infer $p_k$ for other countries. We follow Polzin et al.\supercite{polzin21} that estimates WACC across Europe using linear interpolation between geographically close locations and a fixed ratio between solar and wind markups. This can further be applied to North America and East-Asia. For Africa, we apply the same method based on data from Thornton. \supercite{thorntonAfricaRenewableEnergy2018} For South America and the Middle East, where data is particularly limited, we use the theoretically derived data in Ondraczek et al.\supercite{ondraczekWACCDogEffect2013}

There are two important assumptions underlying equations \eqref{eqn:crf} and \eqref{eqn:wacc}.
Firstly, we assume that an asset's lifetime is equal to the financing period. This assumption is adequate from an investor's perspective but has some limitations. In particular, the asset lifetime may be longer than the financing period, i.e., debt may be repaid before the asset ceases operation. Our CRF figures can be re-scaled to take this into account if our forecasts are to be applied to such assets.
Secondly, we assume that renewable financing is done through individual projects. This means that the WACC is independent of corporate finances unrelated to a specific renewable project. Steffen\supercite{steffen18} shows that this is appropriate for current and historical projects. When assets are not funded through project finance, the WACC needs to be adjusted depending on the credit rating of the corporate funding the asset.

\newpage
\section{LCOE model development}\label{app:model}
The model development consists of two key steps: 
\begin{enumerate}
    \item Decompose the asset-level LCOE $\cL_{ijt}$ into its components $A_{ijt} \in \{I_{ijt}, \cC_{ijt}, O_{ijt}, \cY_{ijt}\}$.
    \item For each component $A$, develop a model that outputs forecasts for $A_{jt}$, the capacity-weighted average of $A_{ijt}$ over assets $i$ in country $j$ and year $t$.
\end{enumerate}
We describe each step in detail, including the model validation performed to confirm the reliability of our model outcomes and reject model alternatives.

We first describe the LCOE decomposition (Step 1). The subsequent sections detail the component models (Step 2). Sections \ref{app:model_module} and \ref{app:model_turbine} describe the PV module and wind turbine model development, respectively. Sections \ref{app:model_bos_pv} and \ref{app:model_bos_wind} are significantly more involved in developing our models for solar and wind BOS costs. In particular, we explore the role of local learning in BOS costs and how to use it in forward-looking forecasts. We aggregate module, turbine, and BOS costs to total investment costs in Section \ref{app:model_inv}. Sections \ref{app:model_onm} provides additional details on the operating and maintenance cost model.

\subsection{LCOE decomposition and aggregation}\label{app:model_LCOE}
{\bf Summary:}
\emph{We decompose the non-linear LCOE into its components using a linear approximation. For each component, we make use a statistically validated forecasts. Using an out-of-sample test, we further show that the aggregate LCOE model does not overfit the empirical data. 
}
\newline

\noindent
Our starting point is the asset-level LCOE $\cL_{ijt}$. Re-stating the LCOE decomposition from the main text, it is given as
\begin{equation}\label{eqn:app_lcoe}
    \cL_{ijt} \equiv \frac{I_{ijt} * \cC_{ijt} + O_{ijt}}{\cY_{ijt}}\ \text{ (USD / kWh),}
\end{equation}
for asset $i$ in country $j$ and year $t$.
We are ultimately interested in the capacity-weighted average LCOE $\cL_{jt}$. We average over assets $i$, for each country $j$ and year $t$, specific to the technology. Each component model forecasts future values of the capacity-weighted average component ($I_{jt}, \cC_{jt}, O_{jt}, \cY_{jt}$). For simplicity, we drop the index $t$ for the remainder of this section.
Unfortunately, calculating $\cL_{j}$ from the capacity-weighted average components is not straightforward due to the non-linear nature of Equation \eqref{eqn:app_lcoe}. The national weighted average of the LCOE may differ from the LCOE of the weighted average component. 

We, therefore, perform a log-linear approximation of equation \eqref{eqn:app_lcoe} that we can use to make approximate forecasts for $\cL_j$. This generates an additional error term $\log \cE$, which we quantify empirically. Afterwards, we compare the disaggregate model to the global Wright's law model of Lafond et al.\supercite{lafondHowWellExperience2018} to confirm that we have not grossly over-fitted our data.

\subsubsection{Linear LCOE approximation}
In this section, we asses the relationship between the weighted average LCOE $\cL_j$ and the weighted average components $\{I_j, \cC_j, O_j, \cY_j \}$, particularly:
\begin{equation}\label{eqn:avg_vs_avg}
    \cL_j \stackrel{?}{=} \cL(I_j, \cC_j, O_j, \cY_j).
\end{equation}
Here, we develop a theoretical model of this relationship that we later test empirically.

We first consider the LCOE of a single asset $i$, dropping the subscript $t$ for convenience. As discussed in the main text and section \ref{app:model_onm}, we model the operating and maintenance costs as a fixed share $\gamma$ of the total investment costs, i.e., $O_{ij} = \gamma * I_{ij}$. This means we can re-write the asset-level LCOE \eqref{eqn:app_lcoe} as a linear equation:
\begin{equation}\label{eqn:agg_LCOE}
    \log \cL_{ij} = \log I_{ij} + \log \left( \cC_{ij} + \gamma\right) - \log \cY_{ij} = \sum_{A \in \cA}\log A,
\end{equation}
for components $A \in \{ I, \cY^{-1}, O, \cC + \gamma \} =: \cA$. Deriving the relationship between weighted average LCOE and weighted average components thus reduces to assessing the difference between the weighted average of a variable and the weighted average of its logarithm.

Based on Jensen's inequality, we know that the log of the average is greater or equal to the average of the log. Since this applies to both sides of the equation (the LCOE $\cL_{ij}$ and components $A_{ij}\in\cA$) the resulting biases cancel each other out to some degree but not necessarily completely. We can describe this relationship explicitly using a first-order Taylor expansion around the average LCOE and components.

Without going into the details of the Taylor expansion, we can write the average of the logarithm of a random variable $X$ as the logarithm of the average, $\Bar{X}$, such that
\begin{equation}
    \overline{\log X} \approx \log\Bar{X} + \varepsilon_X,
\end{equation}
where the first-order error term $\varepsilon_X$ depends on the variance in the random variable $X$. In particular, if $X$ has a low variance, $\varepsilon_X$ is small, but if the variance is high, $\varepsilon_X$ is large.

Using this identity, we can write Equation \eqref{eqn:agg_LCOE} as
\begin{align}
    \log \cL_j &\approx \log I_j + \log (\cC_j + \gamma) - \log \cY_j + \varepsilon_I + \varepsilon_\cC - \varepsilon_\cY - \varepsilon_\cL\\
    &\approx \log I_j + \log (\cC_j + \gamma) - \log \cY_j + \log\cE_j, \label{eqn:LCOE_E_app}
\end{align}
where $\log\cE_j$ is an aggregate first-order error term from the Taylor expansion.

Unfortunately, we cannot decompose $ \log\cE_j$ further since this requires asset-level data on all LCOE components. In future research, asset-level cost data could be used to better capture the variation around the component average. This is particularly helpful in assessing how policies targeted at different sub-markets of the renewables industry could lower the average national LCOE.

\subsubsection{Aggregate error model}
We can, however, study $\log \cE_j$ empirically as an aggregate error term. In other words, we investigate the difference between $\log \cL_j$ and $\log I_j \cY_j^{-1} (\cC_j + \gamma)$ from Equation \eqref{eqn:LCOE_E_app}.
We first perform a regression analysis on Equation \eqref{eqn:LCOE_E_app}. This tells us how much of the variance of the log-LCOE is explained by the log of each component and $\log \cE_j$.
We then investigate the trend and correlation of $\log \cE_j$. We do not find significant evidence of either, implying that we can reasonably model $\log \cE_j$ as i.i.d.\ random variables. 
Based on our dataset alone, $\log \cE_j$ follows a fat-tailed Cauchy distribution. However, when we consider a more granular dataset for the US,\supercite{barboseTrackingSun20242024} we find that most extreme values in $\log \cE_j$ are caused by outliers. Once we remove the outliers, which we believe to be spurious, the distribution can be well-approximated as a normal distribution.

\paragraph{Regression analysis}
We consider the model
\begin{equation}\label{eqn:lcoe_regression}
    \log \cL_j = \beta_I\log I_j + \beta_\cC\log (\cC_j + \gamma) - \beta_\cY\log \cY_j + \epsilon_j.
\end{equation}
Table \ref{tab:lcoe_regression} shows the results of the regression. Equation \eqref{eqn:lcoe_regression} explains 99\% of the variance in the data, leaving $\log\cE_j$ explaining the remaining 1\%. It's interesting to see that if we include only the investment cost $I$ and capacity factor $\cY$, we can already explain 97\% of the variance.
\newcolumntype{Y}{>{\centering\arraybackslash}X}
\begin{table}
    \centering
    \footnotesize
    \begin{tabularx}{1\textwidth}{>{\raggedright\arraybackslash}p{1.5cm} X X X X X | X X X X X}
        & \multicolumn{5}{c}{Solar} & \multicolumn{5}{c}{Wind} \\ \hline
        $\log I$ & -0.24& & & 0.96 & 1.0& -0.23 & & & 1.0 & 1.0\\ 
        $\log \cY$ & & -0.28 & & -1.33 & -1.0 & &-0.24 & & -1.26 &  -1.0\\ 
        $\log (\cC_j + \gamma)$ & & & 0.83 & & 1.1 & & & 0.93 & & 0.99\\ \hline
        $R^2$ & 0.69 & 0.79 & 0.81 & 0.97& 0.99 & 0.90& 0.93 & 0.95 & 0.97 & 0.99\\ \hline
    \end{tabularx}
    \caption{Regression results for LCOE against its components, according to Equation \eqref{eqn:lcoe_regression}. We omit the indication of statistical significance since all regressors are highly significant above 99\%.}
    \label{tab:lcoe_regression}
\end{table}

\begin{figure}
    \centering
    \includegraphics[clip, trim=2cm 1cm 2cm 1cm, width=\textwidth]{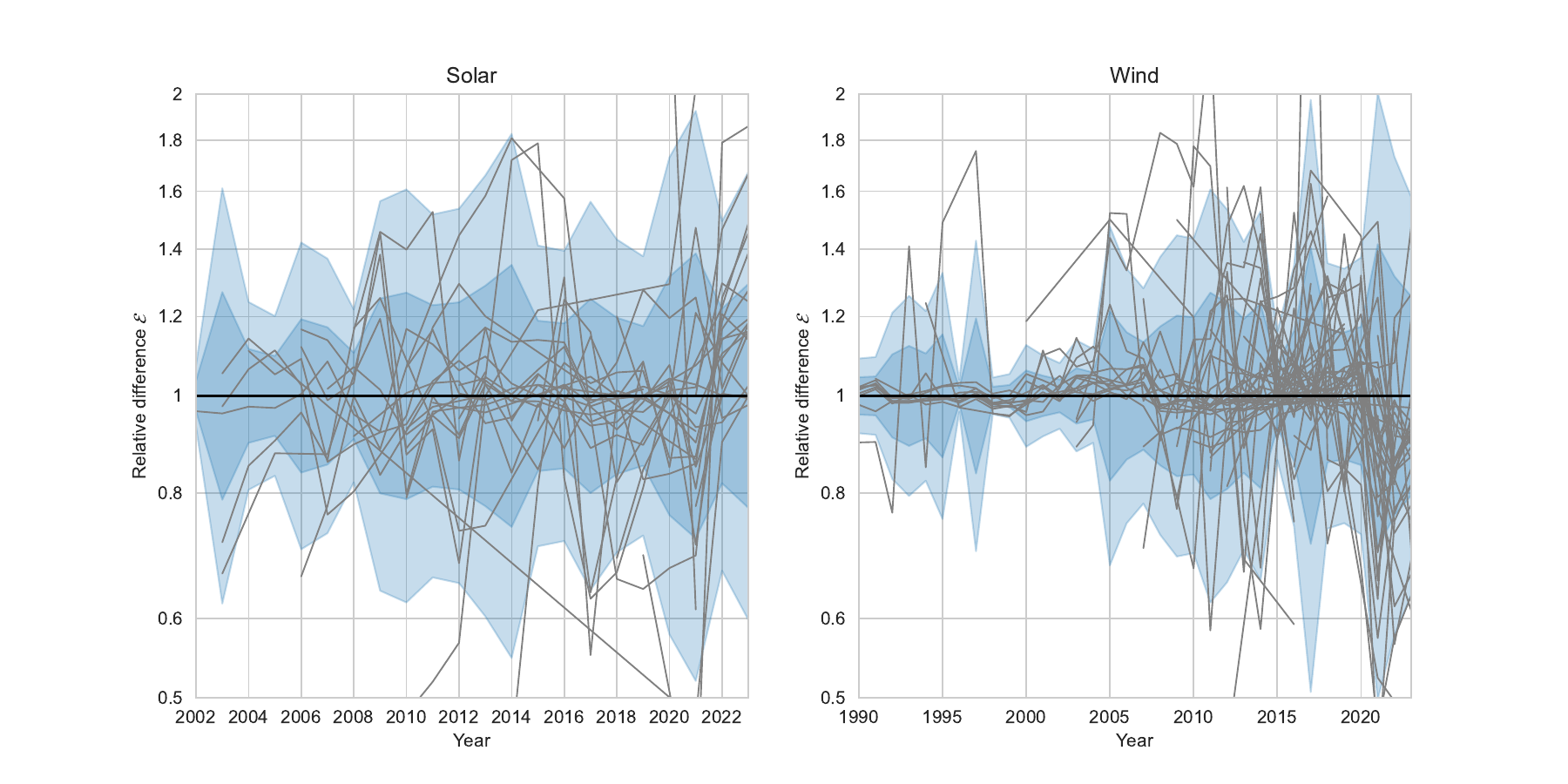}	
    \caption{\textbf{Natural Logarithm of the relative difference between the average LCOE $\E_\omega \cL$ and the LCOE of the averages $\hat{I} \hat{\cY} (\hat{\cC} + \hat{\gamma})$ over time.} 
    Grey lines show $\cE$ for a specific country over time. Blue shading indicates one / two standard deviations of the entire $\cE$ sample within one year. We observe that the distribution of $\log \cE$ is stationary over time, with no significant drift or changes in the standard deviation over time.}\label{fig:avg_timeseries}
\end{figure}

Figure \ref{fig:avg_timeseries} shows $\log \cE_j$ for individual countries $j$ over time. The size of $\log \cE_j$ is comparable for solar and wind. By definition, both distributions are centered around zero since we use a fixed effect model to estimate the capacity factor. The fact that Turkey and Argentina drive the standard deviation increases in 2021 and 2022. In both cases, very large interest rates drive up the CRF, which is not reflected in the LCOE data.

\paragraph{Trend analysis}
For the trend in $\log \cE_j$ we maximize the available data by calculating the first (year-on-year) differences in national data. Based on a two-sided t-test, we do not find a significant positive or negative trend. 
For the trend in the cross-sectional variance of $\log \cE_j$, we also calculate its first (year-on-year) difference. Similarly, a two-sided t-test finds no significant positive or negative trend.

\begin{figure}
    \centering
    \includegraphics[width=\textwidth]{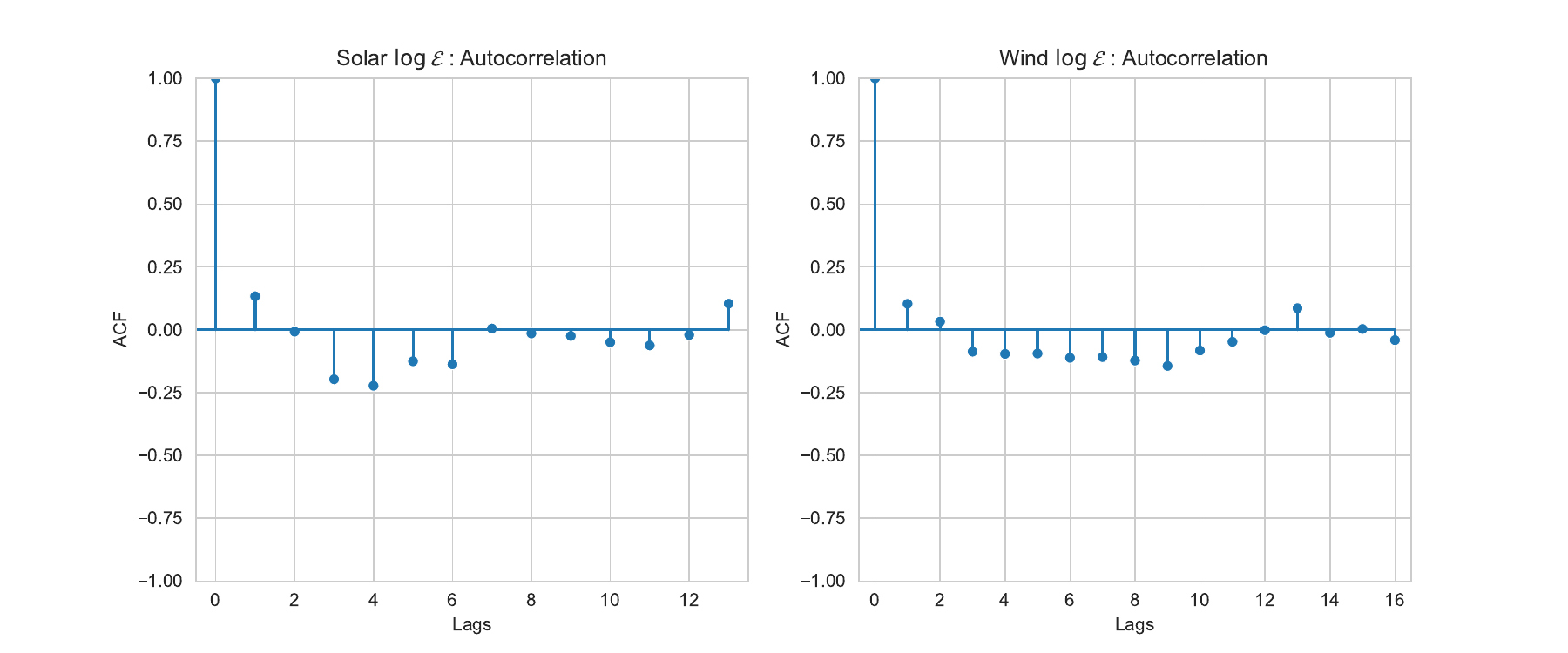}
    \caption{\textbf{Average national autocorrelation function $\log \cE$.} 
    We take the average of the ACF of $\log \cE$ for individual countries for solar PV \textbf{(left)} and wind \textbf{(right)}. Neither technology shows significant autocorrelation for the first lags.}\label{fig:acf_logE}
\end{figure}

\paragraph{Autocorrelation analysis}
We calculate the autocorrelation function (ACF) for each country's $\log \cE_j$ time series and look at the average ACF across countries. The ACF of a time series $(X_t)_{t \in \N}$ and lag $\tau$ is defined as the serial correlation between $(X_t)_{t \in \N}$ and the lagged series $(X_{t-\tau})_{t \in \N}$. While the ACF for individual countries will likely be very noisy, the average thereof should be significantly less so and give us a meaningful indication of the correlation.
The resulting ACF is shown in Figure \ref{fig:acf_logE}, with solar on the left and wind on the right panel. We can see that the ACF takes very low values below 0.25 for all tested lags. We conclude there is no significant correlation between individual years of $\log \cE_j$, and we can sample them independently for our forecast.

\paragraph{Distribution analysis}
The previous findings imply that to forecast future values of $\log \cE_j$ we can reasonably approximate them by i.i.d.\ samples around zero.
To define how we want to sample $\log \cE_j$, we first consider the in-sample distribution based on our dataset.

\begin{figure}
    \centering
    \includegraphics[clip, trim=2cm 1cm 1cm 1cm, width=\textwidth]{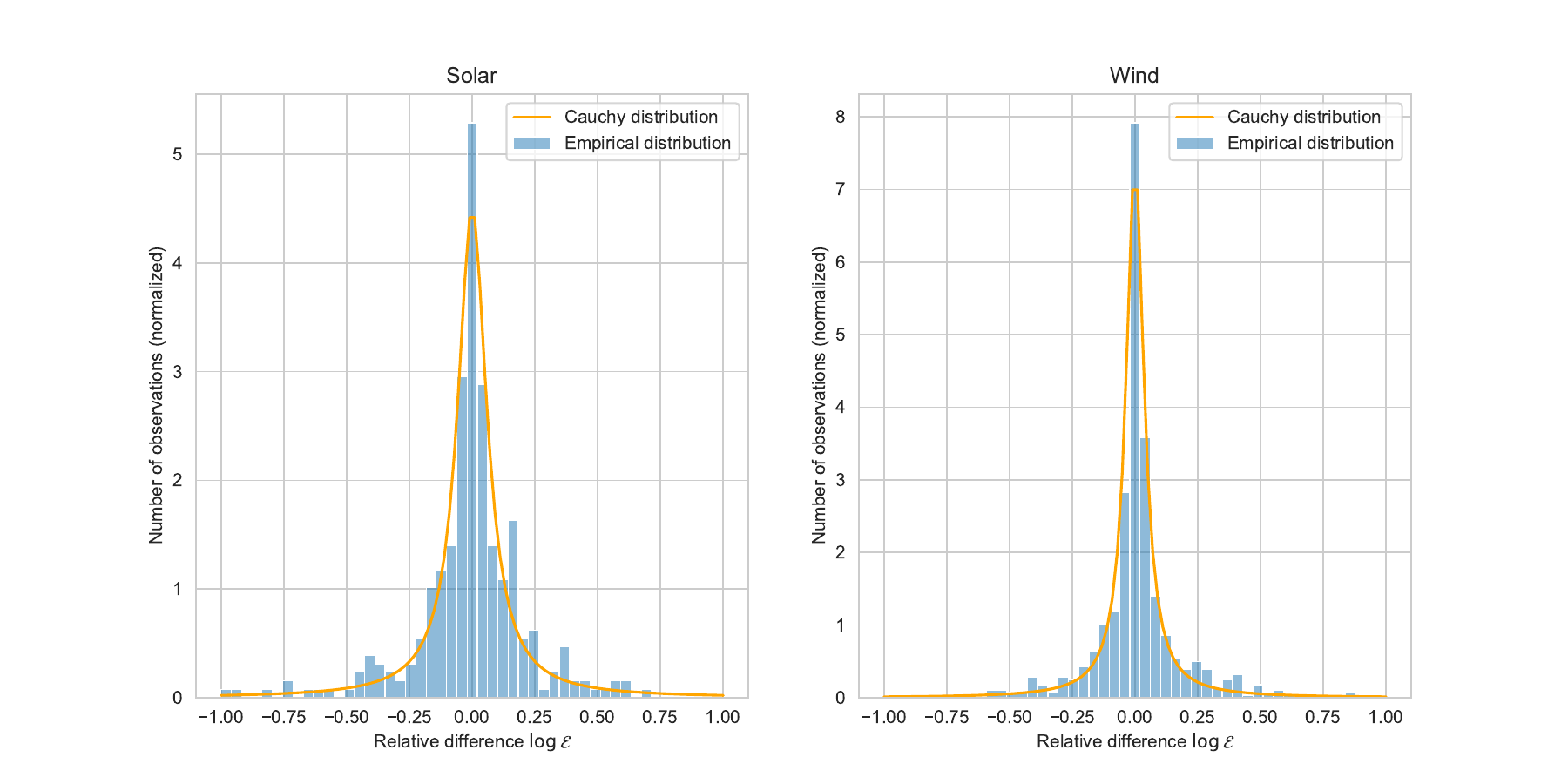}
    \caption{\textbf{Empirical distribution of $\log \cE$.} 
    We calculate $\log \cE$ for solar \textbf{(left)} and wind \textbf{(right)} and look at the resulting distribution. 
    In both cases, we observe a large number of very positive or negative values of $\log \cE$, indicating a heavy-tailed distribution.}\label{fig:dis_logE}
\end{figure}

The in-sample distribution is shown in Figure \ref{fig:dis_logE}, with solar on the left and wind on the right panel. We see that both distributions are symmetric around zero. Both distributions show extremely positive and negative values. This indicated that $\log \cE_j$ follows a distribution with two heavy tails.
A Cauchy-Lorenz distribution with heavy tails fits this distribution well, as indicated by the orange line in Figure \ref{fig:dis_logE}. Similarly, a Cauchy distribution based on a maximum likelihood estimator can be used to forecast $\log \cE_j$, with appropriate out-of-sample accuracy.

Nevertheless, the large errors and short-term effects raise the question of whether extreme values of $\log \cE_j$ are actually caused by economic effects or rather issues with the underlying datasets. While asset-level component data is, unfortunately, not widely available, we can study the log-difference effects for the investment cost $\log I_{ijt}$ in the US utility-scale solar sector using an asset-level dataset from Berkley Lab.\supercite{barboseTrackingSun20242024}
The Berkley dataset consolidates the investment costs for over 2.4 million solar photovoltaic assets in the United States between 2000 and 2023. It covers around 5\% of total U.S. solar capacity in the US, although coverage varies substantially by year.

\begin{figure}
    \centering
    \includegraphics[clip, width=\textwidth]{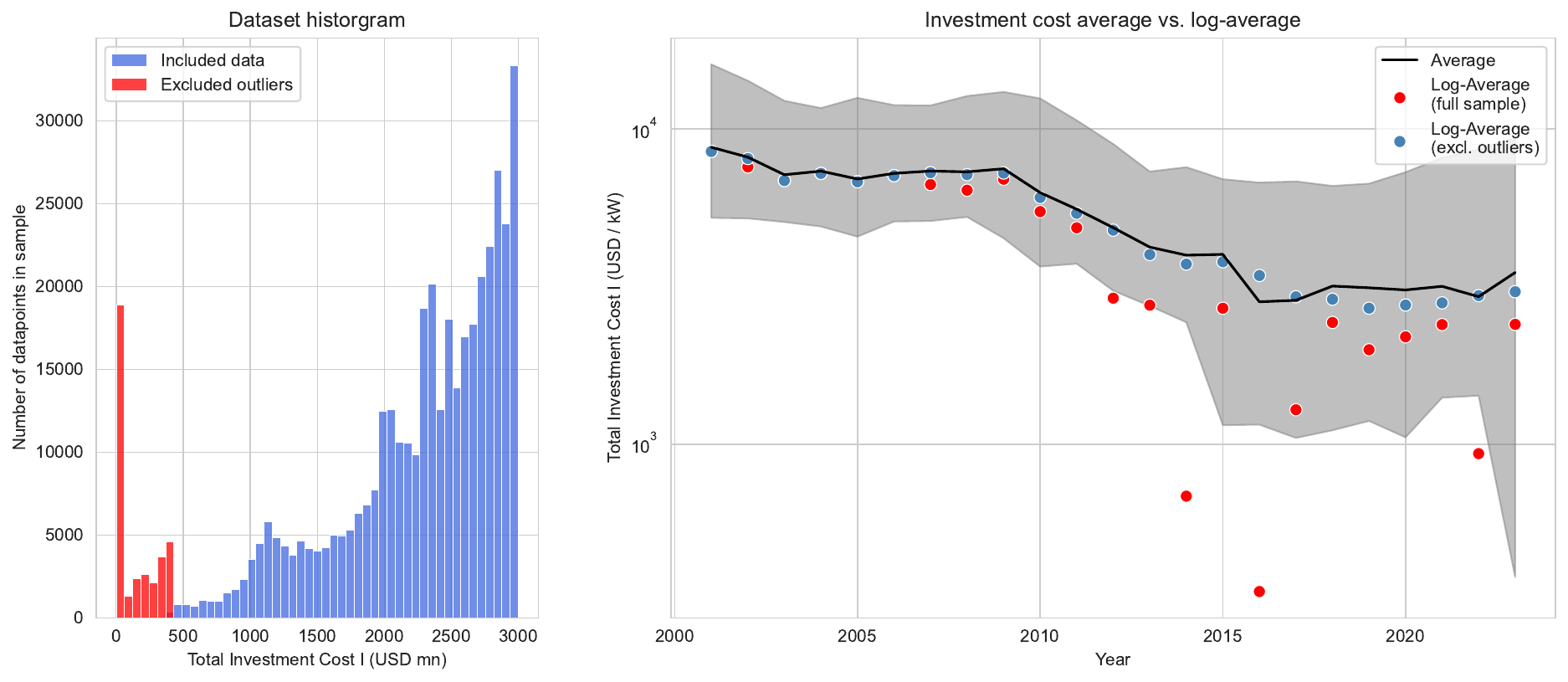}
    \caption{\textbf{Average and log-average of the total investment cost $I$ in the US.} 
    \textbf{(Left panel)} The distribution of total investment cost from the Berkley Lab dataset.\supercite{barboseTrackingSun20242024} We exclude datapoints with costs above USD 3mn for better visualization. There are a number of datapoints with costs below USD 400 tsd that are outliers with respect to the full data distribution.
    \textbf{(Right panel)} The black line shows the average $I_{jt}$ in the US over time, with the gray area indicating the 95th quantile of asset-level observation. 
    The red dots show the exponential of the average of $\log I_{ijt}$ over time. For most years, it falls close to $I_{jt}$, within the 95th quantile. There are, however, multiple outliers with an order of magnitude lower log-average costs. The blue dots show the average of $\log I_{ijt}$ if we exclude the outliers with costs below USD 400 tsd.}\label{fig:US_costs}
\end{figure}

Figure \ref{fig:US_costs} (right panel) shows the average and log-average total investment costs in the US over time. In most years, average and log-average are close to each other. However, in 2014, 2016, 2017, and 2022, the two values differ substantially (red dots), with a log-difference of up to 20\%. The reason for this is individual renewable assets with large capacity and particularly low costs. The left panel of Figure \ref{fig:US_costs} shows the distribution of the total investment cost across the Barkley Lab sample, excluding datapoints with costs above USD 3 million. There are a number of utility-scale assets with costs below USD 400 thousand that are visible outliers to the sample distribution.
If we remove utility-scale assets with unreasonably low investment costs below USD 400 thousand, average and log-average closely align (blue dots on the right panel).

Including these outliers in our forecasts inflates the errors of our LCOE forecast. We, therefore, exclude extreme $\log \cE_{jt}$ values when estimating the size of this error term. 
Based on the previous total investment cost analysis for the US, we exclude $\log \cE_{jt}$ differences exceeding 20\%. This is approximately equal to excluding the 10th and 90th percentile. The remaining errors are forecast with a normal distribution, based on a standard deviation estimate of roughly 9\% for solar and 7\% for wind. This is a rather simple analysis that requires further asset-level justification when such data becomes available.

\subsubsection{Forecast accuracy of the LCOE model}
Conceptually, forecasting the LCOE from its components should work well since we include more information about the underlying dynamics. In practice, overfitting and parameter estimation errors in each component model are a real concern, so it may be better to forecast the aggregate directly.\supercite{grunfeldAggregationNecessarilyBad1960} The choice between the disaggregate and a simpler top-down model depends on the sample size used to estimate parameters, the noise levels in the model, and the model dynamics.\supercite{lutkepohlForecastingContemporaneouslyAggregated1984}
We answer this question based on empirical and theoretical considerations.

Empirically, we perform an out-of-sample cross-check between the aggregate model and simpler alternatives. The results are not statistically significant. This insignificance stems from the fact that we do not have sufficient data for the out-of-sample test results to be significant. This is because we can only use a subset of our data for this test. As we will discuss in section \ref{app:model_inv}, we still expect the disaggregate model to outperform the aggregate when training it on our full dataset if Wright's or Moore's law continue to hold for the LCOE components.

\subsection{Module cost forecasts}\label{app:model_module}
{\bf Summary:}
\emph{We model global average solar photovoltaic module costs and do not differentiate national costs. We use the stochastic Wright's law model of Lafond et al.\supercite{lafondHowWellExperience2018} to forecast future global average module costs, conditional on global solar deployment.}
\newline

\noindent
Most countries have had very similar module costs in the past.\supercite{ieapvpsTrendsPhotovoltaicApplications2015} This is expected with such a highly consolidated market; China has around 80\% global market share in module manufacturing.\supercite{ieaphotovoltaicpowersystemsprogrampvpsTrendsPhotovoltaicApplications2022}
One exception to this is Japan. Protectionism in the Japanese market resulted in the decoupling of local prices from the global average after 2010.\supercite{raupach-sumiyaRegionalEconomicEffects2015} We thus exclude Japan from the module cost forecast.
In line with most energy technology literature, we do not treat this decoupling explicitly but consider a global module cost. 
We thus drop the country-subscript $j$ from our notation for the remainder of this section.

For global average manufacturing costs, we can use an existing probabilistic forecast model. Lafond et al.\supercite{lafondHowWellExperience2018} show that one can model Wright's law as a geometric random walk with nonlinear drift and autocorrelated residuals. This holds for a large number of technologies, includingn solar modules. We therefore apply this model to project future solar module costs:
\begin{equation}\label{eqn:module_model}
    \text{Module model: } \log \cM(t) = \log \cM(t-1) - \hat{\omega} \log \frac{Z(t)}{Z(t-1)} + u_t + \rho u_{t-1},
\end{equation}
where $Z(t)$ is the global solar capacity until time $t$, $\omega$ is the estimated Wright's law exponent with $\hat{\omega} \sim \cN(\omega, \hat{\sigma}_\omega^2)$, $u_t \sim \cN(0, \hat{\sigma}_u^2)$ is i.i.d.\ noise, and $\rho = 0.19$ is the autocorrelation parameter. The parameters $\hat{\omega}, \hat{\sigma}_\omega^2, \hat{\sigma}_u^2$ are estimated using a linear regression on the first difference of the module costs.

Since we do not alter the model used by Lafond et al., we do not need to perform a separate validation. The model in Equation \eqref{eqn:module_model} has been extensively studied on over 50 technologies, as well as applied to broader energy system scenarios.\supercite{way22}

\subsection{Turbine cost forecasts}\label{app:model_turbine}
{\bf Summary:}
\emph{We observe different costs for Chinese- and Western-manufactured wind turbines. 
However, due to the small sample size, we can not tell with statistical significance if a model that distinguishes manufacturing-location-specific learning is is more accurate than a global model.
We show that the stochastic Wright's law model of Lafond et al.,\supercite{lafondHowWellExperience2018} based on global learning provides appropriate uncertainty estimates.}
\newline

\begin{figure}
    \centering
    \includegraphics[width=\textwidth]{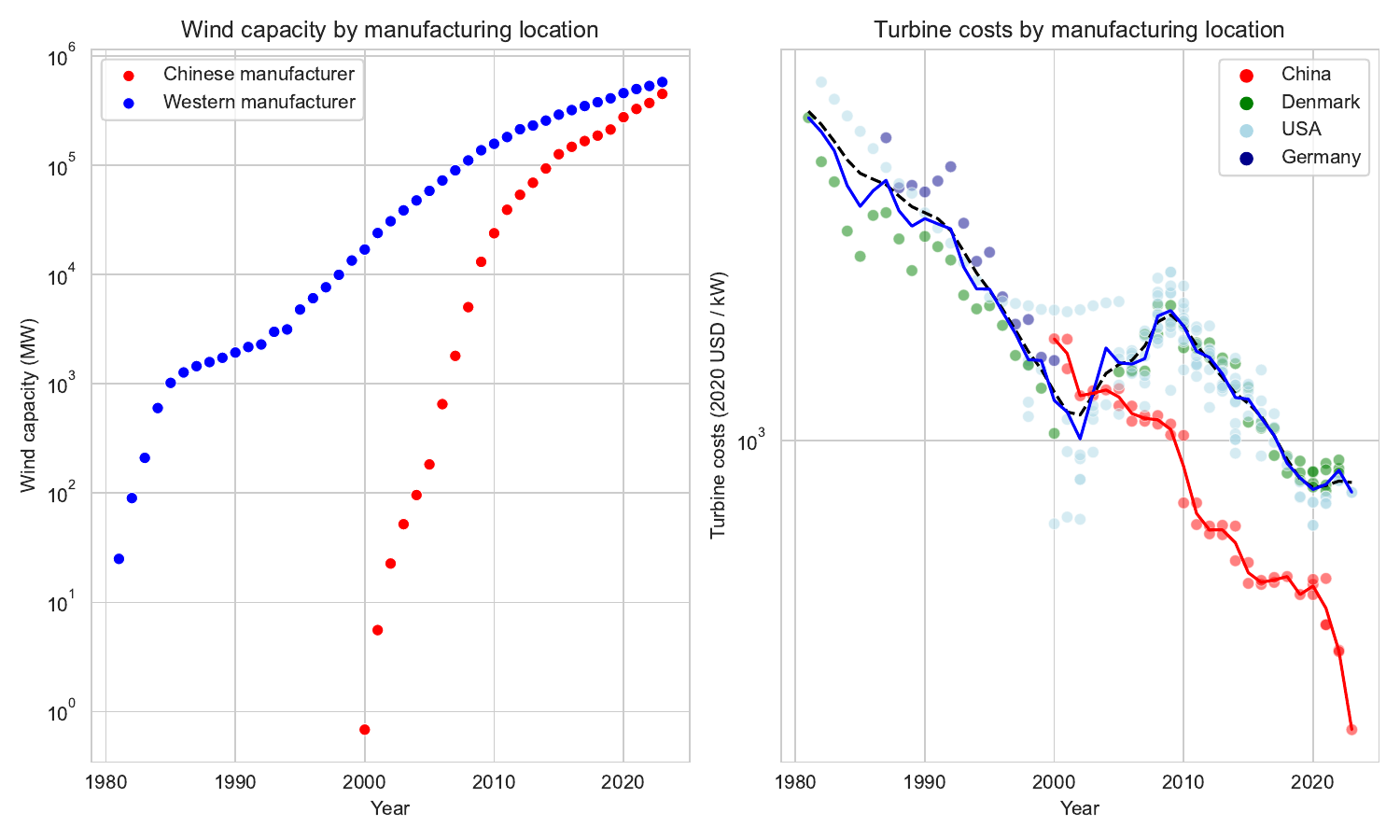}
    \caption{\textbf{Wind turbine costs vs cumulative deployment and year.} 
    The left panel shows the increase in wind turbine capacity manufactured by Chinese and Western companies between 1982 and 2021. The right panel shows the decrease in the respective turbine costs over the same period.
    The red and blue lines show a (weighted) average cost for Chinese/Western turbines. The black dotted line is a smoothed interpolation of the western turbine cost used to calculate wind BOS costs in subsequent sections.
    }\label{fig:turbines}
\end{figure}

\noindent
Unlike solar modules, wind turbines have shown substantially different costs in different locations. As shown in Figure \ref{fig:turbines}, Western-manufactured (Europe and the US) and Chinese-manufactured turbines used to have similar costs in the early 2000s. However, since around 2005, Chinese turbines have shown significantly lower costs than their Western alternatives.
While Western turbines have been around for longer, Chinese turbines have also grown more quickly. At the moment, they are mainly sold in China and, to some extent, India and Brazil, but foreign market share is continuously increasing, and Chinese turbines have recently overtaken Western alternatives in terms of global market share. \supercite{ihsmElectricPlantUnit2023, gosensProspectsGlobalMarket2014, hayashiGoneWindLearning2018, woodmackenzieChinaLeadsGlobal2024}

This raises the question of whether the lower costs of Chinese turbines are due to local experience effects and if we should differentiate between manufacturing locations in our forecasts. Unfortunately, with only two short time series, we do not have sufficient data to assess with statistical significance if a differentiated model is more appropriate.

Nevertheless, we must select one of these four alternatives to make forecasts. Here, we choose a global experience curve model in line with Lafond et al,\supercite{lafondHowWellExperience2018} due to its simplicity. A global model simplifies the IAM forecast comparison performed later.
In some applications, such as investigations of trade policies and disjoint markets, it may be more insightful to use a model with regional differentiation. Researchers will have to be careful to investigate the impact of their model choice on their results since we cannot clearly state a preference for one model over the other.

To validate our model on regional turbine costs, we need a full specification of the residuals associated with our forecasts. This way, we can identify whether our forecast residuals are consistent with the model.

We can use the same model as Equation \eqref{eqn:module_model} for each market $j$ (either Western or Chinese turbines):
\begin{equation}\label{eqn:turbine_model}
    \text{Turbine model: } \log \cM_j(t) = \log \cM_j(t-1) - \omega \log \frac{Z(t)}{Z(t-1)} + \eta^j_{t} + \rho \eta^j_{t-1},
\end{equation}
where $\eta^j_t \sim \cN(0, \hat{\sigma}_\eta^2)$ and $\omega - \hat{\omega} \sim t_m * \hat{\sigma}_\omega$. Here $\hat{\sigma}_\eta$ denotes estimated variances of the noise, $\hat{\omega}$ denote the estimated Wright's law exponent, $\hat{\sigma}_\omega$ is the estimated standard errors of the parameters, and $\omega$ is the true Wright's law exponent. $t_m$ is a Student t-distribution with $m$ degrees of freedom, equivalent to the number of data points used to estimate the model.
We are also including the previous autocorrelation of our noise with $\rho = 0.19$. \supercite{lafondHowWellExperience2018} 

We estimate the model parameters $\hat{\omega}$ and $\hat{\sigma}_\eta$ with the same method as Lafond et al.\supercite{lafondHowWellExperience2018} on the global average turbine costs $\cM(t)$. Since $\cM(t)$ is the weighted average between the two $\cM_j(t)$ time series, we need to correct for this effect when estimating the variance $\hat{\sigma}_\eta$. Since we have two time series following the same random process, we can do so with an additional factor of 2. So, $\hat{\sigma}_\eta  = 2 \cdot \Tilde{\sigma}_\eta$, where $\hat{\sigma}_\eta$ is the estimated variance of the two $\cM_j$ processes, and $\Tilde{\sigma}_\eta$ is the estimated variance of their average $\cM$.

Note that neither the Wright's law exponent $\omega$ nor the exogenous shock scale $\sigma_\eta^2$ are specific to the turbines being manufactured in China or the West. Both manufacturing markets share universal parameters since they follow the same learning dynamics. Secondly, the bias-variance trade off means that fewer model parameters are preferred in cases with little data.
Smart, Lafond \& Farmer \supercite{smartMacroMicroPrediction2024} show this explicitly for Wright's law. With short time-series and similar model parameters, applying the same parameters to both technologies is preferred over technology-specific parameters.

We validate this model using backtesting.
We consider all 5-year moving average windows between times $t$ and $t-4$ to estimate our model parameters $\Theta_t := (\hat{\omega}, \hat{\sigma}_\eta, \hat{\sigma}_\omega$). Based on these parameters, we make stochastic forecasts for the future of our training data $t + \tau$, using Monte Carlo sampling. This gives us the median point forecast and associated theoretical uncertainty distribution for each forecast horizon $\tau$. 
We can now compare the theoretical uncertainty distribution to the empirically observed residuals for each value of $t$, $\tau$, and market $j$.

\begin{figure}
    \centering
    \includegraphics[clip, width=\textwidth]{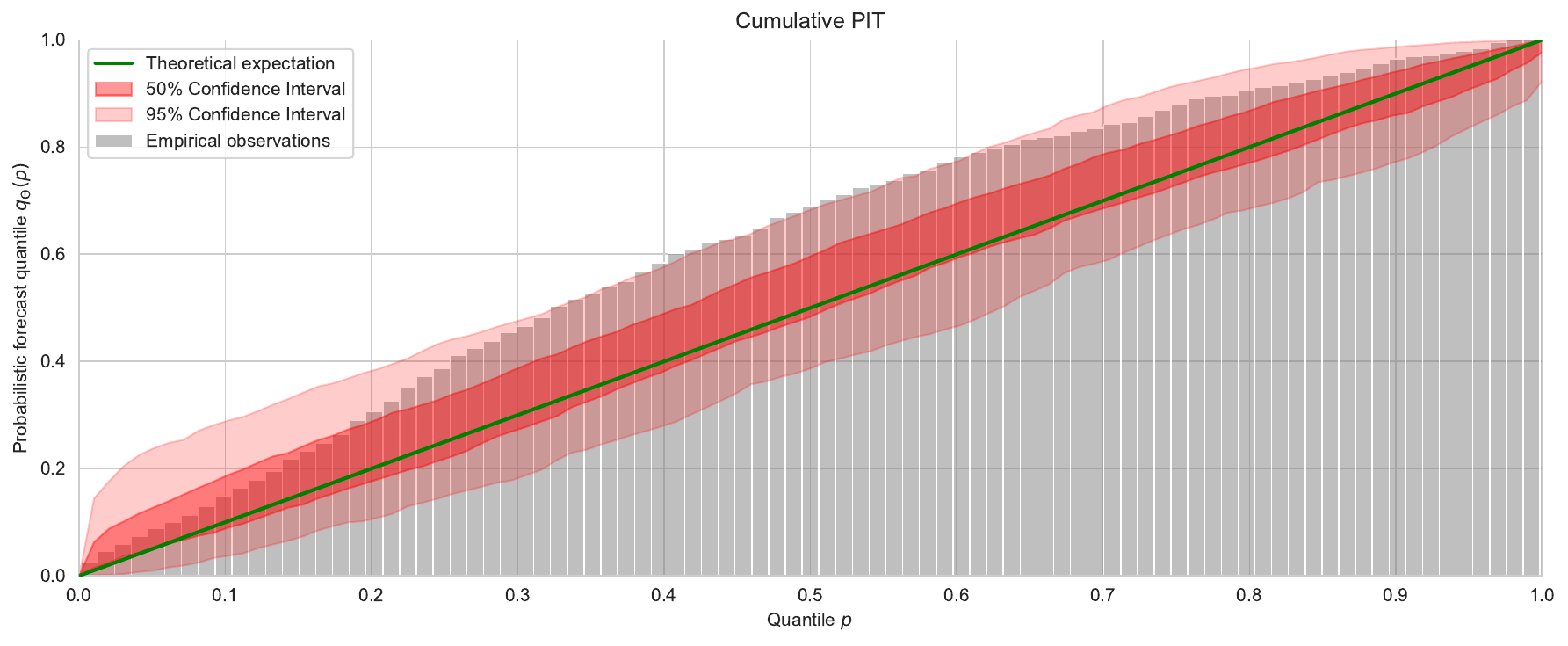}
    \caption{\textbf{Out-of-sample validation of the turbine cost forecast error.} 
    We estimate model parameters on 5-year moving windows of our data. We make stochastic forecasts for the future of this window with a Monte Carlo approach.
    The empirical cumulative PIT $q_\Theta(p)$ is shown in gray. The green line is the theoretically expected PIT. The red areas are the 50\% and 95\% confidence intervals under a surrogate random walk model. For almost all values of $p$, the empirical distribution falls within the 95\% percentile, indicating that our model is a reasonable representation of reality.
    Between $p = 0.5$ and $p=0.6$, the empirical distribution is marginally outside of the 95\% quantile but still well within the 96\% quantile.
    }\label{fig:turbines_error}
\end{figure}

To compare these probabilistic forecasts to the true data, we compute the cumulative {\it probability integral transform} (PIT) of our forecasts.\supercite{gneitingProbabilisticForecasting2014, gneitingProbabilisticForecastsCalibration2007} The reason for using the cumulative PIT is to better examine statistical deviations thereof, as explained below. 
To calculate the PIT, consider a quantile $p \in [0,1]$ of our probabilistic forecast. If our quantiles are correct, we expect the share of empirical observations that lie below this quantile, $q_\Theta(p)$, to be equal to $p$, i.e., $q_\Theta(p) = p$. 
Figure \ref{fig:turbines_error} shows the observed quantiles $q_\Theta(p)$ against $p$. The gray bars are the empirical observations. Theoretically, we expect them to follow the identity line indicated in red. Graphically, we see that the empirical observations are above the identity. This means that our probabilistic forecasts tend to be larger than the empirical observations instead of symmetrically scattered around them. This difference, however, is not statistically significant.

To assess if the empirical observations are significantly different from the theoretical model we apply a surrogate data method. We assume the same data structure as in our empirical observations and simulate 500 local geometric random walks conditional on future cumulative deployment, using the same Wright's law parameters of our model. We then compute the cumulative PIT for each simulation run, i.e.\ estimate $q_\Theta(p)$. This provides confidence intervals for each value of $p$, shown as red areas in Figure \ref{fig:turbines_error}. Since for each surrogate dataset, the $q_\Theta(p)$-values are not independent from each other, we plot the cumulative PIT. This way, we can compute meaningful confidence intervals that help the graphical analysis.
We see that our gray bars (the empirical data) fall just within the 95\% confidence interval of the surrogate data, with one exception between the 50\% and 60\% quantile, which is within the 96\% quantile. We conclude that there is no significant difference between our forecasts and an ideal model and that the model is appropriate.

While this shows that we can use the \emph{Wright, Global (U)} model to make forecasts for turbine costs, it does not mean that the \emph{Wright, Local (U)} model is inappropriate. In fact, when we repeat this exercise for the \emph{Wright, Local (U)} model (not shown here), we also find that the observations fall within the respective 95\% confidence interval.

\subsection{Solar BOS cost forecasts}\label{app:model_bos_pv}

{\bf Summary: }
\emph{BOS costs display substantial national correlation. We study local and global learning models to forecast future national BOS costs. A global learning model is more appropriate than its local alternative due to the cross-sectional correlation. We validate a model that forecasts national deviation from the global average as a mean-reverting AR(1) process.}
\newline

\noindent
To develop our solar BOS cost model, we first perform an in-sample analysis, followed by a similar validation exercise as for wind turbines. In addition, we illustrate how we have excluded other model alternatives.

\subsubsection{In-sample analysis}

We first perform a descriptive analysis of the solar BOS costs data. National BOS data display a significant temporal trend, cross-correlation between national BOS costs, and autocorrelation within each national BOS cost. 

\paragraph{Trend analysis}
As evident from Figure 1 in the main text and confirmed by a panel regression analysis, there is a significant decline in the logarithm of the BOS costs with increasing cumulative deployment. We are thus looking for a learning model that includes this trend in our cost forecasts.

\paragraph{Cross-sectional correlation analysis}
We use a cross-sectional dependence test for unbalanced panel data, proposed by Pesaran et al.\supercite{pesaranGeneralDiagnosticTests2021, dehoyosTestingCrossSectionalDependence2006} We estimate the statistic
\begin{equation}\label{eqn:cross_sec_dep}
    CD = \sqrt{\frac{1}{N(N-1)}}\left(\sum_{j=1}^{N-1}\sum_{j'=j+1}^{N}\sqrt{T_{jj'}}\hat{\rho}_{jj'} \right),
\end{equation}
where $N$ is the number of countries in our sample, $j$ and $j'$ are the country-pairs we iterate over, $T_{jj'}$ is the number of years for which we have data in both countries and $\hat{\rho}_{jj'}$ is the estimates pairwise Pearson correlation. Under the null hypothesis of no cross-sectional correlation, $CD \sim N(0,1)$.
We find that $CD$ significantly differs from zero or that the national time series are significantly correlated.

If the cross-sectional correlation is large, the statistical significance of our analysis is compromised, even though this will not show up in many standard statistics packages. One way of assessing the limits of the statistical significance is to calculate the {\it Effective Number of Spatial Degrees of Freedom} (ESDOF), a measure originating in Metrology.\supercite{brethertonEffectiveNumberSpatial1999} The ESDOF shows that out of the 42 national BOS time series (59 for wind), only 5 (17 for wind) are statistically independent and can effectively be used for model validation.

To calculate the ESDOF for $N$ time series, we normalize our data by taking the first differences and calculate the $N\times N$ covariance matrix {\bf C}. Using a Singular Value Decomposition, we find the eigenvalues $(\lambda_i)_{i\in[1, N]}$ of {\bf C}. The ESDOF $N^*$ is then given by\supercite{brethertonEffectiveNumberSpatial1999}
\begin{equation}
    N^* = \frac{\left(\sum \lambda_i\right)^2}{\sum \lambda_i^2}.
\end{equation}

Our model validation method has to reflect this correlation. For the remainder of this section, we deal with this challenge by subtracting the cross-sectional average BOS costs from the time series. Specifically, we calculate
\begin{equation}\label{eqn:relative_bos}
	\log \overline{\cJ}_j(t) = \log \cJ_j(t) - \log \cJ(t),
\end{equation}
where $\log\cJ(t)$ is the unweighted global average across all $\log \cJ_j(t)$. The reason for using an unweighted average (compared to a capacity-weighted average) is that we do not want to prioritize accuracy for countries that deploy a lot of solar in our subsequent model validation.

For solar, the resulting time-series $\log \overline{\cJ}_j(t)$ are no longer cross-sectionally correlated, as confirmed by another cross-sectional dependence test with p-value below 1\%. The ESDOF has increased from 5 to 15. This is still significantly below the number of countries in the sample due to the sparse nature of our dataset. Out of the 42 countries in our dataset, only 22 have more then 10 years of available data. 
For wind, the resulting time-series $\log \overline{\cJ}_j(t)$ has 22 ESDOF since more data is available. The cross-sectional dependence test still shows that the time-series are correlated, althouth this is likely from the autocorrelation in $\log \overline{\cJ}_j(t)$.

\begin{figure}
    \centering
    \includegraphics[clip, trim= 2cm 0cm 2cm 0cm,width=\textwidth]{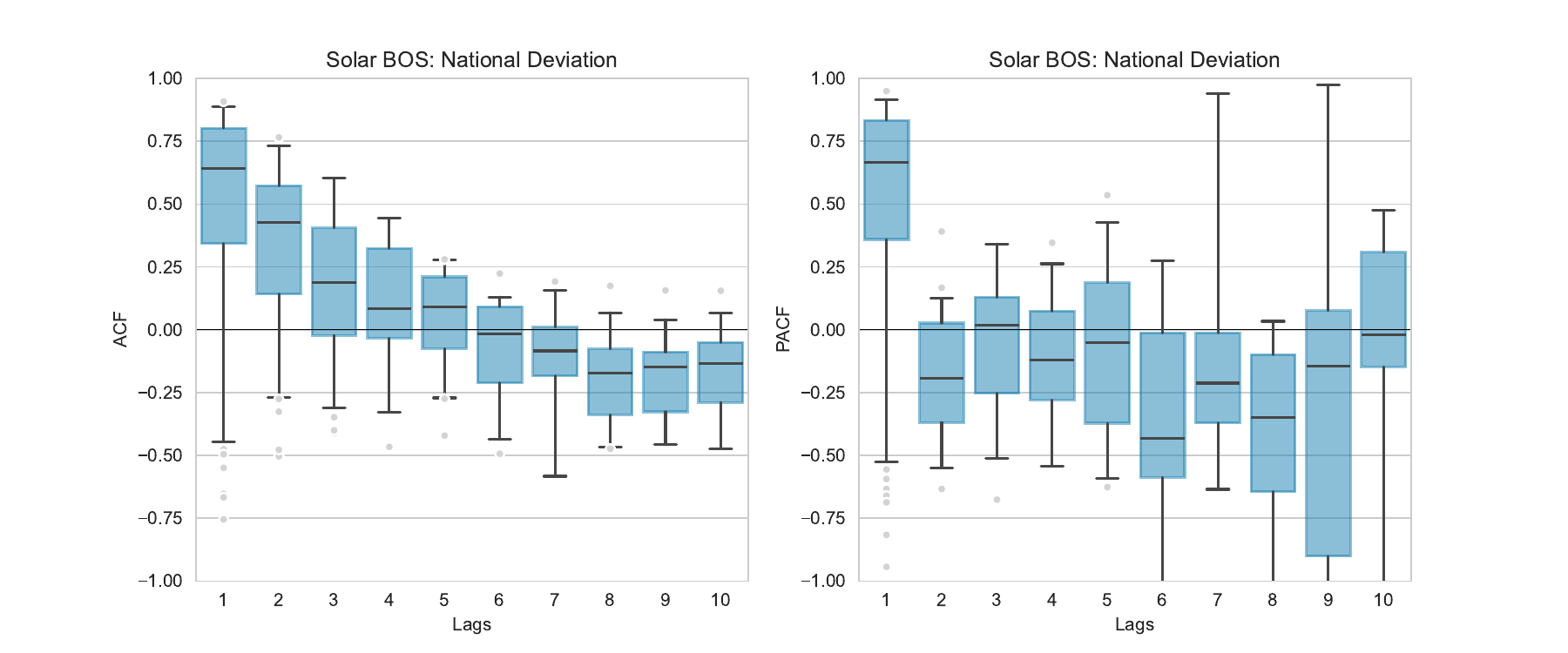}
    \caption{\textbf{Cross-sectional weighted average ACF and PACF for the national solar BOS costs.}
    We calculate national ACF and PACF functions and take their cross-sectional median and confidence intervals. The weights are based on the length of each national time series. \textbf{(left panel)} The cross-sectional weighted average ACF plot. \textbf{(right panel)} The cross-sectional weighted average PACF plot. 
    The line shows the (weighted) median ACF / PACF. The box shows the (weighted) interquartile range. The whiskers show the 90\% confidence range. The dots show outliers exceeding this range.
    The ACF shows a marginally significant first lag. Similarly, the PACF shows a marginally significant first lag. These plots indicate a first-order autocorrelation.
    }\label{fig:BOS_n_acf_solar}
\end{figure}

\paragraph{Autocorrelation analysis}
We calculate the Autocorrelation function (ACF) and Partial Autocorrelation function (PACF) for each national time series. The ACF of a time series $(x_t)_{t \in \N}$ for lag $h$ is estimated as the Pearson correlation between all pairs $x_t$ and $x_{t-h}$. The PACF of the time series is the correlation between $x_t$ and $x_{t-h}$, correcting for any linear dependence on $x_{t-1}, ..., x_{t-h+1}$.
The ACF and PACF of the national time series are very noisy due to the limited number of data points in each country. To reduce the noise, we calculate a weighted median and confidence intervals of the (P)ACF. The weights are given by the length of each national time series since longer time series have a lower measurement error for the (P)ACF.
The resulting average functions are shown in Figure \ref{fig:BOS_n_acf_solar}. Both ACF and PACF show one significant lag. This indicates that our national time series are significantly autocorrelated.

\subsubsection{Solar BOS model validation}\label{app:pv_bos_val}

To validate the global forecasting model for solar BOS costs described in the main text, we require a full description of the errors associated with our forecasts.
\begin{align}
    \log \cJ(t) &= \log \cJ(t-1) - \omega \log \frac{Z(t)}{Z(t-1)} + u_t + \rho u_{t-1},\label{eqn:solar_global_bos} \\ 
    \log \cJ_j(t) - \log \cJ(t) =: \log\overline{\cJ}_j(t) &= \varphi \log\overline{\cJ}_j(t-1) + \eta_{jt}. \label{eqn:solar_nat_bos}
\end{align}
We model the global average BOS costs $\cJ(t)$ as a Wright's law model conditional on global solar capacity $Z(t)$, with i.i.d.\ normal shocks $u_t \sim \cN(0,\sigma^2_u)$, Equation \eqref{eqn:solar_global_bos}.
The national deviations from the global average $\log\overline{\cJ}_j$ are calculated by subtracting the global average from the national time series. The resulting time series is modelled with an AR(1) process, Equation \eqref{eqn:solar_nat_bos}. 
We estimate $\varphi$ as the cross-sectional weighted average of estimated lag-1 correlation parameters, such that $\varphi \approx 0.85$. This is similar to the average lag-1 coefficient for wind, as discussed later.
We estimate the scale of the local noise terms $\eta_{jt} \sim \cN(0, \sigma^2_\eta)$ based on the cross-sectional sample variance $\hat{K}^2_\eta$, similar to Farmer \& Lafond, where \supercite{farmer16}
\begin{equation}
    \hat{K}^2_\eta = \hat{\sigma}^2_\eta / (1 - \varphi^2).
\end{equation}

\begin{figure}
    \centering
    \includegraphics[clip, width=\textwidth]{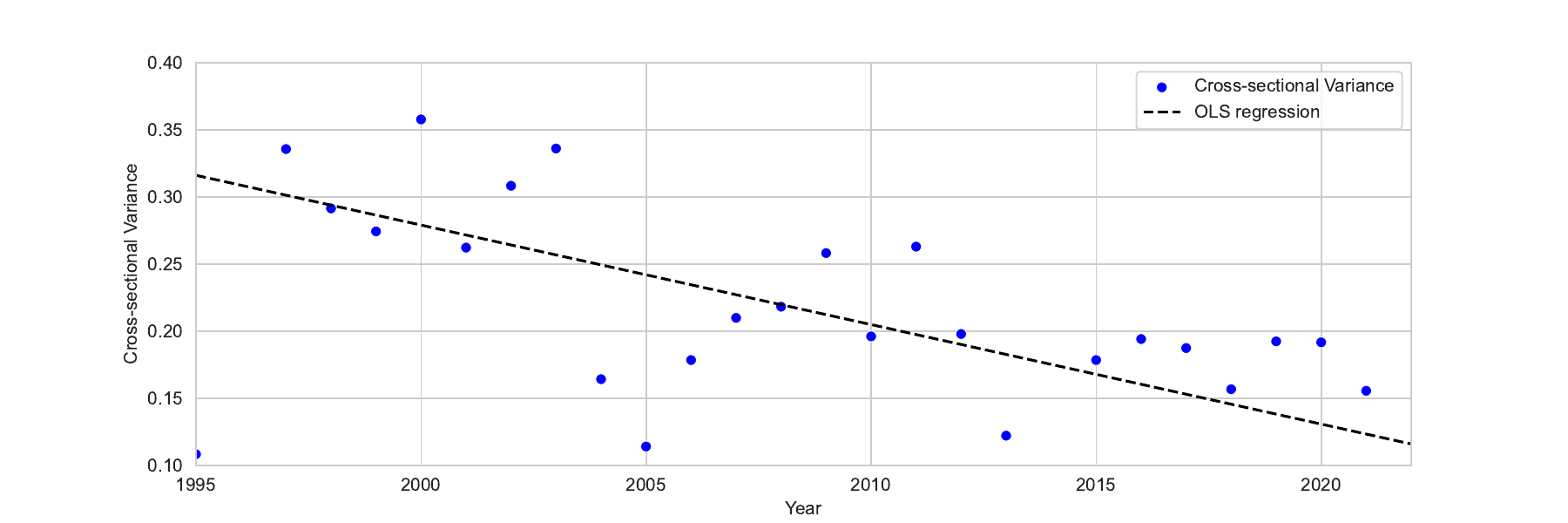}
    \caption{\textbf{Cross-sectional variance of the logarithm of the solar BOS over time.} 
    \textbf{(blue dots)} For each year, we compute the cross-sectional variance in the logarithm of national solar BOS costs, $\log\overline{\cJ}_j(t)$. 
    The black line shows a linear regression of the resulting data. It shows a statistically significant decline.
    }\label{fig:bos_var}
\end{figure}

As shown in Figure \ref{fig:bos_var}, the cross-sectional variance $\hat{K}^2_\eta$ is decreasing over time. Note that this variance is that of $\log \overline{\cJ}_j$, such that $\hat{K}^2_\eta$ is relative and its decrease is not due to the decrease in BOS costs $\cJ_j$.
One possible explanation of the variance decrease is our use of a universal solar module cost $\cM$ when calculating $\cJ_j$; we do not account for national differences in module costs. This effect has been less pronounced in recent years due to the increasing consolidation of the module market.
Nevertheless, we do not know if or how this trend will continue. Therefore, we model the national deviations conditional on the cross-sectional variation $\hat{K}^2_\eta$ and assume that they remain constant in our forecasts.

We focus our model validation on the national deviations from the global average, i.e., $\log \overline{\cJ}_j$ in Equation \eqref{eqn:solar_nat_bos} since the global model has been validated by Lafond et al.\supercite{lafondHowWellExperience2018}
Equivalent to the turbine model described in Appendix \ref{app:model_turbine}, we validate this model using backtesting. For each point in time, we make stochastic forecasts for future values of $\log \overline{\cJ}_j$, conditional on $K_\epsilon^2(t)$. We then calculate the cumulative PIT $q(p)$ and compare it to what we would expect from a surrogate data simulation.

\begin{figure}
    \centering
    \includegraphics[clip, width=\textwidth]{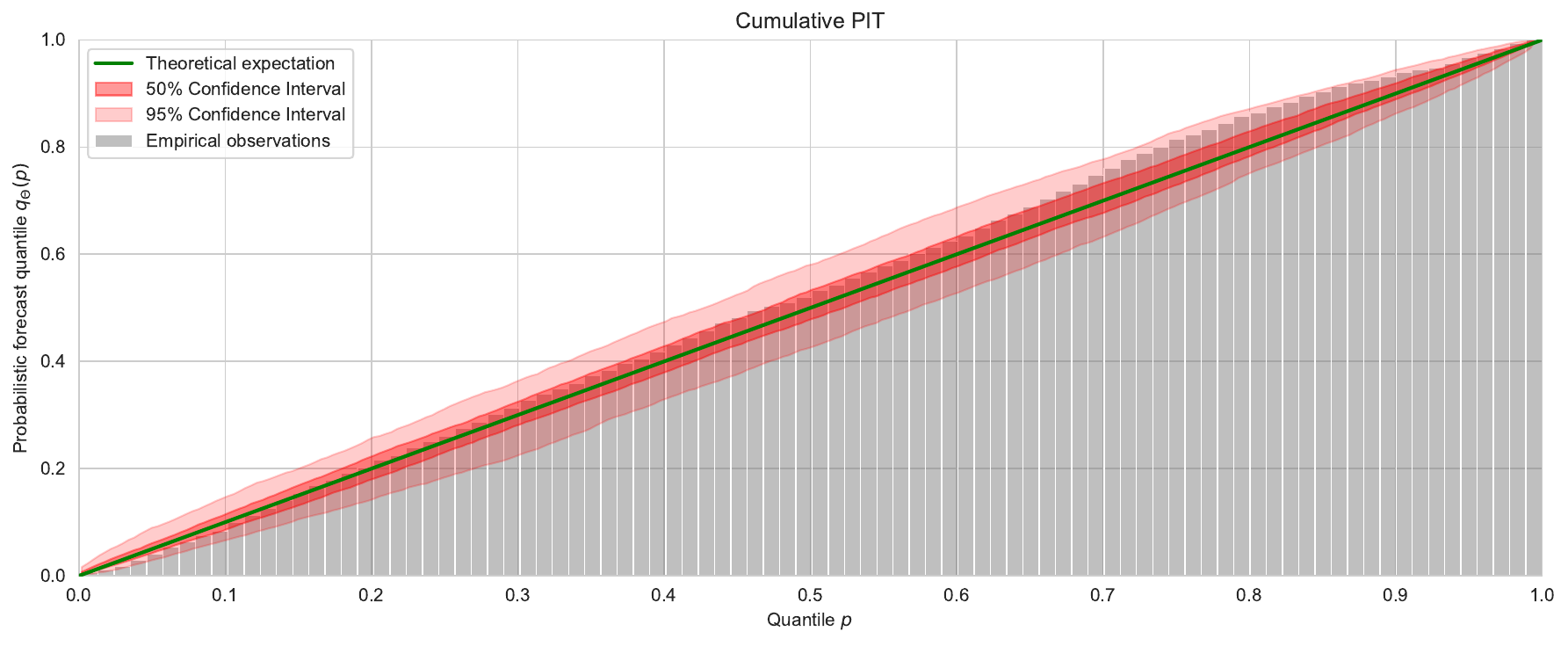}
    \caption{\textbf{Out-of-sample validation of the solar BOS cost forecast $\log\overline{\cJ}_j$, relative to the true global average cost.} 
    We consider only the national solar BOS cost relative to the global average BOS cost $\log\overline{\cJ}_j$. We make stochastic forecasts for the relative costs using the AR(1) model described in the main text, conditional on $K_\epsilon^2(t)$. 
    The empirical cumulative PIT $q(p)$ is shown in gray. The green line is the theoretically expected PIT. The red areas are the 50\% and 95\% confidence intervals under a simulated AR(1) model. We see that the forecast distribution is statistically identical to the AR(1) model.
    }\label{fig:bos_error_solar_dev}
\end{figure}

Figure \ref{fig:bos_error_solar_dev} shows the observed quantiles $q(p)$ against $p$.
We see that the empirical quantiles are close to the identity. While there are some deviations from the universal distribution, they are not statistically significant.

\subsubsection{Solar BOS model comparison}
To illustrate our validation process, we further outline the empirical and theoretical analyses performed to exclude alternative forecasting models.
We compare models that consider BOS costs a purely local component subject to local learning,\supercite{elshurafa18} to models that consider BOS costs subject to global learning. 
Intuitively, we expect local BOS components to be independent from one another and subject to local learning. However, a global model may provide better forecasts for correlated time series, as we have here. 

Table \ref{tbl:pv_bos Models} describes the four model alternatives we compare in this section. We consider three Wright's law models, where costs decline as a function of cumulative deployment, and one Moore's law model, where costs decline over time. 
The \emph{Wright, Global (U)} model is the chosen model described in equations \eqref{eqn:solar_global_bos}-\eqref{eqn:solar_nat_bos}. Global learning is the dominant effect, and national BOS costs depend strongly on what costs are observed in other countries. 
The other three models focus on local learning as the dominant effect, whereby future costs decline independent of other countries deployment. The difference between the \emph{Wright, Local (U)} and \emph{Wright, Local} model is that the former assumes the same learning rate across countries. While cumulative deployment may differ between countries, they all learn at the same rate.

\begin{table}
\begin{centering}
\footnotesize
\begin{tabularx}{\textwidth}{>{\raggedright\arraybackslash}p{2cm} X >{\raggedright\arraybackslash}p{4cm} X}
	\textbf{Name} & \textbf{Description} & \textbf{Functional Form} & \textbf{Learning implications}\\ \toprule
    \rowcolor{green!25}
    Wright, Global (U) & We forecast the global weighted average $\cJ$ based on Wright's law, conditional on global cumulative deployment $Z(t)$. We use the same forecast in all countries. The learning parameter $\omega$ is estimated using a linear regression on the global average. \supercite{way22} & $\log \cJ_j(t) = \newline \phantom{sp}\log \cJ(t-1) - \newline \phantom{spacespace}\omega \log \frac{Z(t)}{Z(t-1)}$ & All countries learn with increasing global experience. There is significant learning spillover through shared experience, i.e., countries depend on each other's deployment.
    \\ \midrule
    \rowcolor{red!25}
    Wright, Local (U) & We forecast each country independently, using Wright's law. We use the same universal Wright's law exponent $\omega$ for all countries based on a panel regression. & $\log \cJ_j(t) = \newline \phantom{sp} \log \cJ_j(t-1) - \newline\phantom{spacespace}\omega \log \frac{Z_j(t)}{Z_j(t-1)}$ & All counties learn at the same rate, with increasing local experience. There is no spillover between countries and costs are independent of deployment in the other countries. Costs are likely to diverge if the cumulative deployment grows at different rates in different countries.
    \\ \midrule
    \rowcolor{red!25}
    Moore, Local (U) & We forecast each country independently, using Moore's law. We use the same universal Moore's law exponent $\theta$ for both countries based on a panel regression. & $\log \cJ_j(t) = \newline\phantom{sp}\log \cJ_j(t-1) - \theta$ & All counties learn at the same rate, independent of their local experience. There is no spillover between countries and costs are independent of other countries. Costs are likely to diverge across countries over time.
    \\ \midrule
    \rowcolor{red!25}
    Wright, Local & We forecast each country independently, using Wright's law. We estimate a country-specific Wright's law exponent $\omega_i$ for each country based on a regression model. \supercite{elshurafa18} & $\log \cJ_j(t) = \newline \phantom{sp}\log \cJ_j(t-1) - \newline \phantom{spacespace}\omega_j \log \frac{Z_j(t)}{Z_j(t-1)}$ & All counties learn with increasing local experience but at different rates. There is no spillover between countries; costs are independent of deployment in the other countries. Costs are likely to diverge even if the cumulative deployment grows at the same rate across countries.
    \\ \bottomrule
\end{tabularx}
\caption{\textbf{Overview of compared forecast models for solar BOS costs.} We compare 4 different forecast models forecasting future solar costs in different countries. $\theta$ and $\omega$ are Moore's and Wright's law parameters, respectively, independent of the market. $\omega_j$ is a Wright's law parameter specific to one market $j$. We choose the \emph{Wright, Global (U)} model as most appropriate for our forecasts. We reject the other models on an empirical and a theoretical basis
}\label{tbl:pv_bos Models}
\end{centering}
\end{table}

\paragraph{Wright, Local model}
We reject the \emph{Wright, Local} model based on one empirical and one theoretical argument. Empirically, we perform a backtesting exercise where the \emph{Wright, Local} model performs worse than its alternatives. We then apply novel theoretical method\supercite{smartMacroMicroPrediction2024} to identify individual countries for which the \emph{Wright, Local} model would outperform its alternatives. For around half the countries, the \emph{Wright, Local (U)} model is preferred due to data limitations. For the remaining half, there is no statistically significant difference between the \emph{Wright, Local} and \emph{Wright, Local (U)} model. This means that we can just as well use the \emph{Wright, Local (U)} model and reject the \emph{Wright, Local} model.

We perform an out-of-sample test based on backtesting:
\begin{enumerate}
    \item Estimate all model parameters on every 5-year window, i.e., between times $t$ and $t-4$
    \item Use the model parameters to forecast future BOS costs in times $t+\tau \ge t$
    \item Compare forecast residuals between models using an aggregate measure, the so-called \emph{scoring rule};\supercite{gneitingStrictlyProperScoring2007} the model with the lowest scoring rule is considered the most accurate.
\end{enumerate}
Here, we use the {\it Root Mean Squared Error} (RMSE) as our scoring rule. The RMSE of model $l$ and forecast horizon $\tau$ is defined as
\begin{equation}\label{eqn:rmse}
    RMSE(\tau) = \sqrt{\frac{1}{n} \sum_{t,j} \left( \log \hat{\cJ}^l_j(t + \tau) - \log \cJ^l_j(t + \tau) \right)^2 },
\end{equation}
where $n$ is the number of times $t$ and respective country $j$ in our sample, and $\hat{\cJ}^l_j(t + \tau)$ is the forecast of $\cJ_j(t + \tau)$ for model $l$.

\begin{figure}
    \centering
    \includegraphics[clip, trim= 2cm 0cm 2cm 1cm,width=\textwidth]{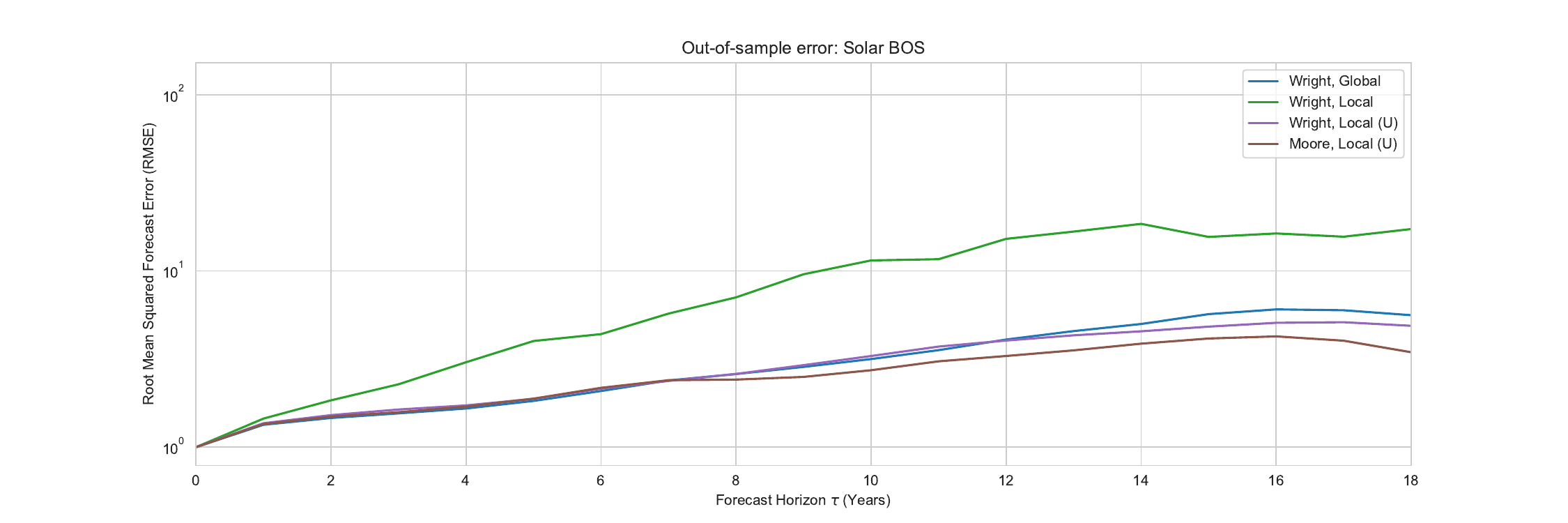}
    \caption{\textbf{Out-of-sample residuals for different forecasting models of solar BOS.}
    BOS costs are forecast under different models described in Table \ref{tbl:pv_bos Models}. We estimate the model parameters on 5-year moving average training data and compare the forecasts with the actual data. 
    The x-axis shows the forecast horizon. The y-axis shows the RMSE of the logarithm of the BOS costs for our testing samples. The RMSE is shown as the relative error on the (linear) BOS cost.
    Except for the \emph{Wright, Local} model, all models show similar forecast accuracy.
    }\label{fig:BOS_mse_solar}
\end{figure}

The results of this analysis are shown in Figure \ref{fig:BOS_mse_solar}. The \emph{Wright, Local} model performs visibly worse than the other models. This difference is significant at 99\% based on an F-test, Barlett-test, and Levene-test. However, none of these tests account for the correlation in our data. These results, therefore, need to be treated with care. 
Furthermore, the test results are sensitive to the size of the training data (5 years). For longer training data, we expect the difference in RMSE between the \emph{Wright, Local} and \emph{Wright, Local (U)} to approach zero or switch sign. \supercite{smartMacroMicroPrediction2024, pesaranForecastingPanelData2024} 
The reason for the lower accuracy of the \emph{Wright, Local} model is that this model is prone to over-fitting--an issue that decreases with a larger training data set. Based on this empirical test, it is not clear if the \emph{Wright, Local (U)} model still performs better than the \emph{Wright, Local} model if we use all available data points to estimate the model parameters.

Since we cannot perform a backtesting exercise whilst also estimating our model parameters on all available data, we rely on the theoretical investigation by Smart \& Farmer \supercite{smartMacroMicroPrediction2024} that compare geometric random walk forecast models with and without parameter heterogeneity. Under the assumption that the national learning parameters $\omega_i$ follow a gamma distribution with parameters $\alpha_\omega$ and $\beta_\omega$, each random walk is subject to i.i.d.\ noise shocks $\epsilon_{it} \sim \cN(0, \sigma_\epsilon^2)$, and each parameter is estimated based on $m$ national data points, the \emph{Wright, Local (U)} model has lower forecast errors than the \emph{Wright, Local} model iff $m > \sigma_\epsilon^2 \beta_\epsilon^2 / \alpha_\epsilon$.
In our case, we find that $\sigma_\epsilon \approx 0.42$. This noise level is significantly higher than in the global model due to national fluctuations. For the distribution of $\omega_i$, we find that a gamma-distribution with $\alpha_\omega \approx 27.0$ and $\beta_\omega \approx 47.9$ describes observed in-sample national learning parameters well (excluding three data points with $\omega_i$ exceeding 1.5). This means that conditional on \emph{Wright, Local} being the true model for $m \ge 15$, the \emph{Wright, Local} model is appropriate to forecast future national BOS costs.

\begin{figure}
    \centering
    \includegraphics[clip,width=\textwidth]{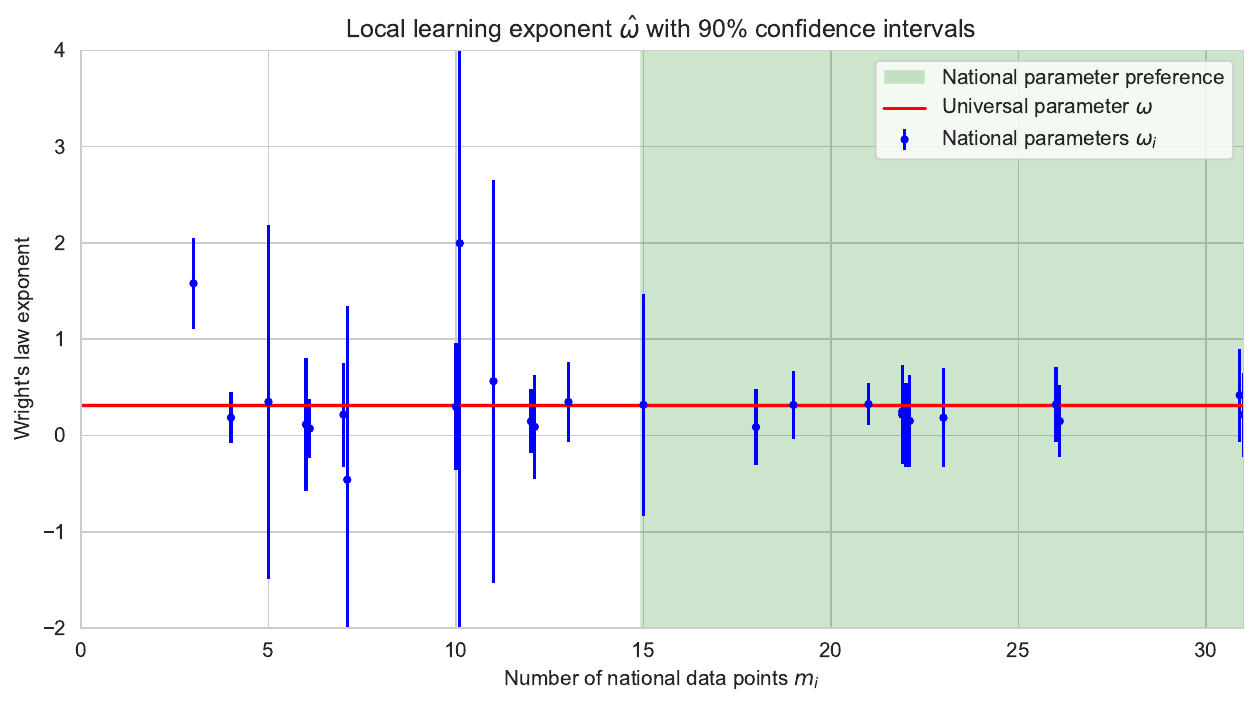}
    \caption{\textbf{National Wright's law parameters $\omega_j$ against the number of available data points.}
    We show the estimated national Wright's law exponents $\hat{\omega}_j$ depending on the number of national data points available ($m_j$). Blue dots indicate the parameters with the associated 90\% confidence intervals. The confidence intervals are not symmetric due to a logarithmic y-scale.
    The red line shows the universal Wright's law parameter $\hat{\omega}$ from the \emph{Wright, Local (U)} model. The green shaded area indicates parameters with $m \ge 15$. For these parameters, a national model would be preferred. 
    We see that for the parameters in the green shaded area, the universal parameter $\hat{\omega}$ always falls into the individual parameter's 90\% confidence intervals of $\hat{\omega}_j$.
    }\label{fig:BOS_solar_param}
\end{figure}

Figure \ref{fig:BOS_solar_param} shows the national parameters $\hat{\omega}_i$ against the number of data points $m$ available for each parameter estimation. The green shaded area indicates $m \ge 15$, i.e., parameters for which the \emph{Wright, Local} model is preferred.
We observe that for those parameters with $m \ge 15$, the universal parameter $\hat{\omega}$ is always included in the 90\% confidence interval of $\hat{\omega}_i$. This has important practical implications since it means that, for those cases where a national parameter is justified, it ceases to make a large difference in our forecasts. This means that in both cases, $m < 15$ and $m \ge 15$, the \emph{Wright, Local (U)} model is appropriate and statistically equivalent to the \emph{Wright, Local} model when a switch could be made. Based on this, we exclude the \emph{Wright, Local} model from further analysis.

\paragraph{Wright, Local (U) \& Moore, Local (U) model}
Empirically, as we have seen in Figure \ref{fig:BOS_mse_solar}, there is no statistically significant difference in point-forecast accuracy between the \emph{Wright, Local (U)}, \emph{Moore, Local (U)}, and \emph{Wright, Global} model. This is not surprising. We expect the \emph{Wright, Local (U)} and \emph{Moore, Local (U)} models to perform similarly a priori due to the exponential growth in cumulative national solar deployment. We also expect the \emph{Wright, Local (U)} and \emph{Wright, Global} models to perform similarly due to the high colinearity between national and global solar deployment.

There are, however, two additional empirical arguments we use to reject the \emph{Wright, Local (U)} model. Since these are equivalent for the \emph{Moore, Local (U)} model, we only provide details on the \emph{Wright, Local (U)} model.
Firstly, we consider the full stochastic forecast that this model implies. Based on a backtesting exercise, we show that the uncertainty associated with the \emph{Wright, Local (U)} model is larger than what we observe in the data. In other words, the uncertainty bands of the \emph{Wright, Local (U)} model are unnecessarily large. 
Secondly, the \emph{Wright, Local (U)} implies an increase in the cross-sectional variance over time since independent random walks naturally disperse. This assumption is not supported by the data. 
We investigate both of these facts in turn.

We first need to define the of out-of-sample residuals under the different models. 
The model used to forecast residuals in the \emph{Wright, Global} model is described in the main text and Section \ref{app:pv_bos_val}. We apply the random walk model of Lafond et al.\ \supercite{lafondHowWellExperience2018} to forecast the global average and use an AR(1) model to describe the national deviations from that global average.
For the \emph{Wright, Local (U)} model, we consider the random walk representation from Lafond et al.\ \supercite{lafondHowWellExperience2018} on the national BOS costs:
\begin{equation}
    \log \cJ_j(t) = \log \cJ_j(t-1) - \hat{\omega} \log\frac{Z_j(t)}{Z_j(t-1)} + \eta_{jt} + \rho \eta_{jt-1},
\end{equation}
where $\eta_{jt} \sim \cN(0, \hat{\sigma}^2)$ i.i.d.\ and $\hat{\omega}$ follows a student distribution around the true value of $\omega$. $\hat{\omega}$ is estimated from a panel regression with fixed effects. $\hat{\sigma}^2$ is estimated on the first log-differences of the cost and cumulative deployment.

Our previous scoring rule, the RMSE, does not capture the full forecast accuracy since it only considers the (median) point forecast. An alternative scoring rule that considers the full distribution of stochastic forecasts is the \emph{Continuous Ranked Probability Score} (CRPS). The CRPS is commonly used to compare stochastic forecasts on continuous random variables. It is defined as
\begin{equation}\label{eqn:crps}
    CRPS(F, y) = \int_\R \left( F(x) - \I_{(x \le y)} \right)^2 dx,
\end{equation}
where $F(\cdot)$ is the cumulative density function of our forecast, and $y$ is the empirical observation. If our forecast is a delta function that matches the true data exactly, i.e., we produce perfect forecasts and are certain that they are correct, then $CRPS = 0$. If either the forecast is off, or we are unsure how accurate the forecast is (or both), then $CRPS > 0$. In our case, we want to estimate the CRPS based on an ensemble of discrete stochastic point forecasts such that we can use Monte Carlo methods to approximate the CRPS. 

There are different discrete estimators for the CRPS. In our case, we apply the so-called ''energy formulation``. For each data point $y$ that we make stochastic forecasts for, we write
\begin{equation}\label{eqn:crps_energy}
    \widehat{CRPS}(\{x_i\}_{i = 1,..., M}, y) = \frac{1}{M}\sum_{i=1}^M |x_i - y| - \frac{1}{2M^2}\sum_{i, j=1}^M |x_i - x_j|,
\end{equation}
where $\{x_i\}_{i = 1,..., M}$ is the stochastic forecast ensemble. $y$ is the empirical out-of-sample data we.\supercite{zamoEstimationContinuousRanked2018} We choose $M=500$ in the estimation below. It is non-trivial to show that Equation \eqref{eqn:crps} and \eqref{eqn:crps_energy} are equivalent. The first term in \eqref{eqn:crps_energy} is the mean absolute error (MAE). The second term is independent of the true data $y$ and, instead, describes the width of the forecast distribution. A narrow and accurate forecast will have low MAE and low forecast width. A narrow yet inaccurate forecast may have low width yet high MAE. A wider forecast will have a large width and high MAE, such that the CRPS quantifies the trade-off between having a very uncertain forecast (but captures the true value) to one that is less uncertain but may not capture the true value. 
While this estimator \eqref{eqn:crps_energy} is biased, it outperforms unbiased alternatives for large values of $M$ and few observations $y$. \supercite{zamoEstimationContinuousRanked2018} 

As for the RMSE, we compare the average CRPS of our two models for different forecast horizons $\tau$. As explained earlier, we focus on the forecast residual relative to the global average cost, $\log\overline{\cJ}_j$ defined in Equation \eqref{eqn:relative_bos}, since this significantly increases the statistical significance of our comparison.

\begin{figure}
    \centering
    \includegraphics[clip, width=\textwidth]{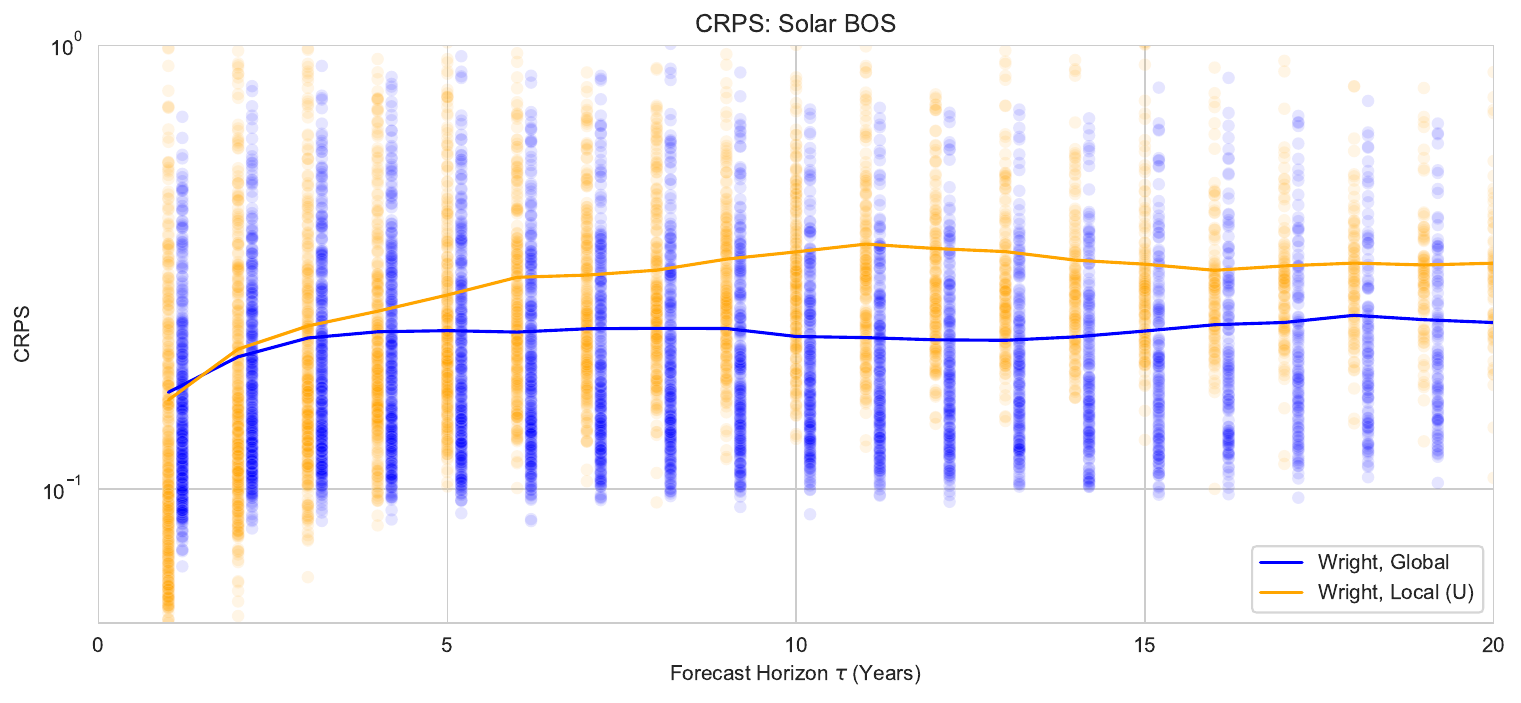}
    \caption{\textbf{Out-of-sample comparison of the solar BOS cost forecast errors under the \emph{Wright, Local (U)} and \emph{Wright, Global (U)} model relative to the global average solar BOS costs, in terms of CRPS.} 
    We consider only the national solar BOS cost relative to the global average BOS cost. We estimate model parameters on 5-year moving average windows on our data and forecast the costs relative to the true global average using the \emph{Wright, Global (U)} (orange) and \emph{Wright, Local (U)} model (blue). We make stochastic forecasts for the future of this window with a Monte Carlo approach. For each forecast, we estimate the CRPS using the energy formulation.
    The dots represent individual forecasts, and the lines are the average overall forecasts with each model. 
    We see that, with the exception of $\tau = 1$, the \emph{Wright, Global (U)} model outperforms the \emph{Wright, Local (U)} model. 
    }\label{fig:BOS_error_solar_crps}
\end{figure}

The results of this analysis are shown in Figure \ref{fig:BOS_error_solar_crps}. Except for $\tau = 1$, the \emph{Wright, Global (U)} model shows a lower CRPS than the \emph{Wright, Local (U)} model. The average CRPS is smaller, and there are several individual data points for which the forecasts are more accurate under the \emph{Wright, Global (U)} model. Since we do not find a different RMSE of both forecasts, this means that while both point forecasts have similar accuracy, the \emph{Wright, Global} model has a smaller confidence interval around these forecasts. The \emph{Wright, Local (U)} model over-estimates the future uncertainty around the point forecast.

The reason for this difference between the models lies in their forecast for cross-sectional averages. The \emph{Wright, Global (U)} model predicts the deviation from the global average BOS costs with an AR(1) model. This means that the cross-sectional variance remains approximately constant over time. The \emph{Wright, Local (U)} model, on the other side, predicts an increase in cross-sectional variance over time. Differences in national solar deployment infer different expected costs. Additionally, the stochastic noise is independent for each country, resulting in further divergence of national costs.
In reality, we have seen in \ref{fig:bos_var} that the cross-sectional variance has declined historically. 
Since this is contrary to what we expect under the \emph{Wright, Local (U)} model, we reject it and work with the \emph{Wright, Global (U)} model going forward.

\subsection{Wind BOS cost forecasts}\label{app:model_bos_wind}
{\bf Summary:}
\emph{We do not find significant historical cost declines in national wind BOS costs. Instead of an experience curve model, we forecast future global BOS costs as a mean-reverting process. Equivalent to solar, we forecast national deviations from the global mean as a mean-reverting process.}

\subsubsection{In-sample analysis}
Before developing our forecast model, we perform an in-sample analysis of the wind BOS costs. This demonstrates that wind BOS costs follow a constant trend and are significantly cross-correlated. These findings are robust with respect to the lag we use to calculate national BOS costs.

\paragraph{Trend analysis}
We perform a regression analysis on national BOS costs using the model
\begin{equation}\label{eqn:bos_regression}
    \log \cJ_j(t) = \alpha_j + \omega * \log Z_j(t) + \Tilde{\omega} * \log Z(t).
\end{equation}
$\alpha_j$ is a country fixed effect, $\omega$ is a Wright's law type learning parameter with respect to the local cumulative deployment $\log Z_j(t)$, and $\Tilde{\omega}$ is a Wright's law parameter with respect to the global cumulative deployment $\log Z(t)$.

The results of this analysis (and derivatives thereof) are shown in Table \ref{tab:bos_regression}. For solar, we see highly significant global and local learning. This is expected given the large cost decline and take up of solar electricity. For wind we do not see any significant global or local learning (except a 10\% significance for local learning).

\newcolumntype{Y}{>{\centering\arraybackslash}X}
\begin{table}
    \centering
    \footnotesize
    \begin{tabularx}{1\textwidth}{>{\raggedright\arraybackslash}p{3cm} Y Y Y | Y Y Y}
        & \multicolumn{3}{c}{Solar PV} & \multicolumn{3}{c}{Wind} \\ \hline
        Local learning $\omega$ & -0.3114$^{***}$& &  -0.1692$^{***}$ & -0.0218$^{*}$& & -0.0328\\ 
        Global learning $\Tilde{\omega}$ & & -0.3666$^{***}$ & -0.1785$^{***}$ & &-0.0255 & 0.0175\\ \hline
        Entity Effects & X & X & X & X & X & X \\ \hline
        $R^2$ & 0.7895 & 0.7912 & 0.8151 & 0.0059 & 0.0044& 0.0065\\ \hline
        \multicolumn{2}{l}{\tiny $^{*}\, p<0.1 $ ;$^{**}\, p<0.05 $ ;$^{***}\, p<0.01$ }
    \end{tabularx}
    \caption{Regression results for BOS costs, according to Equation \eqref{eqn:bos_regression}. For Solar, all regression parameters are highly significant. 
    For wind, the local learning parameter is significant below 10\% if we include only local learning.}
    \label{tab:bos_regression}
\end{table}

This absence of learning has been noted previously by Key et al.\ \supercite{keyScalingTrendsBalanceofsystem2022} but has otherwise gotten little attention in the literature.
In some sense, the BOS costs are comparable to price development in the fossil fuel sector, where inflation-adjusted prices have remained approximately constant for more than a century despite improvements in extraction technologies. \supercite{pindyckLongRunEvolutionEnergy1999}
As a result, we expect further learning effects in onshore wind to be substantially below those in solar. Turbine costs are already decreasing slower (with respect to their deployment) than solar modules and BOS costs, leading to significant short- and long-term differences in cost reduction.

\paragraph{Cross-sectional correlation analysis}
We apply the same cross-sectional correlation test to our data as we did for solar. Specifically, we observe significant cross-sectional correlation based on a cross-sectional dependence test from Pesaran et al.,\supercite{pesaranGeneralDiagnosticTests2021} see Equation \eqref{eqn:cross_sec_dep}. We need to take this into account in our forecast model. 

\begin{figure}
    \centering
    \includegraphics[clip, trim= 2cm 0cm 2cm 0cm,width=\textwidth]{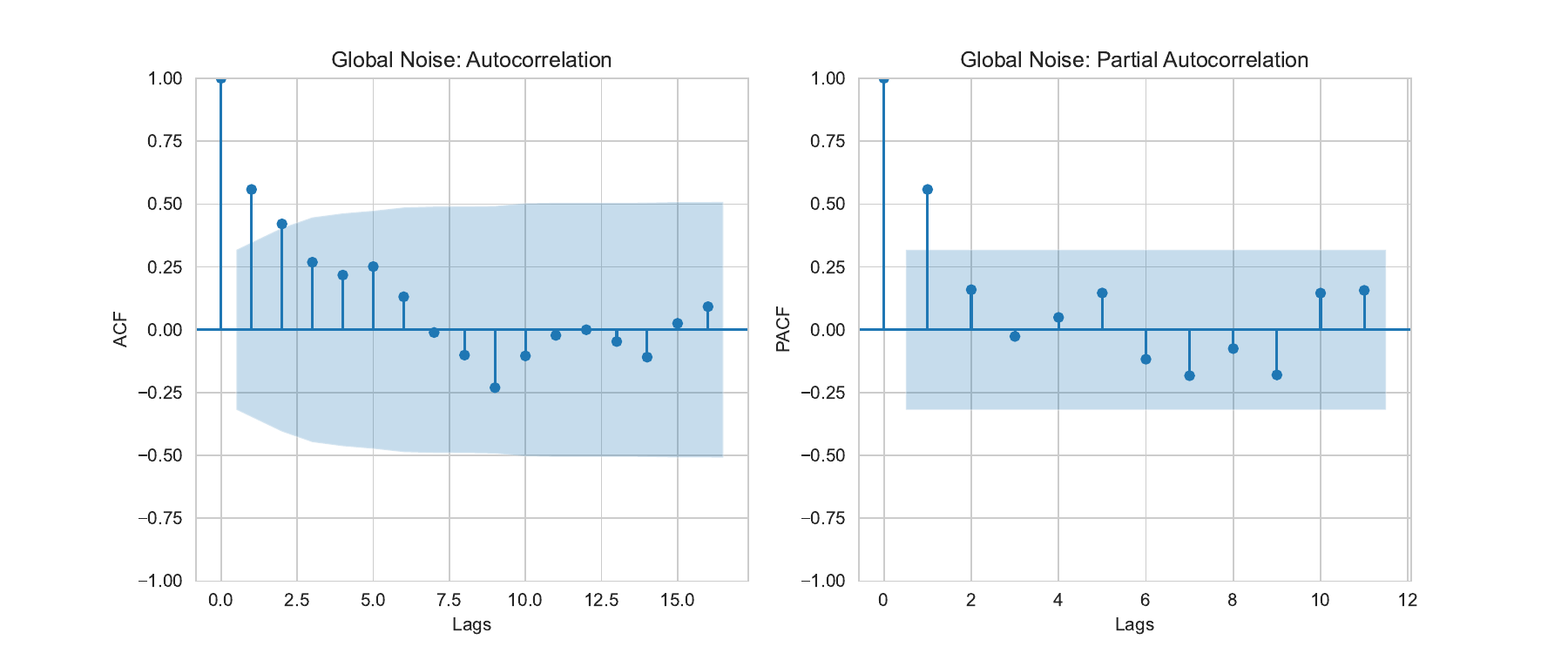}
    \caption{\textbf{ACF and PACF for the cross-sectional average wind BOS costs.} 
    We consider the cross-sectional (national) average BOS costs for wind.
    \textbf{(left panel)} The ACF plot for the resulting time series. \textbf{(right panel)} The partial PACF plot for the resulting time series.
    The ACF is declining gradually, although only the first three lags are significant, indicated by the blue shading. The PACF is declining more abruptly, with only the first lag being significant.
    These two plots indicate an AR process at the origin of our data.}\label{fig:BOS_g_acf}
\end{figure}

\paragraph{Autocorrelation analysis}
We first investigate the autocorrelation in our global average BOS costs $\log\cJ(t)$. As for solar, $\log\cJ(t)$ is calculated as the unweighted average across all $\log\cJ_j(t)$, where $j$ is the country identifier and $t$ is the year of observation.
We then compute the autocorrelation function (ACF) and partial autocorrelation function (PACF) for the resulting time-series $\log\cJ(t)$. We plot the autocorrelation and partial autocorrelation function of the resulting time series in Figure \ref{fig:BOS_g_acf}.
The autocorrelation function (left panel) declines gradually. However, only the first two lags are significant, as indicated by the blue shading. The partial autocorrelation function (right panel) declines rapidly after the first significant lag.
These functions indicate that the global average can be approximated by an AR(1) process.

\begin{figure}
    \centering
    \includegraphics[clip, trim= 2cm 0cm 2cm 0cm,width=\textwidth]{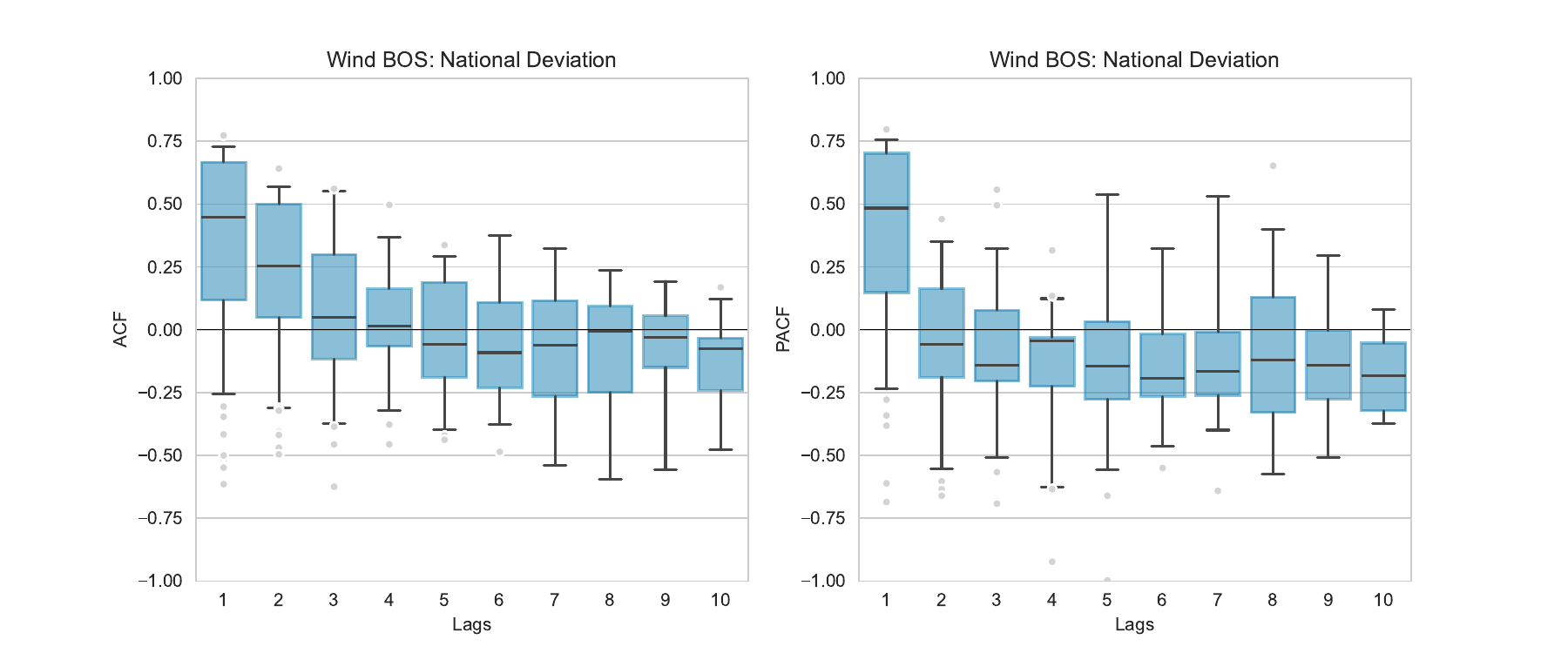}
    \caption{\textbf{Cross-sectional weighted average ACF and PACF for the national wind BOS cost deviations.}
    We consider the national BOS costs deviation from the global average. We calculate national ACF and PACF functions and take their cross-sectional averages. \textbf{(left panel)} The cross-sectional weighted average ACF plot. \textbf{(right panel)} The cross-sectional weighted average PACF plot. The weights are based on the length of each national time series.
    The line shows the weighted median ACF / PACF. The box shows the weighted interquartile range. The whiskers show the 90\% confidence range. The dots show outliers exceeding this range.
    The ACF shows a marginally significant first lag. Similarly, the PACF shows a marginally significant first lag. These plots indicate a first-order autocorrelation.
    }\label{fig:BOS_n_acf}
\end{figure}

We repeat a similar analysis for the national BOS costs. For each country in our sample, we normalize the time-series by subtracting the global average BOS costs, according to equation \eqref{eqn:relative_bos}. We then calculate the ACF and PACF for $\log\overline{\cJ}_j(t)$ in each country $j$ and take the respective cross-sectional weighted averages. The weighting depends on the length of the national time-series since longer time-series have a lower measurement error for the ACF and PACF. The reason for taking the average is the high level of noise in individual countries. The cross-sectional average reduces that noise to provide reasonable autocorrelation functions. 
The resulting average functions are shown in the left and right panels of figure \ref{fig:BOS_n_acf}. We find one significant lag in both the ACF or PACF.
This means that we can approximate the $\log\overline{\cJ}_j(t)$ time-series with a similar first-order autocorrelation process.

\paragraph{Sensitivity analysis in BOS estimation}
As described in the main text, we calculate the wind BOS costs using a 1.5-year lag for turbine costs, such that
\begin{equation}
    \cJ_j(t) = I_j(t) - \omega_{t-1}\cM_j(t-1) - \omega_{t-2}\cM_j(t-2),
\end{equation}
where $I(t)$ is the total investment cost and $\cM(t)$ the turbine costs in year $t$, and $\omega_{t} = 1/2$ denotes the weights applied to individual years.
While the length of this lag is based on previous literature, it is important to investigate how robust our results are against this assumption. We repeat the previous regression analysis of wind BOS costs for different values of $\omega_t$ or, equivalently, different lags.

Table \ref{tab:bos_sensitivity} shows our previous regression results for different lags. We focus on the 1-dimensional regressions with respect to the global and local cumulative deployment. The results show that for different lags, there is no significant learning. All regression results are insignificant at 5\%.
This means that our regression results are robust with respect to the lag applied to the turbine costs and there is no significant learning effect.

\begin{table}
    \centering
    \footnotesize
    \begin{tabularx}{1\textwidth}{>{\raggedright\arraybackslash}p{3cm} Y Y Y Y Y Y Y}
        \hline
        Lag (Years) & 0 & 0.5 & 1 & 1.5 & 2 & 2.5 & 3\\ \hline
        Only Local learning $\omega$ & -0.0112 & -0.0164 & -0.0196$^{*}$ & -0.0218$^{*}$ & -0.0216$^{*}$ & -0.0208$^{*}$ & -0.0145\\
        Only Global learning $\Tilde{\omega}$ &  -0.0006 & 0.0120 & -0.0194 & -0.0255 & -0.0285$^{*}$ & - 0.0310$^{*}$ & -0.0195\\ \hline
        Entity Effects & X & X & X & X & X & X & X\\ \hline
        \multicolumn{2}{l}{\tiny $^{*}\, p<0.1 $ ;$^{**}\, p<0.05 $ ;$^{***}\, p<0.001$ }
    \end{tabularx}
    \caption{Regression results for wind BOS costs, according to Equation \eqref{eqn:bos_regression}, under different lags. We see that independent of the lag applied to the turbine costs, the cost improvement in the wind BOS costs is not significant.}\label{tab:bos_sensitivity}
\end{table}

\subsubsection{Wind BOS model validation}
Based on these in-sample statistics, we develop a model that reflects the absence of learning effects and correlation structures.
We take from the approach in Way et al.\ \supercite{way22} for fossil fuel prices and model future wind BOS costs as an AR(1) process around the historical average.
\begin{align}
    \log \cJ(t) &= \phi \log \cJ(t-1) + \epsilon_t + \kappa, \label{eqn:wind_global_bos} \\
    \log \cJ_j(t) - \log \cJ(t) \equiv \log\overline{\cJ}_j(t) &= \varphi \log \overline{\cJ}_j(t-1) + \eta_{jt}. \label{eqn:wind_nat_bos}
\end{align}
As for solar PV, Equation \eqref{eqn:wind_global_bos} describes the global average wind BOS cost $\log \cJ(t)$. Equation \eqref{eqn:wind_nat_bos} describes the national deviation from the global average $\log \overline{\cJ}_j(t)$.
Both the global average BOS costs \eqref{eqn:wind_global_bos}, as well as the national deviations thereof \eqref{eqn:wind_nat_bos} follow an AR(1) process. The global average is calculated as a cross-sectional average of national BOS costs in each year $t$. The national deviations are calculated as the difference from that average in each year $t$.

For both global average \eqref{eqn:wind_global_bos} and national deviations \eqref{eqn:wind_nat_bos} we need to determine the autocorrelation parameters $\phi$ and $\varphi$, and the size of the exogenous shocks $\epsilon_t \sim \cN(0, \sigma_\epsilon^2)$ and $\eta_{jt} \sim \cN(0, \sigma_\eta^2)$. Furthermore, the global average model \eqref{eqn:wind_global_bos} depends on the constant $\kappa$.
For $\phi$, we only have one (global) short time series available, making estimation very difficult. Therefore, we use the average value for fossil fuels in Way et al.\supercite{way22} of $\phi = 0.88$.
For $\varphi$, time series are also short, but we have multiple independent cross-sectional (national) time series. We can, therefore, estimate $\varphi_j$ based on a maximum likelihood estimation in all countries $j$ and take the average thereof. This leaves us with $\varphi \approx 0.7$, similar to our model for solar BOS costs.
For $\hat{\kappa}$, we use the long-term mean cost to estimate the constant
\begin{equation}
    \hat{\kappa} = \frac{1}{m}\sum_t \log \cJ(t) (1 - \phi).
\end{equation}
The estimation error of $\hat{\kappa}$ can be estimated based on the usual standard error in the mean.
Finally, we estimate the variance of the i.i.d.\ shocks $\epsilon$ and $\eta$ based on the residuals of Equations \eqref{eqn:wind_global_bos} and \eqref{eqn:wind_nat_bos}.

As before, statistical significance is a challenge when it comes to validating this model due to the limited number of independent data points. For the global model \eqref{eqn:wind_global_bos}, an AR(1) process is the natural choice given its mean-reverting nature. Our choice of $\phi$ is also a natural choice that offers a good match to the empirically observed forecast accuracy. We focus our validation on the national deviation model \eqref{eqn:wind_nat_bos} where we can make statistically significant statements. We further do not include the estimation of $\varphi$ in our validation but take it as a given input. Estimating the autocorrelation parameter on a short time series is challenging and limits our ability to validate the more general model and noise parameter $\sigma_\eta$.\supercite{farmer16}

Our procedure is the same as previously described; we consider 5-year moving average windows to estimate $\sigma_\eta$ and make stochastic forecasts for the respective future of $\log\overline{\cJ}_j(t)$. We compare the distribution of our forecasts to that of the residuals across quantiles $p$ using the cumulative PIT.
We limit this analysis to forecasts made from 2003 onward. The reason for this is the limited data on national investment costs before that. Prior to 1999, we only have data for 6 countries, resulting in only 5 independent time series since one degree of freedom is used to estimate the global average BOS cost. Moreover, prominent data sources like IRENA\supercite{irenaRenewablePowerGeneration2021} show exactly equivalent costs for some countries, such as Germany, Sweden, Italy, Spain, and France, in 1993. It is unclear if the reason for this anomaly is due to issues with the underlying data or reflects an actual dependence of costs between these countries.

\begin{figure}
    \centering
    \includegraphics[clip, width=\textwidth]{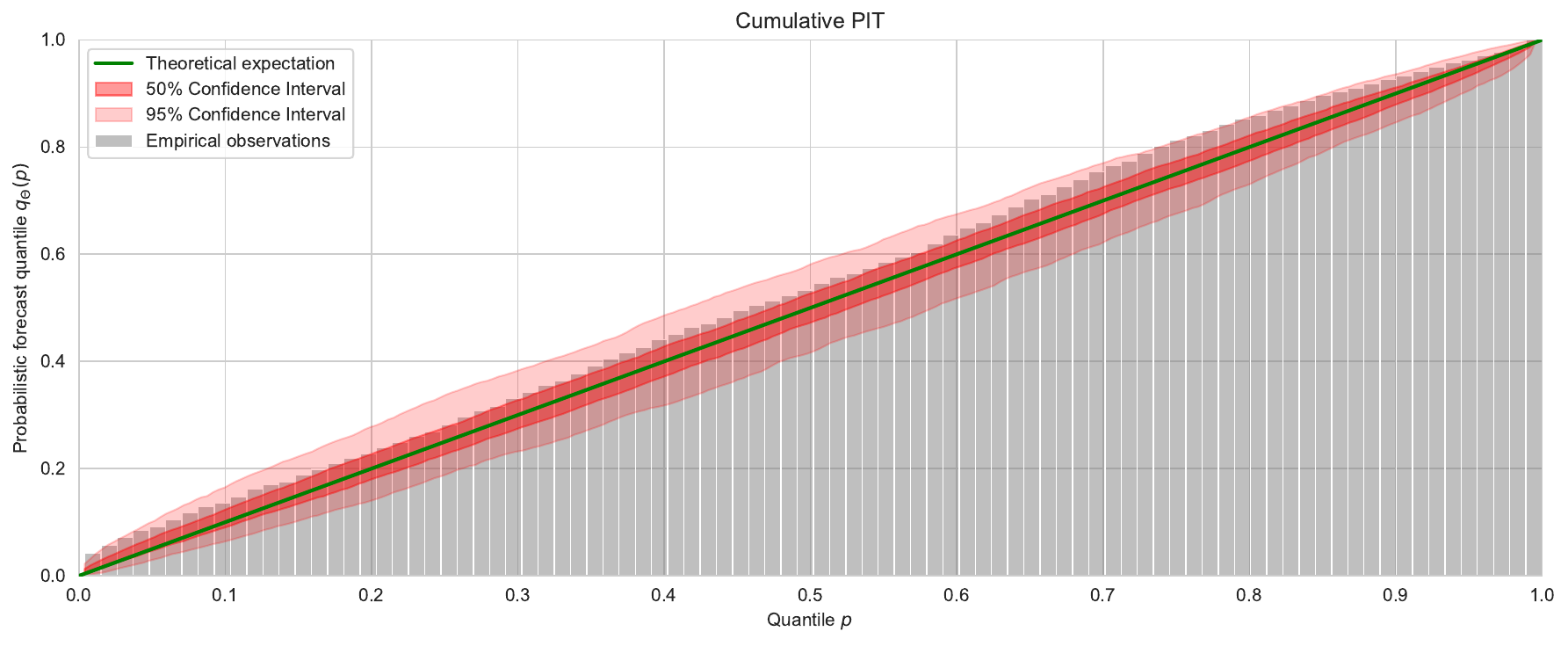}
    \caption{\textbf{Out-of-sample validation of the national wind BOS cost forecast error, relative to the true global average cost.} 
    We consider only the national deviation of wind BOS cost relative to the global average BOS cost \eqref{eqn:wind_nat_bos}. We estimate model parameters $\sigma_\eta$ on 5-year moving average windows on our data and make stochastic forecasts for the relative costs using the AR(1) model described in the main text. 
    The empirical cumulative PIT $q_\Theta(p)$ is shown in gray. The green line is the theoretically expected PIT. The red areas are the 50\% and 95\% confidence intervals under a simulated AR(1) model. We see that the forecast distribution is statistically identical to the AR(1) model.
    }\label{fig:bos_error_wind}
\end{figure}

Figure \ref{fig:bos_error_wind} shows the observed cumulative PIT $q_\Theta(p)$ against $p$. The gray bars are the empirical observations, and the red areas are the associated confidence intervals. As they were for the previous models, these are based on surrogate data. We see that the empirical quantiles are close to the identity line across all percentiles without any statistically significant differences.

\subsection{Investment cost forecasts}\label{app:model_inv}
{\bf Summary:}
\emph{We aggregate our forecasts for the modules/turbines and BOS costs to obtain a forecast for the total investment cost. For wind, the difference in learning rates between turbines and BOS costs implies that our aggregate model for investment cost forecasts is preferred over a pure Wright's law model.For solar, due to the comparable learning rates of photovoltaic modules and BOS costs, our aggregate model performs similarly to a simpler Wright's law model on the investment cost. }
\newline

When we are interested in the total investment costs $I$, we can sum the disaggregate module/turbine and BOS forecasts according to
\begin{equation}
    I_{jt} = \cM_{jt} + \cJ_{jt}.
\end{equation}

Since the disaggregate model (based on $\cM$ and $\cJ$) contains more information, we would expect it to always outperform alternative formulations on the aggregate $I$. However, if we have little data to estimate our model parameters, using Wright's law on the total investment cost $I$ directly could be preferred over forecasting turbine/modules and BOS costs independently.\supercite{grunfeldAggregationNecessarilyBad1960, berminghamUnderstandingForecastingAggregate2014, hendryCombiningDisaggregateForecasts2011, lutkepohlForecastingContemporaneouslyAggregated1984} The reason for this is the bias-variance trade-off discussed earlier.\supercite{baumgartnerNeedBetterStatistical2024}
Furthermore, turbine, module, and BOS costs may be correlated with one another. In this case, the disaggregate model may overstate the future uncertainty on the total investment cost $I$.

This topic warrants further study outside the scope of this work.\supercite{ferioli09_2} Here, we perform two simple analyses to identify a model. Firstly, we investigate the cross-correlation between solar modules, wind turbines, and respective BOS costs. We then perform a heuristic test to identify which of the two investment cost model alternatives would perform better, conditional on the true underlying process.

\paragraph{Cross-correlation analysis}
We consider the first difference of global average log costs for modules, turbines, and BOS costs. We then compute the correlation coefficient between the three time series using a Pearson test. We find that neither wind turbines and BOS nor solar modules and BOS costs are correlated with statistical significance. This indicates that we can use both forecast models independently and combine their results into an aggregate forecast.

\paragraph{Simulation analysis}
Using a simulation-based test, we show that the aggregate/disaggregate model choice most likely will not make a large difference for solar. For wind, the disaggregate model is likely the preferred choice.

We choose a simplified Moore's law setup to simulate surrogate data for two time series $X$ and $Y$. Both of them follow an equivalent process with different parameters, i.e.,
\begin{align}
    \text{solar module / wind turbine costs: } X_t &= X_{t-1}\ee^{\theta_X} \ee^{\epsilon_t} \\
    \text{BOS costs: } Y_t &= Y_{t-1}\ee^{\theta_Y} \ee^{\epsilon'_t},
\end{align}
where $\epsilon_t,\ \epsilon'_t \sim \cN(0, \sigma^2)$ are i.i.d.\ noise shocks.
We consider two settings. In the first, we set both learning parameters $\theta_X = \theta_Y = -0.1$, close to what we observe for solar modules and BOS costs. In the second, we consider $X$ evolving as a geometric random walk with $\theta_X = -0.1$ and $Y$ evolving as an AR(1) process ($\theta_Y = 0$ plus mean-reversion). 
In both settings, we simulate time series $X$ and $Y$ and make forecasts using both disaggregate and aggregate model parameters. We fit the forecast model on a subset of $m$ years and use this to forecast 15 years into the future and calculate the Root Means Squared Error (RMSE) of these forecasts over a sample of 5000 simulation runs.

\begin{figure}
    \centering
    \includegraphics[clip,trim=0cm 0cm 0cm 0cm,width=\textwidth]{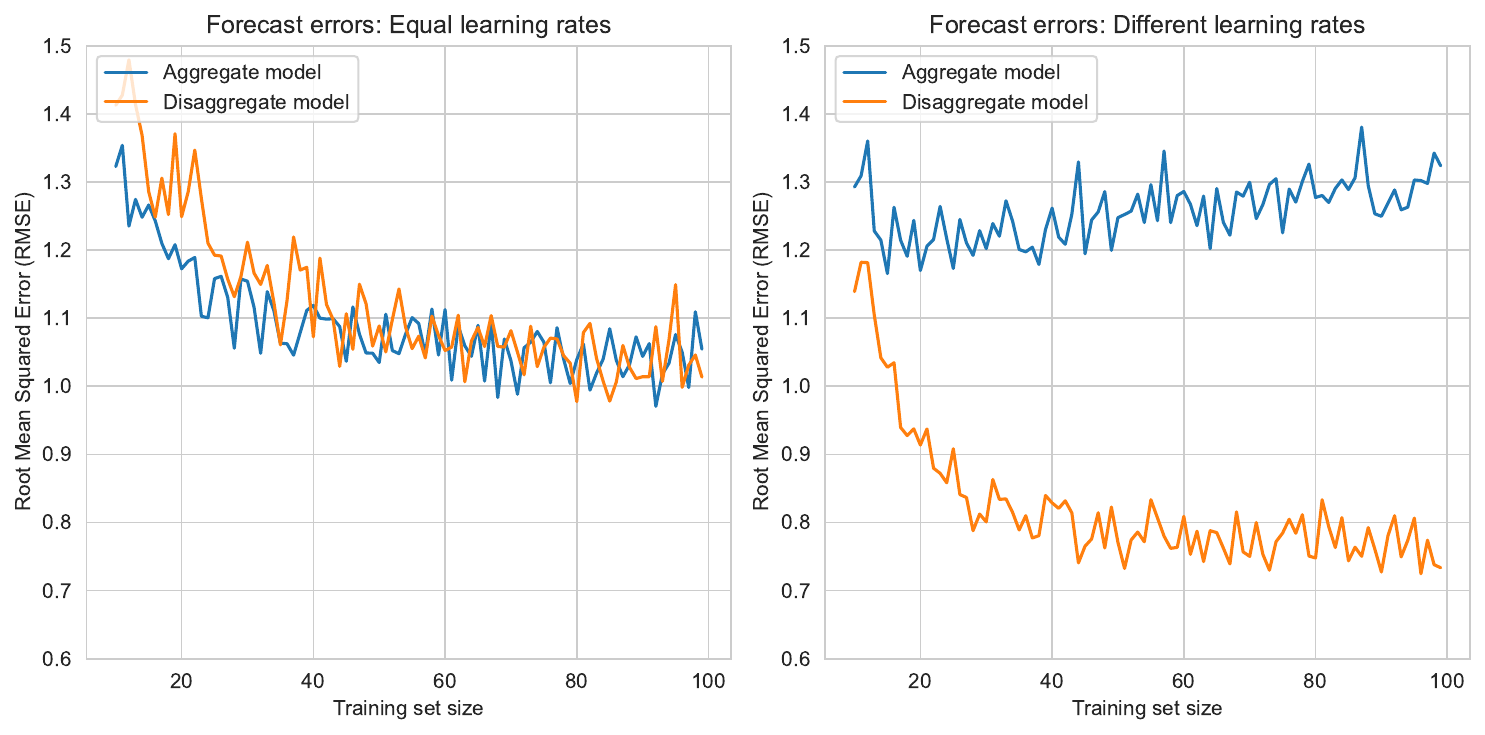}
    \caption{\textbf{Surrogate out-of-sample residuals for aggregate and disaggregate models, depending on the underlying processes.} 
    We simulate the sum of two geometric random walks \textbf{(left, equivalent to solar)} and the sum of a geometric random walk and an AR(1) process \textbf{(right panel, equivalent to wind)}. In each case, we apply a disaggregated (orange) and aggregated (blue) Moore's law model with different sizes of the training set. We compare the out-of-sample residuals of both models for 15-year ahead forecasts using the RMSE.
    We see that for two geometric random walks, both models perform very similarly, particularly if more than 25 years of data are used to estimate the parameters. If one model evolves according to an AR(1) process, the disaggregate model significantly outperforms the aggregate, even with very few data being used to estimate model parameters.}\label{fig:inv_sim}
\end{figure}

The results of this simulation are shown in Figure \ref{fig:inv_sim}. 
With equal learning, both models perform very similarly. The aggregate model outperforms the disaggregate for small training sets, but the difference is small and disappears if more than 25 data points are available. 
For the process with different learning rates, the two models perform differently. Even with few data points being used to estimate parameters, the disaggregate model outperforms the aggregate. This means that modellers may use either the aggregate or disaggregate model for solar but should use the disaggregate model in the case of wind.

\subsection{Capacity Factor forecasts}\label{app:cf}
{\bf Summary:}
\emph{We show that national capacity factors for solar and wind are approximately constant over time. We model capacity factors as i.i.d\ random variables. We compare two ways of estimating the capacity factor based on the annual electricity generation and the LCOE data. Using an out-of-sample test, we show that the LCOE-based estimation provides more accurate forecasts.}
\newline

\noindent
We first investigate the change of national capacity factors over time. The literature disagrees on this topic, with some researchers depicting steep increases in capacity factor, \supercite{irenaRenewablePowerGeneration2021, nrelAnnualTechnologyBaseline2022} while others describe capacity factors as historically constant. \supercite{borettiTrendsPerformanceFactors2020, bolingerUtilityScaleSolarEmpirical2018, boccardCapacityFactorWind2009}
We investigate trends in the capacity factor based on historical electricity generation and capacity.

\subsubsection{Capacity Factor stationarity}

To assess the stationarity of a country's capacity factor, we follow the approach by Bolson et al.\ \supercite{bolsonCapacityFactorsElectrical2022} and consider the actual national electricity generated per installed capacity.
This approach differs from bottom-up modelled capacity factors, such as the one calculated by the Global Solar Atlas,\supercite{esmapGlobalSolarAtlas} which explicitly considers solar irradiation and other environmental factors. While the bottom-up approach is helpful for developing new solar sites and comparing technological choices, it can lead to significant estimation errors when compared with actual capacity factors. \supercite{bolsonCapacityFactorsElectrical2022}

To identify trends in national capacity factors, we need to find a closed-form expression for the capacity factor of the newly built assets in a given time period. To do so, we must consider the large year-on-year growth in renewable capacity as well as the degradation $D_j(t)$ of existing assets.
Accrued over an asset's lifetime, degradation of equipment has a large effect on the capacity factor. Assuming a 20-year asset lifetime, solar equipment degrade by around 21\% and wind assets by 28\%.\supercite{staffellHowDoesWind2014, kimReviewDegradationPhotovoltaic2021}

Unfortunately, asset-level capacity and generation data are not widely available. Instead, our data is limited to the total annual electricity generation $Y_j(t)$, and the end-of-year electric capacity $X_j(t)$. One way to solve this constraint is to use a weighted average capacity factor $\cY'_j(t)$ that we derive below.

For a given country $j$, we start by writing
\begin{equation}
    Z_j(t) = Z_j(t-1) +X_j(t),
\end{equation}
where $Z_j(t)$ denotes the total national capacity at the end of year $t$, and $X_j(t)$ is the additional capacity built throughout year $t$.

The total electricity generation $Y_j(t)$ in year $t$ depends on the installed capacity. Additionally, we need to account for the degradation $D_j(t)$ of existing assets. We therefore write
\begin{equation}
    \Delta Y_j(t) \equiv Y_j(t) - Y_j(t-1) = \frac{1}{2}\left[ \cY_j(t)X_j(t) + \cY_j(t-1)X_j(t-1)\right] - D_j(t).
\end{equation}
$\cY_j(t)X_j(t)$ is the average electricity generated by $X_j(t)$, dependent on the specific capacity factor $\cY_j(t)$. The factor of $1/2$ accounts for the fact that $X_j(t)$ is built throughout the year, assuming it is built halfway between each year.
The degradation is taken as
\begin{equation}
    D_j(t) = \delta * Y_j(t),
\end{equation}
where we assume a fixed annual decline in power output $\delta$. For wind, this degradation is around 1.6\% annually, \supercite{staffellHowDoesWind2014} while for solar, it is around 1.2\% annually.\supercite{kimReviewDegradationPhotovoltaic2021}
If we consider the weighted average capacity factor $\cY'_t$ for newly built resources in years $t$ and $t-1$, as
\begin{equation}
    \cY_j^{'}(t) = \frac{\cY_j(t) X_j(t) + \cY_j(t-1)X_j(t-1)}{X_j(t) + X_j(t-1)}.
\end{equation}
Instead of investigating $\cY_j(t)$, we consider $\cY_j^{'}(t)$ since it is mathematically easier, and any long-term trend in $\cY_j(t)$ will translate into $\cY_j^{'}(t)$. We write
\begin{equation}
    \cY_j^{'}(t) = \frac{2 \Delta_1 Y_j(t)}{X_j(t) + X_j(t-1)} + \frac{2 D_j(t)}{X_j^k(t) + X_j(t-1)},
\end{equation}
where $\Delta_1$ indicates that we are taking the difference between time $t$ and $t-1$.

\begin{figure}
    \centering
    \includegraphics[clip, trim = 2.5cm 0.5cm 2.5cm 0.5cm ,width=\textwidth]{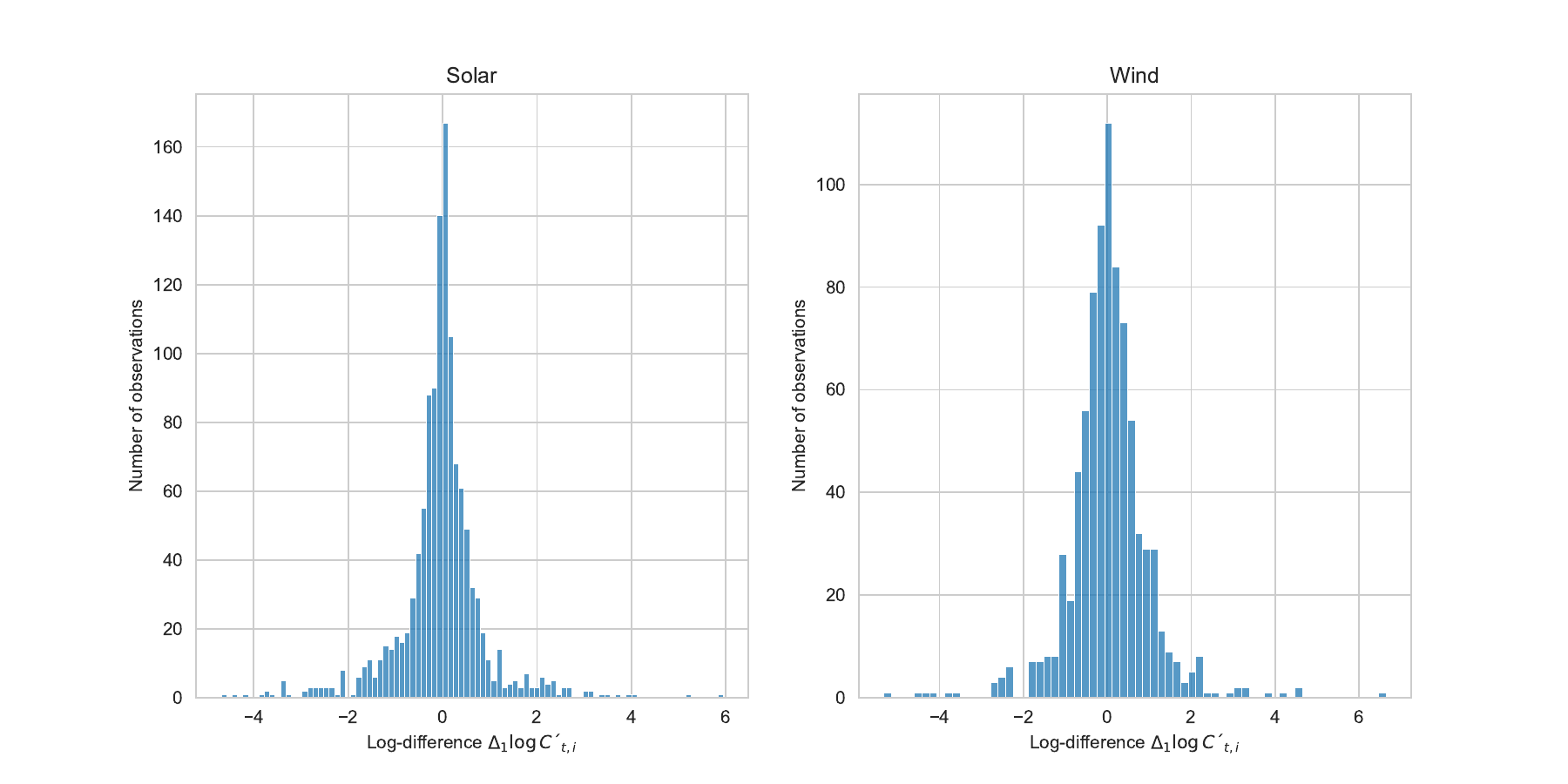}
    \caption{\textbf{Observed distribution for the normalized log-difference of the capacity factor for newly installed capacity.} Blue bars show the distribution of observed values for $\Delta_1 \log \cY_j^{'}(t) = \log \cY_j^{'}(t) - \log \cY_j^{'}(t-1)$. For missing country-year data points, we take the next best higher-order difference in $\log \cY_j^{'}(t)$ and normalize by the number of years included.}\label{fig:ws_df}
\end{figure}

Based on this formulation, we can now test for trend-stationarity in $\cY_j^{'}(t)$. We assume a multiplicative lognormal error since the electricity generation can not be negative, and it is more natural to assume the error to scale with the deployed capacity. We also consider the autocorrelation between $\cY_j^{'}(t)$ and $\cY_j^{'}(t-1)$ since both values depend on the electricity produced in year $t-1$. This means that
\begin{equation}
    \log \cY_j^{'}(t) = \log \left[ \cY_j(t)X_j(t) + \cY_j(t-1)X_j(t-1) \right] - \log \left[ X_j(t) + X_j(t-1)\right] + \epsilon_t + \rho_t * \epsilon_{t-1},
\end{equation}
where $\epsilon_t \sim \cN(0, \sigma_t^2)$ and $\rho_t$ is the autocorrelation parameter. By the law of large numbers, this serves as a reasonable approximation of the mean $0$. We can use this expression to consider the normalized difference between the logarithms of $\cY_j^{'}(t)$ and $\cY_j^{'}(t+\tau)$. Under the null hypothesis, the difference is centered around $0$, i.e.
\begin{equation}
    H_0\text{: } \Delta_\tau \log \cY_j^{'}(t) \equiv \frac{1}{\tau} \left( \log \cY_j^{'}(t) - \log \cY_j^{'}(t+\tau) \right) \sim \cN(0, \frac{\sigma^{'2}_t + \sigma^{'2}_{t-2}}{\tau^2}),
\end{equation}
where we discard the autocorrelation term $\rho$ in place of $\sigma'$.
For $\tau = 1$, this simply corresponds to taking the first difference in the logarithm of our adjusted capacity factor. Maximizing the data available to us, we consider all first differences for all countries. The distribution of $\Delta_{1} \log \cY_j^{'}(t)$ (taking $\tau = 1$) then behaves as a Gaussian mixture.
Figure \ref{fig:ws_df} shows the observed distributions. Visually, we find that both distributions are symmetric and centred around zero. Using the result of Bakinov \& Székely, \supercite{bakirovStudentTtestGaussian2006} we can apply a one-sided Student t-test to find that both for solar and wind, the deviation from zero is not significant at 90\% confidence, with p-values of 61\% for solar and 16\%) for wind.
We note that this p-value provides an upper bound to the true statistical significance since we have not considered the likely high level of autocorrelation between the $\log \cY_j^{'}(t)$.

This analysis shows that, while there has been some increase in the capacity factor in some countries, there is no statistically significant global trend towards higher capacity factors. This means that for both solar and wind we should forecast future national capacity factors as trend-stationary. For solar, this further means that global changes in the inverter loading ratio (ILR)\supercite{irenaRenewablePowerGeneration2024, bolingerUtilityScaleSolar20222022,deschampsOptimizationInverterLoading2019} have not had a significant impact on the capacity factor.

The global average capacity factor, as in the average across national capacity factors, may still appear to be increasing. This analysis shows that that is spurious. Any change in the global average capacity factor is most likely driven by sample bias. Over time, more national capacity factor data becomes available. When we add this to the global average, the sample average changes even though national capacity factors are stationary.

There are two important caveats to our statistical analysis. Firstly, some countries may still exhibit changes in capacity factor due to regulation or land ownership. If better geographical sites are made available to renewable sources, that could improve the capacity factor. Vice versa, as the best sites fill up, the capacity factor could decrease. Our analysis shows that neither of these effects have had a large impact on national capacity factors so far. Secondly, the growing use of tracking modules and future use of bifacial and multi-junction modules may have some impact on the ILR and solar capacity factor.\supercite{solarworldCalculatingAdditionalEnergy2021, desouzasilvaImpactBifacialModules2021} These technologies allow solar assets to capture more sunlight over the course of a day, which could drive an increase in capacity factor. So far, this has not been the case but that might change if these technologies decrease in cost.

\subsubsection{Capacity Factor model comparison}
As we saw in the previous section, we can use the actual electricity generation and capacity to estimate the capacity factor. Alternatively, we can estimate the capacity factor from the LCOE data. In this section, we describe and compare both of these methods to identify the one best suited to our forecast application. Since we are using the electricity-based capacity factor to harmonize some of our wind LCOE data, the test is less meaningful than it is for solar. For completeness, we nevertheless report on both technologies here.

For the electricity model, we can use the average annual capacity factor (excluding outliers with an annual capacity factor of zero or exceeding 100\% utilization). Specifically, we can use the unbiased estimator
\begin{align}\label{eqn:elec_cf}
   \text{Electricity Model: } \hat{\cY}_j = \frac{1}{m} \sum_{t=1}^{m} \frac{Y_j(t)}{Z_j(t)},
\end{align}
where $m$ is the number of observations.\footnote{Unlike before, we no longer need to worry about correcting for capacity growth and asset deterioration since we have already shown that capacity factors are approximately constant over time.}

For the LCOE-based model, we can use Equation \eqref{eqn:LCOE_E_app}. We have seen in Section \ref{app:model_LCOE} that $\log \cE$ follows an i.i.d.\ symmetric distribution. If we combine $\log\cE$ and $\log\cY$ into a single random variable we can also estimate the capacity factor as
\begin{align}\label{eqn:fe_cf}
   \text{Fixed-effect Model: } \log \hat{\cY}_j = \frac{1}{T_j} \sum_{t} \left( \log \left( I_j(t) * \cC_j(t) + O_j(t)\right) - \log \cL_j(t)\right)
\end{align}
where we sum over the $T_j^k$ available data points for country $j$ and technology $k$.

To choose between the electricity and fixed-effect models, we compare out-of-sample forecast errors in both models using backtesting. Since we are ultimately interested in forecasting the LCOE we compare our capacity-factor forecasts to the capacity factors observed in the LCOE:
\begin{equation}
     \text{Observed data: } \log\cY_j(t) = \log \left( I_j(t) * \cC_j(t) + O_j(t)\right) - \log \cL_j(t).
\end{equation}
We consider different training data in the form of moving average windows of varying sizes to estimate the capacity factor $\hat{\cY}_j$ for wind and solar based on the two models. 

\begin{figure}
    \centering
    \includegraphics[clip, width=\textwidth]{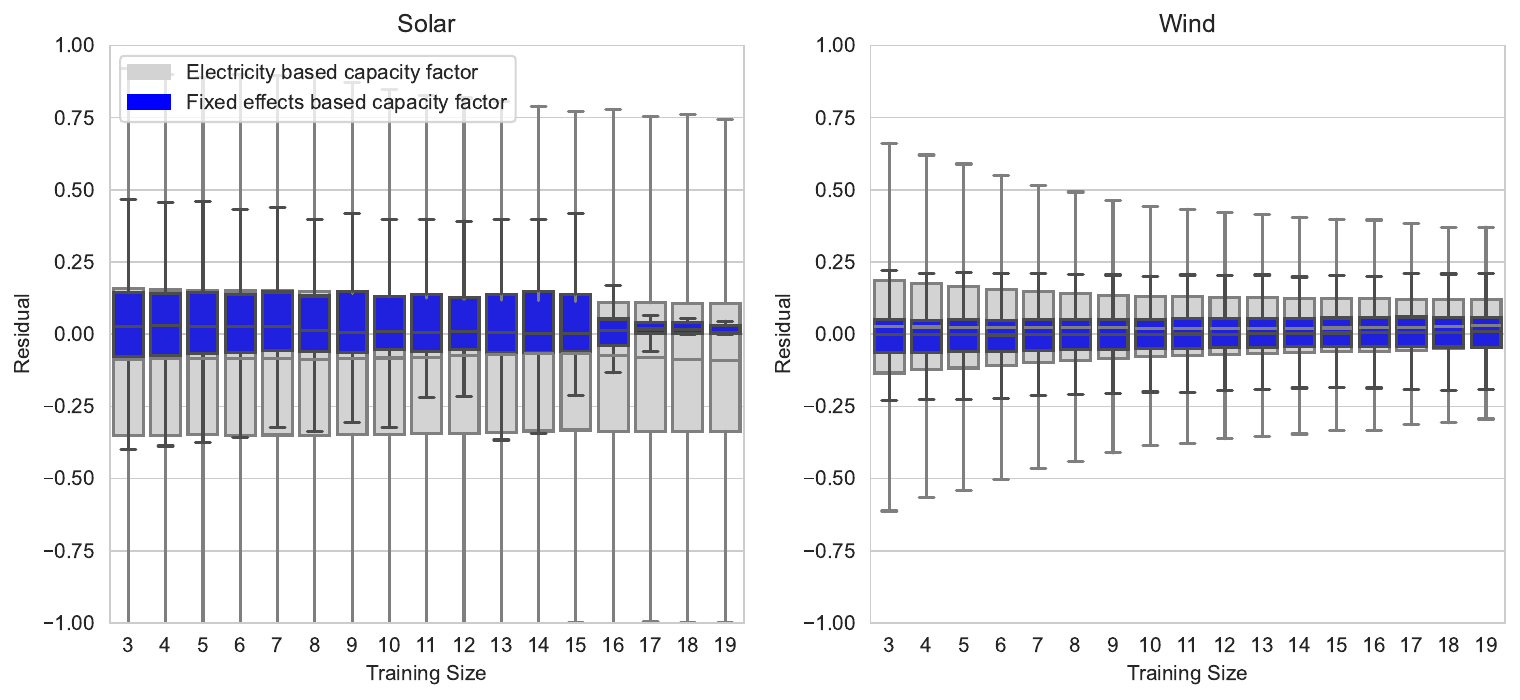}
    \caption{\textbf{Out-of-sample forecasts error for the capacity factor based on electricity generation and capacity, and a fixed-effect model for solar and wind.} 
    We compare the out-of-sample forecast errors for the capacity factor in solar and wind. The x-axis shows the number of training years used in a moving window approach. The y-axis shows the error of the resulting forecasts. The box indicates the IQR, the whiskers extend this to 1.5 times the IQR. The gray boxes represent the electricity-based estimation, and the blue boxes represent the fixed effects estimation.
    We see that the fixed-effects forecasts outperform the electricity-based forecasts in both technologies and all training sizes.
    }\label{fig:cf_perf_comp}
\end{figure}

The results are displayed in Figure \ref{fig:cf_perf_comp}. We first notice that for both models, the error distribution is centred around zero. We also see that beyond 10 training data points, the errors do not decrease significantly as the training size increases. This means that we can estimate the capacity factor well, even if there are only a few data points available. Most importantly, the error is smaller for the fixed-effects model than it is for the electricity-based model.
We, therefore, use the fixed-effects model in both technologies.

\subsection{Operating \& Maintenance cost forecasts}\label{app:model_onm}
{\bf Summary:}
\emph{Based on the limited data available, we show that operating \& maintenance costs are approximately constant relative to the total investment costs. This means that we can use a constant to forecast the relative operating \& maintenance costs.}
\newline

\begin{figure}
    \centering
    \includegraphics[width=\textwidth]{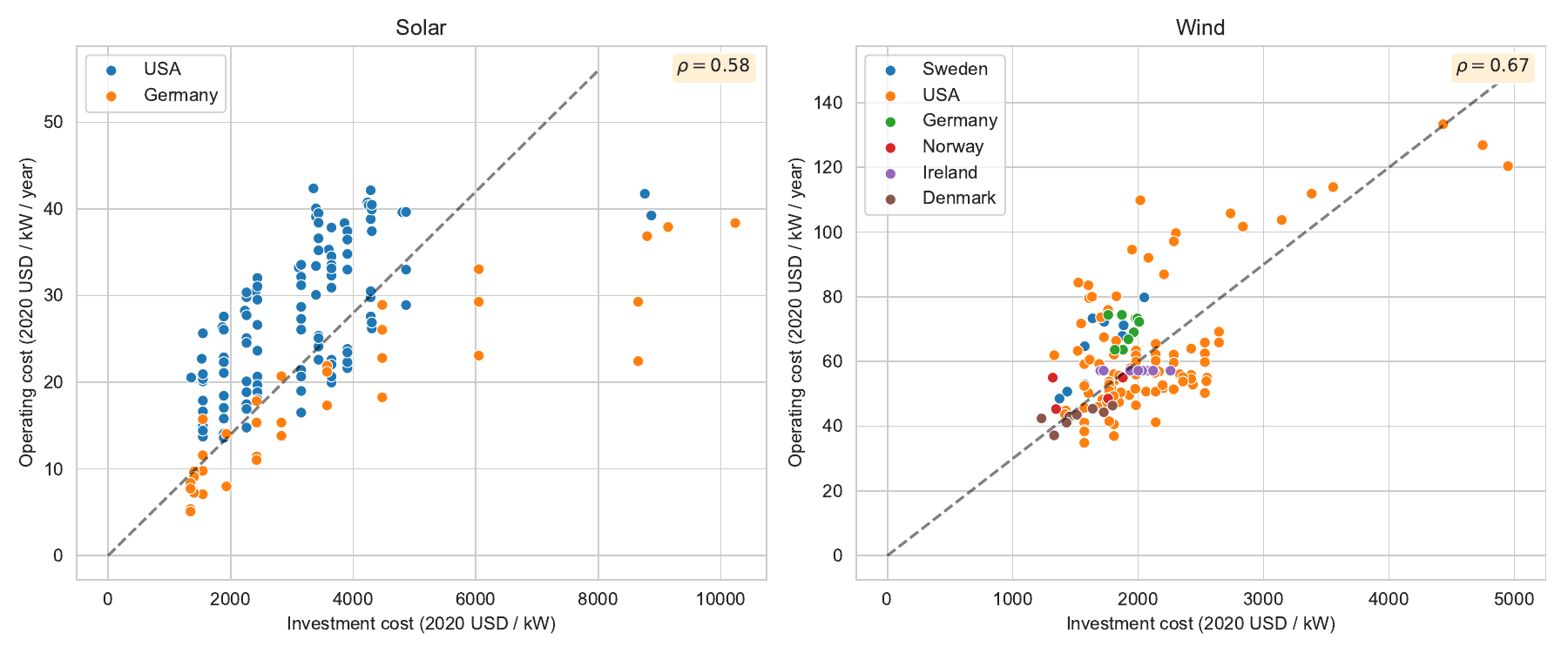}
    \caption{\textbf{Annual operating costs for wind and solar electricity vs investment costs.} Total investment costs are plotted against annual operating costs per kW capacity for solar \textbf{(left)} and wind \textbf{(right)}. The dotted lines indicate a fixed 3\% relationship for wind and 1\% for solar. Solar displays a Pearson correlation coefficient of $\sim0.58$, wind $\sim0.67$. \supercite{rivaIEAWindTCP2018, bolingerUtilityScaleSolar20222022, wiserAssessingWindPower2019}}\label{fig:onm}
\end{figure}

\noindent
Figure \ref{fig:onm} shows O\&M costs against the investment costs. National O\&M cost is very limited, with only two countries available for solar (Germany, US)\supercite{steffen20a, irenaRenewablePowerGeneration2024} and six countries available for wind.\supercite{rivaIEAWindTCP2018, bolingerUtilityScaleSolar20222022, wiserAssessingWindPower2019} Statistical significance is therefore a key limitations.

We see that the observations are distributed symmetrically around the 3\% O\&M share for wind and 1\% O\&M share for solar. There is no clear differentiation between countries. Investment and O\&M costs correlate positively with Pearson correlation coefficients of $\sim0.58$ for solar and $\sim0.67$ for wind.

\subsubsection{Stationarity relative to total investment cost}
We hypothesize that the relative O\&M cost $\gamma_j = O_j / I_j$ is approximately constant across time and countries $j$.
Since both investment and O\& M costs differ across countries, we consider the average O\& M cost in each country and year. This gives us a national time series of relative O\& M costs. For each time series, we compute the first differences $\Delta\gamma_j = \gamma_j(t) - \gamma_j(t-1)$.
We confirm that there is no significant trend in $\gamma_j$ by applying a two-sided t-test. At p-values of 66\% for wind and 9.4\% for solar, neither show significant trends at 95\% confidence.

For wind, $\gamma_j$ is very similar across countries. For solar, $\gamma_j$ differs between Germany and the US. As evident in Figure \ref{fig:onm}, this is primarily driven by 8 US assets with investment costs exceeding USD 6 mn per kW. If we exclude this data, the $\gamma_j$ values are again very similar. Given that solar investment costs are decreasing in the US and elsewhere, these data points are not representative of future O\&M costs. Instead, we consider $\gamma_j$ to be equivalent and stationary across countries and time for both solar and wind.

For our forecasts, we thus apply a fixed relative O\&M cost $\gamma_i(t) = \gamma$ for wind and solar. This is consistent with the Steffen et al.\supercite{steffen20a} who estimate learning rates to forecast O\& M costs. Here, O\&M costs follow Wright's law, decreasing with cumulative electricity production at a similar rate to total investment costs. If both O\&M costs $O$ and total investment costs $I$ decline at similar rates, the relative costs $\gamma = O/I$ is approximately constant.

\subsubsection{Operating \& Maintenance cost model validation}
Next, we estimate the average relative O\&M share $\log \hat{\gamma}$ and associated variance $\hat{\sigma}$ from the data. We assume that the logarithm of the relative O\&M share follows a normal distribution:
\begin{equation}\label{eqn:onm_model}
    \log \frac{O_{ij}}{I_{ij}}  \sim \cN\left(\log \hat{\gamma}, \hat{\sigma}^{2}\left(\frac{1}{m} + 1\right)\right),
\end{equation}
where $m$ is the number of observations.

Based on this, we can estimate $\log \hat{\gamma}$ as the sample mean and $\hat{\sigma}$ as the sample variance of $\log \frac{O_{ij}}{I_{ij}}$:
\begin{align}
    \log \hat{\gamma} &= \frac{1}{m} \sum_{i=1}^{m} \log \frac{O_{ij}}{I_{ij}}, \\\label{eqn:OnM_para}
     \hat{\sigma}^{2} &= \frac{m^{k-1}}{m-1}\sum_{i = 1}^{m} \left( \log \frac{O_{ij}}{I_{ij}} - \log \hat{\gamma}\right)^2.
\end{align}
Our stochastic model of $\hat{\gamma}$ does not include the asset-specific exogenous shocks $\epsilon_i$. Since we are modelling the national weighted average relative O\&M costs, these shocks are balanced out by the law of large numbers.
We estimate $\gamma \approx 3.1\%$ for wind and $\gamma \approx 0.8\%$, similar to earlier studies. We estimate the sample variance $\sigma \approx 0.13$ for solar and $\sigma \approx 0.05$ for wind.

We validate this approach out-of-sample using a cross-sectional bootstrapping approach. Each country-year combination constitutes one data point. We repeatedly split this dataset randomly into disjoint training and testing data, with $m=10$ data points in the training set and the remainder in the test set.
We use the training data to estimate the model parameters $\hat{\gamma}$ and $\hat{\sigma}$
For each observation in our testing data, we calculate the model error:
\begin{equation}
    \cE_i = \log \frac{O_{ij}}{I_{ij}} - \log \hat{\gamma}.
\end{equation}\label{eqn:onm_student}
Since our training-samples will have higher or lower variance $\hat{\sigma}$, we normalize the error by the estimated standard deviation
\begin{equation}
    \frac{1}{\sqrt{\frac{1}{m} + 1}} \frac{\cE_i}{\hat{\sigma}} \sim t(m-1).
\end{equation}
We expect the resulting normalized error distribution to follow a Student t-distribution with $m-1$ degrees of freedom. 

\begin{figure}
    \centering
    \includegraphics[clip, trim = 2.5cm 0cm 2.5cm 0.5cm ,width=\textwidth]{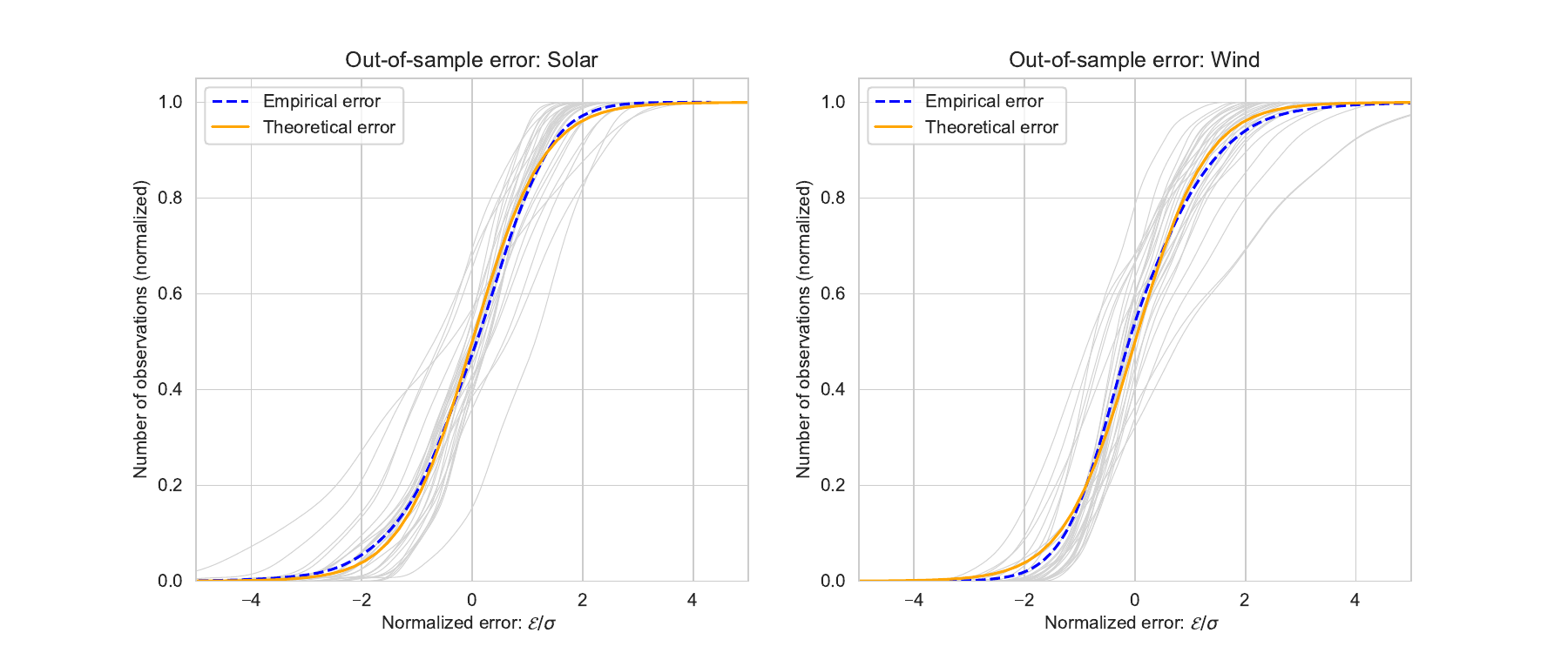}
    \caption{\textbf{Observed out of sample error of the relative O\&M share $\log \gamma$ compared to the theoretical error.} 
    The gray lines show the normalized out-of-sample errors of the observed relative O\&M costs $\log \gamma$. The blue line shows the average over the gray lines.
    The orange line shows the theoretical distribution of the out-of-sample error, as per Equation \eqref{eqn:onm_student}. According to the error model, we expect the normalized out-of-sample error to follow a Student-t distribution. We observe a close match between the model and empirical data.
    }\label{fig:onm_out_sample}
\end{figure}

The result of this analysis is shown in Figure \ref{fig:onm_out_sample}.
We see that the t-distribution describes the out-of-sample error well and, thus, serves as an appropriate stochastic error model.

\section{IAM scenario comparison}\label{app:IAM}
In this section, we provide more detail to the IAM scenario comparison presented in the main methods section.
We first describe the data collection from the AR6 database.\supercite{byersedwardAR6ScenariosDatabase2022} In Section \ref{app:IAM_add}, we show additional country examples. In Section \ref{app:IAM_corr}, we show that if scenarios are corrected for the initial value, the bias in their forecasts is reduced substantially.

\subsection{Scenario data collection}
We construct an ensemble of scenarios from the AR6 database. \supercite{byersedwardAR6ScenariosDatabase2022} These scenarios are the ones that were considered for the sixth assessment report of the IPCC in 2022, although many scenarios were constructed in earlier years. Since not all of these scenarios allow for a fair comparison, we exclude the following subset in our analysis:
\begin{itemize}
    \item \emph{Failed Vetting} These scenarios are identified by the AR6 authors; their input parameters were not considered ``within reasonable ranges for the baseline period – primarily for indicators relating to emissions and the energy sector''
    \item \emph{Negative growth} 5 scenarios for solar and 20 scenarios for wind show negative growth in the operational solar and wind capacity.
    \item \emph{Negative cost} 5 scenarios report negative total investment costs for solar.
\end{itemize}
The resulting ensemble includes 741 scenarios for solar and 516 scenarios for wind. We use 2020 as the base year of our scenario comparison and account for inflation by scaling all cost data to 2020 USD (from 2010 USD as reported in the scenario database).

\subsection{Comparison of scenarios and our total investment forecasts}
As outlined in the main text, we compare the total investment cost $I$ for solar and wind  in 2040 in the AR6 scenarios to our forecasted values for $I$.
For the scenario costs, we take the total investment costs in 2040 directly from the AR6 database. For our forecasts, we make forecasts for $I$ conditionally on the global solar and wind capacity in 2040.

To make our forecasts, we perform three steps. 
\begin{itemize}
    \item Set the global capacity for solar and wind in 2040. We vary this number between the global capacity in 2020 (731 GW for wind, 713 GW for solar) and 25 TW for solar and 10 TW for wind.
    \item Interpolate linearly between the capacity in 2020 and 2040 to obtain a time series of future capacities.
    \item Conditional on the capacity build out between 2020 and 2040, make forecasts for the total investment cost using our model for turbines, modules and BOS costs.
\end{itemize}
This methods provides stochastic forecasts for the total investment costs in 2040, conditional on the global cumulative capacity.

As explained in the main text, we observe discrepancies between our forecasts and the IAM scenarios, where our median (point) forecasts is mostly below the IAM scenario costs.

\subsection{Initial value correction}\label{app:IAM_corr}
The reason for this discrepancy are the initial values of the total investment cost and global capacity used in the IAM scenario construction. Many IAMs are quickly outdated due to the rapid growth of renewable capacity and associated drop in technology costs. As a result, their 2020 costs, as published in 2022, are almost all too high.

\begin{figure}
    \centering
    \includegraphics[width=\linewidth]{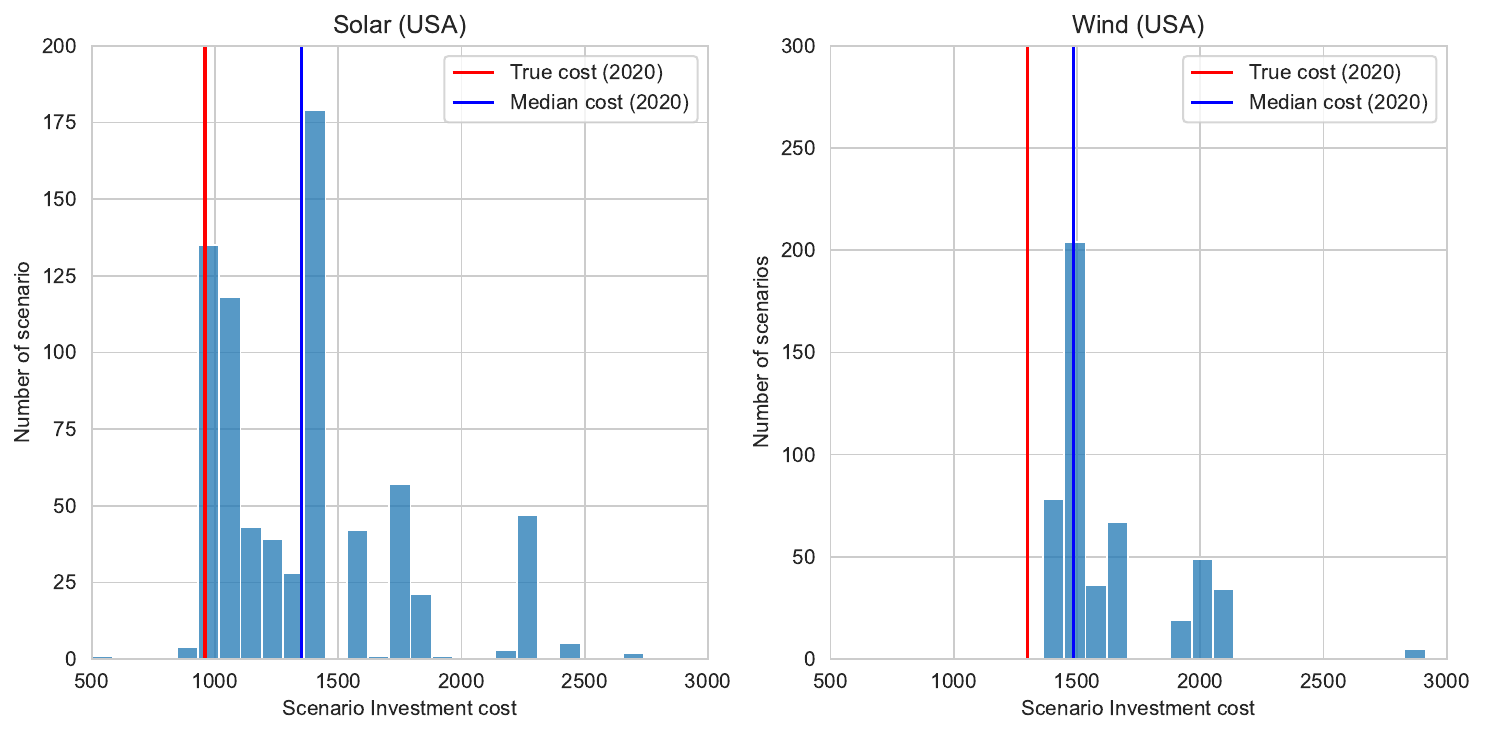}
    \caption{\textbf{Current solar and wind total investment cost for the US as reported in IAM scenarios}
    The blue bars show the solar (left) and wind (right) total investment costs in 2020, as reported by the AR6 scenarios with regionally differentiated capital costs. We exclude scenarios with costs exceeding 3000 USD/kW for better visibility. The blue line indicates the median cost of each technology over the scenario sample. The red line shows the true average solar and wind cost in 2020. We see that almost all scenarios show costs that are larger than what was reported in 2020.}
    \label{fig:IAM_capex}
\end{figure}

Figure \ref{fig:IAM_capex} displays the reported solar and wind capacity for 2020 against the actual data. We see that in both cases, the investment costs assumed in the IAMS are generally higher than what actually occurred. For solar, Remind 2.1 is the only model that reports costs below what was observed. 

\begin{figure}
    \centering
    \includegraphics[width=\textwidth]{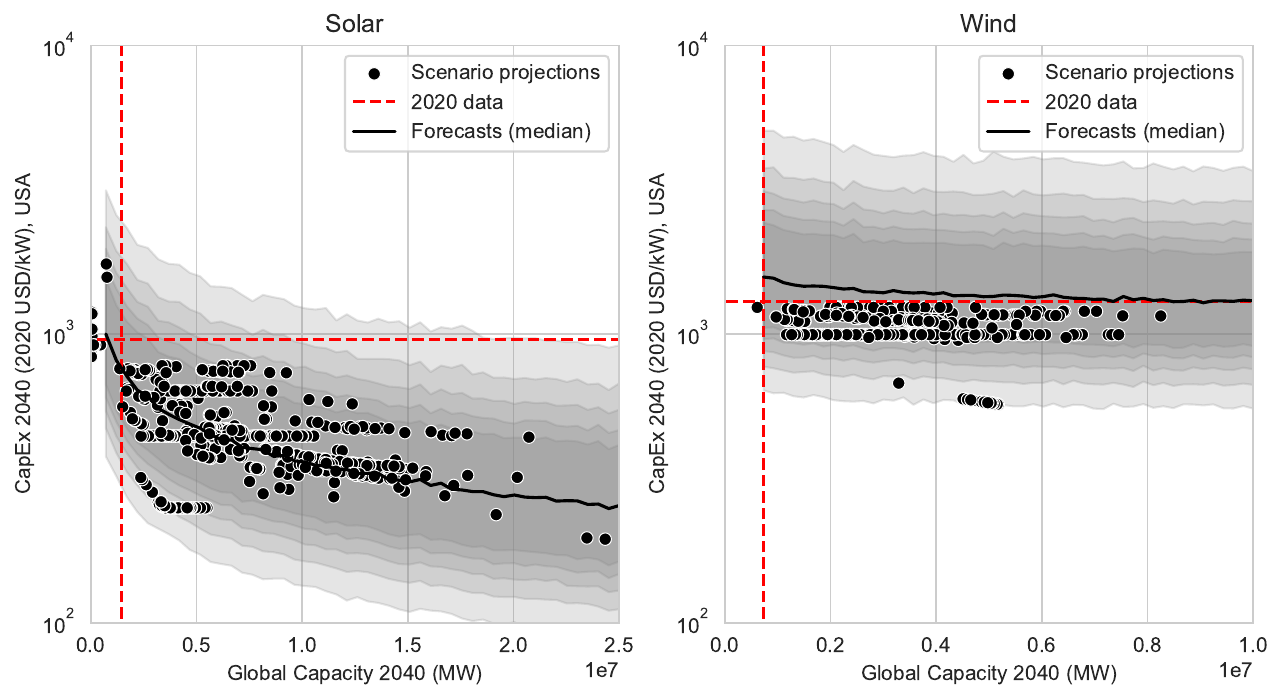}
    \caption{\textbf{Adjusted Solar and wind total investment cost projections reported in IAM scenarios against probabilistic projections from our model.}
    The dots show the total investment cost projections for the US in different scenarios, adjusted for different initial costs. The x-axis shows the associated global solar / wind capacity. We compare them with the investment cost projections obtained from our model. The areas show 50-90\% confidence intervals in 10\% steps. The black line shows the median cost from our model. The red dotted lines show the global capacity and investment cost in the US for 2020.
    In both cases, the new investment cost lowers the expected future costs across almost all scenarios. We see that for solar, the scenarios now align well with our median forecast. For wind, the scenarios show costs that are slightly lower than what we expect. They are, however, still mostly within the 50\% confidence interval.
    }\label{fig:IAM_proj_USA_adj}
\end{figure}

We demonstrate that this initial value is the primary source of the difference between scenarios and our forecasts by re-scaling the IAM scenarios to the same initial values. 
\begin{itemize}
    \item For each scenario calculate the ratio of actual costs in 2020 and scenario costs in 2020,
    \item Multiply the scenario costs in 2040 with the resulting multiplicative factor,
    \item Repeat the previous comparison between our forecasts and the adjusted 2040 scenario costs.
\end{itemize}
The results of this for the US are shown in Figure \ref{fig:IAM_proj_USA_adj}. We see that for solar, the scenario costs now align much better with our median forecast. Now, only 25\% of the scenarios exceed our point forecast. For wind, we observe that scenario costs are slightly below our point forecast but are still mostly within the 50\% confidence interval.

One reason for this discrepancy in 2020 scenario data is that many scenarios in the AR6 database were generated before 2020. For those scenarios, the 2020 cost and deployment are not based on actual data but on projections made in earlier years. In that case, even small biases in the early years of a scenario can lead to significant differences in the respective future cost projections.
These challenges with IAMs have been noted previously, for example, Creutzig et al., \supercite{creutzig17} and Way et al. \supercite{way22}. Consequentially, our results emphasize the need for more stringent scenario validation and more frequent updates to the scenarios as the energy transition evolves.

\section{Estimated model parameters}

The model parameters that were calibrated in this study are reported separately.

\section{Additional Figures}\label{app:IAM_add}
\begin{figure}[H]
    \centering
    \includegraphics[width=0.9\textwidth]{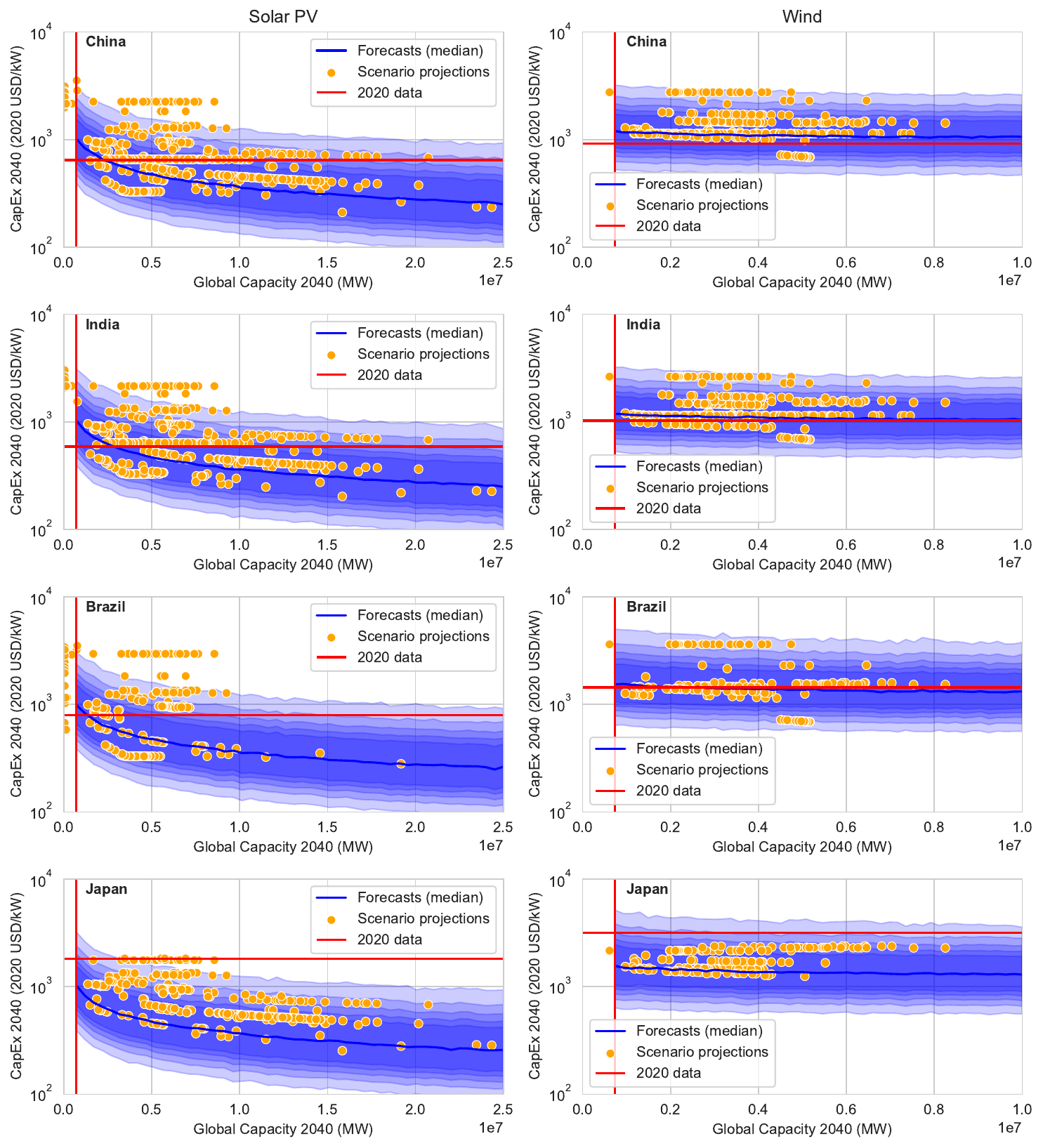}
    \caption{\textbf{Solar (left) and wind (right) total investment cost projections reported in IAM scenarios against probabilistic projections from our model. (not adjusted to initial cost and capacity)}
    The orange dots show the total investment cost projections for the different countries in different scenarios. The x-axis shows the associated global solar / wind capacity. We compare them with the investment cost projections obtained from our model. The blue areas show 50-90\% confidence intervals in 10\% steps. The blue line shows the median cost from our model. The red lines show the global capacity and investment cost in each country for 2020.
    For wind in Brazil, the IAM projections generally align well with the median forecast. For the other countries and solar, we observe some discrepancies. In particular, many IAM scenarios are above the expected future costs, although they generally fall within the 50\% confidence interval of our forecasts. In both cases, the probabilistic forecasts show significantly more uncertainty around the median forecasts than the IAM projections represent.
    }\label{fig:IAM_capacity}
\end{figure}

\begin{figure}
    \centering
    \includegraphics[width=0.9\textwidth]{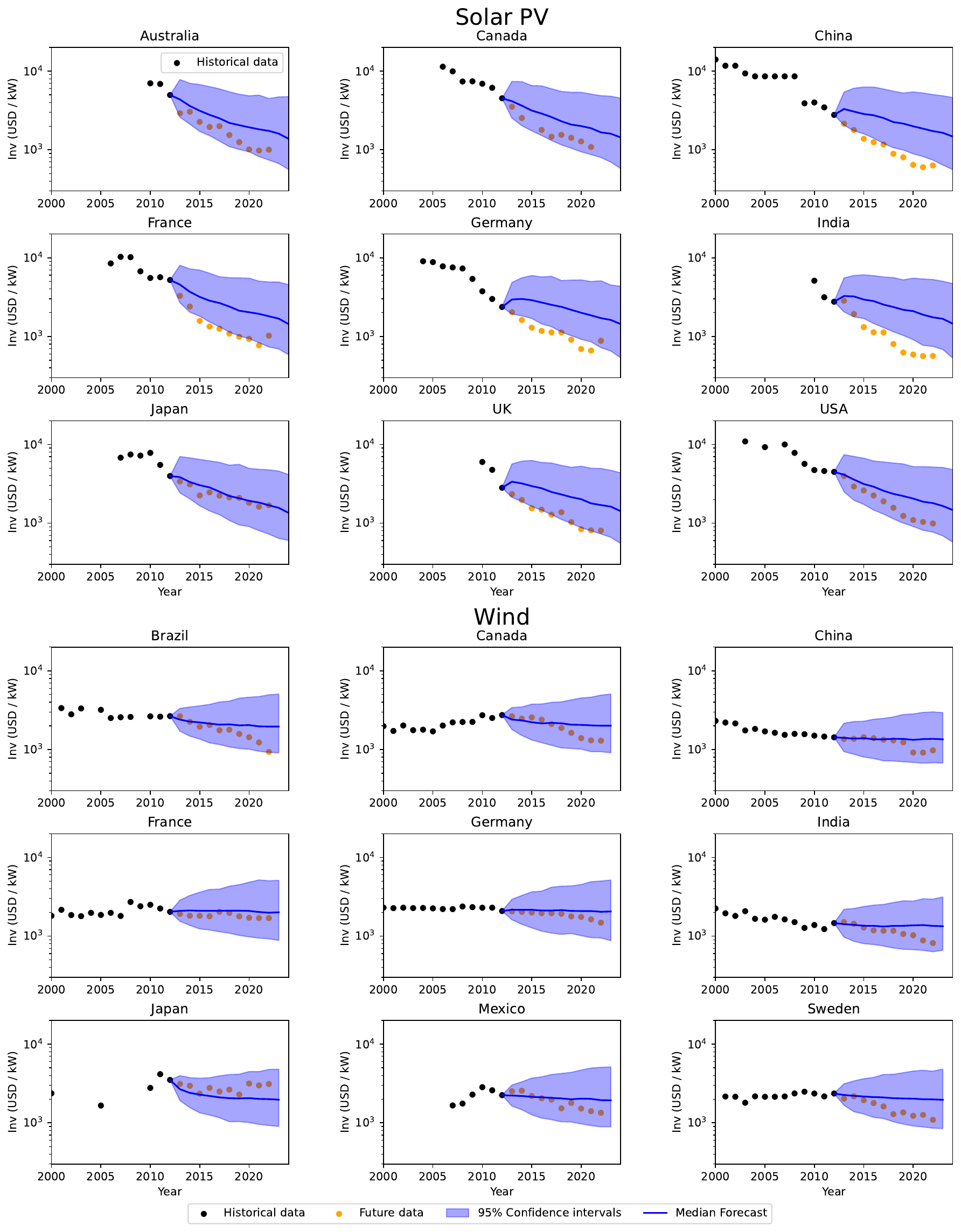}
    \caption{\textbf{Investment cost forecasts compared to true data.} 
    We illustrate the Investment forecast model by comparing the stochastic model outcomes to the actual data for selected countries (in-sample). The top 9 panels show national forecasts for solar, and the bottom 9 panels show national forecasts for wind. The y-axes show the national Investment cost on a logarithmic scale. The black dots indicate historical data, and the orange dots indicate future data. The blue shaded areas show the 95\% confidence interval of our forecasts, and the blue line is the median forecast. 
    With one exception in India, all future data is captured by the 95\% confidence interval. For solar, we see that our forecasts are too high in all countries.
    }\label{fig:example_inv}
\end{figure}
\clearpage

\clearpage

\newpage
\small
\printbibliography[title={Supplementary References}]
\end{refsection}

\end{document}